\documentclass[twocolumn,aps,superscriptaddress,showpacs,floatfix,prd,noshowpacs]{revtex4-2}
\usepackage[body={19.cm,24.cm}]{geometry}     
\usepackage{mathrsfs}
\usepackage{amssymb}
\usepackage{amsmath}
\usepackage{array}
\makeatletter
\@addtoreset{equation}{section}
\renewcommand{\theequation}{\arabic{section}.\arabic{equation}}
\makeatother
\usepackage{graphicx}
\usepackage[normalem]{ulem}
\usepackage{xcolor}
\usepackage{bm}
\usepackage{longtable}
\usepackage{slashed}
\usepackage{enumitem}
\usepackage{empheq}
\usepackage{booktabs}
\usepackage{multirow}

\usepackage{titlesec}
\renewcommand{\thesection}{\Roman{section}}
\titleformat{\section}{\small\bfseries\centering}{\thesection.}{0.5em}{}

\usepackage[T1]{fontenc}
\usepackage[utopia]{mathdesign}

\usepackage[breaklinks=true]{hyperref}
\hypersetup{
  colorlinks=true,
  citecolor=magenta,
  linkcolor=black,
  urlcolor=teal,
}

\renewcommand\sout{\bgroup \color{red} \ULdepth=-.5ex \ULset}

\renewcommand{\v}[1]{\textbf{#1}}
\renewcommand{\rm}[1]{\textrm{#1}}
\renewcommand{\d}{\mathrm{d}}

\usepackage{tikz}

\definecolor{lime}{HTML}{A6CE39}
\DeclareRobustCommand{\orcidicon}{
	\begin{tikzpicture}
	\draw[lime, fill=lime] (0,0) 
	circle [radius=0.16] 
	node[white] {{\fontfamily{qag}\selectfont \tiny ID}};
	\draw[white, fill=white] (-0.0625,0.095) 
	circle [radius=0.007];
	\end{tikzpicture}
	\hspace{-2mm}
}
\foreach \x in {A, ..., Z}{%
	\expandafter\xdef\csname orcid\x\endcsname{\noexpand\href{https://orcid.org/\csname orcidauthor\x\endcsname}{\noexpand\orcidicon}}
}

\DeclareSymbolFont{cmsymbols}{OMS}{cmsy}{m}{n}
\DeclareSymbolFontAlphabet{\mathcal}{cmsymbols}

\begin{document}

\title{Extracting the Central Sound Speed of Neutron Stars through Intrinsic Stellar Structure Scaling\\ with Reduced Equation-of-State Model Dependence}

\author{Jian-Hao Shi\orcidD{}}
\affiliation{Department of Physics and Center for Field Theory and Particle Physics, Fudan University, Shanghai 200438, China} 
\affiliation{Shanghai Research Center for Theoretical Nuclear Physics, NSFC and Fudan University, Shanghai 200438, China}
\author{Bao-Jun Cai\orcidA{}}\email{bjcai@fudan.edu.cn}
\affiliation{Shanghai Research Center for Theoretical Nuclear Physics, NSFC and Fudan University, Shanghai 200438, China}
\affiliation{Key Laboratory of Nuclear Physics and Ion-beam Application (MOE), Institute of Modern Physics, Fudan University, Shanghai 200433, China} 
\author{Yu-Gang Ma\orcidC{}}\email{mayugang$@$fudan.edu.cn}
\affiliation{Shanghai Research Center for Theoretical Nuclear Physics, NSFC and Fudan University, Shanghai 200438, China}
\affiliation{Key Laboratory of Nuclear Physics and Ion-beam Application (MOE), Institute of Modern Physics, Fudan University, Shanghai 200433, China} 
\affiliation{College of Physics, East China Normal University, Shanghai 200241, China}

\date{\today}

\newcommand{\x}{\mathrm{X}}
\newcommand{\y}{\mathrm{Y}}
\newcommand{\hr}{\widehat{r}}
\newcommand{\hP}{\widehat{P}}
\newcommand{\heps}{\widehat{\varepsilon}}
\newcommand{\hrho}{\widehat{\rho}}

\begin{abstract}

Inferring the properties of neutron star (NS) cores from observations remains sensitive to the assumed equation of state (EOS) description: Different dense matter models can satisfy the same nuclear and astrophysical constraints while predicting substantially different properties in the NS core. We demonstrate this ambiguity using the nonlinear Walecka model with a modified $\omega$-meson self-interaction term as an illustrative example, constructing several EOS families and independently generating $10^3$ EOSs for each family under the same constraints. The resulting families exhibit markedly different high-density behavior and core properties, demonstrating that current nuclear and astrophysical constraints do not yet uniquely distinguish among different EOS descriptions. Consequently, core properties inferred within a specific EOS framework may retain substantial information about the assumed model rather than uniquely reflecting intrinsic properties of NS matter. In particular, we show that higher-order $\omega$ self-interactions provide a mechanism for generating interior sound speed peaks within a purely nucleonic framework, illustrating how different interaction prescriptions can produce qualitatively distinct sound speed profiles even under the same observational constraints. Using the intrinsic and perturbative analysis of the dimensionless Tolman--Oppenheimer--Volkoff equations (IPAD-TOV), we derive an estimator $\Theta(\x,\Psi)$ for probing the central sound speed squared (SSS) $s_{\rm c}^2$ without specifying an input EOS, where $\x\equiv P_{\rm c}/\varepsilon_{\rm c}$ is the central EOS-parameter, with $P_{\rm c}$ and $\varepsilon_{\rm c}$ denoting the central pressure and energy density, respectively, and $\Psi$ characterizes the response of the stellar mass to the central energy density. The estimator follows from the stellar structure equations themselves, providing a unified analytical basis for comparing central properties predicted by different EOS descriptions. As a representative application, we apply this framework to canonical $1.4\,M_\odot$ NSs, for which $\Theta$ exhibits tight linear correlations with $s_{\rm{c}}^2$ across the different EOS families, with coefficients of determination of about $0.94$-$0.99$. Remarkably, despite their distinct underlying interactions and high-density behavior, these family-dependent relations converge toward a common optimum. Minimizing their dispersion yields a central SSS of approximately $0.523_{-0.044}^{+0.044}$ and a central EOS-parameter of $0.173^{+0.020}_{-0.018}$ when all families are included. When the family without the $\omega$ self-interaction is excluded, the corresponding values are $0.496_{-0.092}^{+0.092}$ and $0.167^{+0.030}_{-0.029}$, respectively. The inferred canonical NS radii are approximately $11.73^{+1.08}_{-0.92}\,\rm{km}$ and $11.92^{+1.76}_{-1.28}\,\rm{km}$ for the two cases, with the quoted uncertainties corresponding to the $95\%$ confidence intervals. Fitting all EOS samples together also yields a unified $\Theta$-$s_{\rm{c}}^2$ scaling consistent with both extractions, further supporting the consistency of the inferred central properties across the different EOS families. These results demonstrate that, even when current constraints cannot uniquely discriminate among EOS models or their predictions for the NS core, their shared stellar structure behavior can reveal central properties with reduced model dependence. More broadly, such scaling relations provide a bridge between stellar structure and the central EOS, offering a general route to separating robust physical information about NS cores from properties specific to the adopted EOS description and a complementary perspective on the inverse stellar-structure problem.

\end{abstract}

\maketitle

\fontdimen2\font=2.pt

\section{Introduction and Conclusions}

A neutron star (NS) does not reveal its interior uniquely. Different
descriptions of dense matter can produce remarkably similar stellar
masses, radii, and tidal deformabilities while implying substantially
different pressures, energy densities, and sound speeds in the core\,\cite{Lindblom1992,LattimerPrakash2016,Lattimer2021,Baym2018}.
This ambiguity lies at the heart of the inverse problem of NS physics\,\cite{Han2021,Patra2022,Soma2023,Sun2025,Ross2026,Ofengeim2024,Ferreira2025,
ZhangLi2023,Ventagli2025,Wouters2025,Greif2019,Chat25,Alar25,LiA25,Wat16}: {\textcolor[rgb]{0, 0., 0.}{\it Observations constrain the global stellar
structure, whereas the quantities of greatest interest often concern
matter deep inside the star. How much of an inferred core property is
therefore determined by the NS itself, and how much is inherited from
the equation of state (EOS) model used to interpret it?}}

This question has become increasingly important as the observational
information on NSs has accumulated. Measurements of massive
pulsars\,\cite{Demorest2010,Mauviard26,Antoniadis2013,Cromartie2020,Fonseca2021,
Romani2022,Linares2018}, tidal deformabilities from
gravitational wave (GW) events\,\cite{Abbott2017,Abbott2018,Abbott2020} such as GW170817, and NICER measurements of NS masses and radii\,\cite{Miller19,Riley19,Miller21,Riley21,Salmi22,
Salmi24,Ditt24,Choud24,Mauviard25,Miller26-a,Miller26},
together with terrestrial nuclear
constraints\,\cite{Danielewicz2002,Tsang2012,Danielewicz2014,
Dri21,Lovato22,Soren23,Xu2025CPL,Wang2025NST,Fang2026NST,Li2026PRL,MaYG_arxiv}, now
provide increasingly stringent tests of dense matter models. These
developments have substantially narrowed the range of viable EOSs,
but they have not uniquely determined the behavior of matter at the
densities reached in NS cores. In particular, different microscopic
models or phenomenological parametrizations can satisfy the same
available constraints while extrapolating differently beyond the
density regime where nuclear matter is comparatively well constrained\,\cite{Walecka1974,Chin1977,FreedmanMcLerran1977,Akmal1998,LP01,
Alford2008,LCK08,Oertel2017,Li2019,Ann23,CaiLi2025Review,Cui2025NST,Xu2026SCPMA,Xie2026SCPMA,Ding2024NST,XuWang2024SCPMA,Li2026CPL}.

The difficulty becomes particularly transparent through two closely
related quantities that characterize the stiffness of the dense matter
EOS: the speed of sound squared (SSS)\,\cite{CaiLi2025Trace}
\begin{equation}
    s^2={\d P}/{\d\varepsilon},
\end{equation}
and the EOS-parameter\,\cite{Cai2024Front,CaiLi2025Trace,
CaiLiMa2026phi,CaiLiMa2026X}
\begin{equation}
    \phi\equiv{P}/{\varepsilon}
    =
    \varepsilon^{-1}
    \int_0^\varepsilon\d\varepsilon'
    s^2(\varepsilon')=\langle s^2(\varepsilon)\rangle.
\end{equation}
These two quantities probe the EOS from different but complementary
perspectives. The SSS $s^2$ measures the local stiffness of the EOS,
namely how rapidly the pressure responds to an infinitesimal increase
in energy density. By contrast, $\phi$ encodes the accumulated response
of the pressure and may be viewed as an average stiffness of the EOS
up to the state under consideration\,\cite{CaiLiMa2026phi,CaiLiMa2026X,Saes22,Saes24,Marc24}. At the stellar center, this
distinction becomes particularly relevant:
$s_{\rm{c}}^2$ probes the local stiffness of matter at the highest
density reached inside the star, whereas
$\phi_{\rm{c}}=P_{\rm{c}}/\varepsilon_{\rm{c}}\equiv\x=\langle s^2(\varepsilon_{\rm c})\rangle$
characterizes the average stiffness accumulated toward the central
state. Together, $s_{\rm{c}}^2$ and $\x$ therefore provide two
complementary measures of the EOS in the NS core, one local and the
other integrated.

Yet neither the local quantities $s^2$ and $\phi$ nor their central values $s_{\rm{c}}^2$ and $\x$ are direct observables. This
creates a central difficulty in connecting astrophysical observations
to the physics of dense matter. In the conventional inference
procedure, one specifies an EOS or a parametrized EOS family, solves
the Tolman--Oppenheimer--Volkoff (TOV)
equations\,\cite{Tolman1939,Oppenheimer1939,Misner1973}, and uses observations to
constrain the parameters within the assumed EOS description. The
inferred $P_{\rm{c}}$, $\varepsilon_{\rm{c}}$, $s_{\rm{c}}^2$, and
$\x$ therefore inevitably retain sensitivity to the high density
behavior encoded in the adopted model space. This issue is
particularly relevant for $s^2$, whose density dependence, including
the possible development of an interior peak, can vary substantially
among different descriptions of the high density EOS. Consequently,
different EOS models can reproduce similar macroscopic stellar
properties while implying markedly different conditions in their
cores.

This observation suggests on the other hand that improving measurements alone does not
completely remove the conceptual ambiguity. One must also ask whether
there exists information about the NS core that is shared by different
EOS descriptions rather than being specific to any one of them. This
leads to the two central questions of the present work: {\textcolor{black}{\it How
different can the inferred core properties be among EOS models
satisfying the same nuclear and astrophysical constraints? And can
information intrinsic to the stellar structure equations themselves
be used to identify their common behavior and thereby reduce this
model dependence?}}

(1) We address the first question using a generalized nonlinear Walecka
model\,\cite{SW1986} as an illustrative example. The $\omega$-meson self-interaction
provides a particularly useful handle on the high density EOS: by
changing the nonlinear behavior of the repulsive vector field, one can
systematically modify the stiffness and sound speed of dense matter
while preserving the description of nuclear matter around saturation.
We exploit this freedom to construct several distinct EOS families,
with $10^3$ EOSs independently generated for each family under the
same nuclear and astrophysical constraints\,\cite{Tsang2009,WangChen2025,
ZhangChen2023,Huth2021,Tsang24,RichterLi2023,LiGrundler2024,GrundlerLi2025,GrundlerLi2026}. Despite satisfying the same adopted constraints, these
families exhibit substantially different high density behavior,
central properties, and sound speed profiles, making the underlying
model dependence explicit. We further derive their high density
asymptotic behavior analytically, revealing how the nonlinear
$\omega$ interaction changes the dominant contributions to the EOS.
At the finite densities relevant to NS interiors, the competition
between the growing vector repulsion and its nonlinear suppression can
also generate an interior sound speed peak within a purely nucleonic
description. The resulting ensembles provide a controlled setting
for exploring how different high density properties can remain
compatible with current constraints while leading to different
inferences about the NS core.

\renewcommand*\figurename{\small FIG.}
\begin{figure}[h!]
    \centering
    \includegraphics[width=0.78\columnwidth]{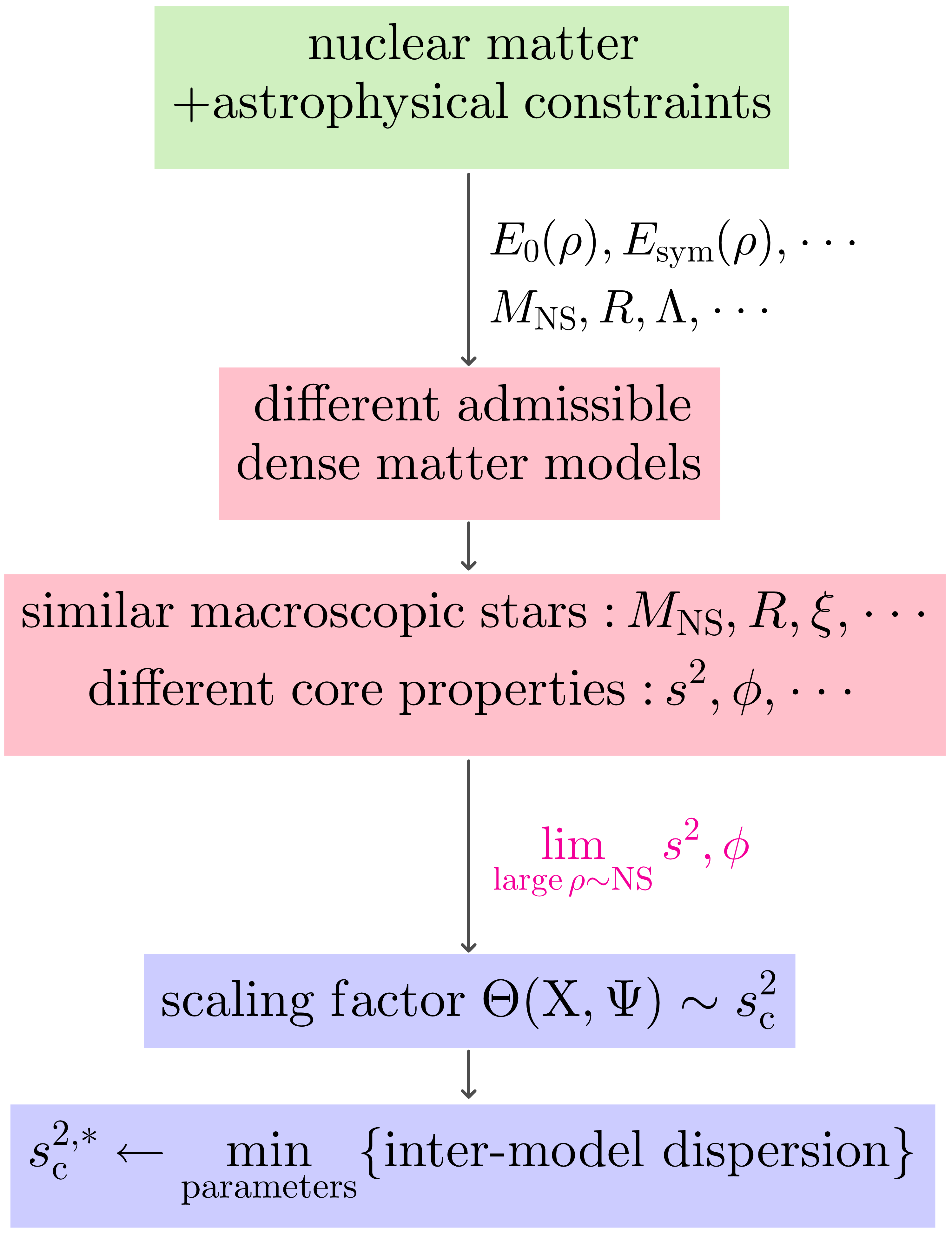}
    \caption{(Color online). Schematic illustration of the strategy
adopted in this work to extract the central SSS of NSs. Nuclear matter
and astrophysical constraints admit different dense matter models that
can describe similar macroscopic NS properties but predict
substantially different core properties. The IPAD-TOV analysis
provides the quantity $\Theta(\x,\Psi)$, which approximates the central
SSS and exhibits tight correlations with $s_{\rm{c}}^2$ within the
different model families. A common estimate $s_{\rm{c}}^{2,\ast}$ is
obtained by minimizing the dispersion among these family relations,
thereby reducing the sensitivity of the extraction to the underlying
dense matter model. See the text and FIG.\,\ref{fig:sss_scaling} for
details.}
    \label{fig:sc2-ext-sk}
\end{figure}

(2) The second question requires a different perspective. Instead of
asking which EOS model best reproduces the observations, we ask what
information about the core is encoded in the stellar structure
equations and persists across different EOS descriptions. The
intrinsic and perturbative analyses of the dimensionless TOV equations
(IPAD-TOV) provide a natural framework for this purpose\,\cite{CaiLiZhang2023ApJ,CaiLiZhang2023PRD,CaiLi2024SSS}.
By expanding the dimensionless stellar variables about the center,
the approach connects the local central EOS to the global response of
the stellar configuration. Previous applications have established relations between central EOS
quantities and stellar observables and explored the SSS and trace
anomaly\,\cite{CaiLi2025Review}. More recently, a tight
scaling was identified between the tidal deformability and a
combination of central EOS quantities, providing a direct approach for
probing the NS core through GW events\,\cite{Shi2026}.

In the present work, we retain a sub-leading contribution to the mass
expansion and derive an improved estimator $\Theta(\x,\Psi)$ for the
central SSS, where $\x=P_{\rm{c}}/\varepsilon_{\rm{c}}$ is the central EOS-parameter and\,\cite{CaiLiMa2026phi,CaiLiMa2026X}
\begin{equation}\label{def-Psi}
    \Psi\equiv
    2\frac{\d\ln M_{\rm{NS}}}{\d\ln\varepsilon_{\rm{c}}}
\end{equation}
is defined in terms of the logarithmic slope (log-slope) of the NS mass $M_{\rm{NS}}$ with
respect to the central energy density $\varepsilon_{\rm c}$. Importantly, $\Theta$ follows
directly from the dimensionless TOV equations rather than from a fit
to any particular nuclear EOS. For canonical $1.4\,M_\odot$ NSs, we find that the approximate estimator $\Theta$ is tightly correlated with the central SSS
$s_{\rm{c}}^2$ calculated directly from the full EOS within each EOS
family, despite their markedly different high density behaviors. The residual differences among these family relations then quantify the remaining sensitivity to the underlying dense matter model.

This structure suggests a way to turn the model diversity itself into
useful information.
To represent this diversity, we consider the power
index $q=3/2$, 2, 3, 4 and 5 of the generalized $\omega$-meson self-interaction, together with a reference family without the interaction characterized by the coupling
$c_\omega=0$. While many more families could in principle be
constructed by varying $q$, these choices already span a representative
range of high density EOS behaviors. Rather than selecting one of these families as the preferred description, even though each can satisfy the relevant nuclear and astrophysical constraints, we search for the value of $\Theta$ at which their individual $\Theta$-$s_{\rm{c}}^2$ relations exhibit
minimum dispersion, as illustrated schematically in
FIG.\,\ref{fig:sc2-ext-sk}. This defines a common extraction of the
central SSS and, through the IPAD-TOV analysis, of the central EOS
parameter. The procedure gives a central SSS of approximately
$0.523_{-0.044}^{+0.044}$ and a central EOS-parameter of
$0.173^{+0.020}_{-0.018}$ when all six families are included. Excluding the
$c_\omega=0$ family gives the corresponding values
$0.496_{-0.092}^{+0.092}$ and $0.167^{+0.030}_{-0.029}$, respectively.
These results further imply canonical NS radii of approximately
$11.73^{+1.08}_{-0.92}\,\rm{km}$ and
$11.92^{+1.76}_{-1.28}\,\rm{km}$ for the two cases, with the quoted
uncertainties corresponding to the $95\%$ confidence intervals.
Fitting all EOS samples together without distinguishing their family
labels also yields a common $\Theta$-$s_{\rm{c}}^2$ scaling
consistent with these extractions, providing an internal consistency
test.

The extracted central SSS also provides a natural reference point for
examining the possible peaked behavior of $s^2$ inside NSs. The
competition between the increasing vector repulsion and the $\omega$
self-interaction provides a new mechanism for generating an interior
sound speed peak within a purely nucleonic EOS, with the power index
$q$ playing an important role in its emergence and location. Within
the IPAD-TOV framework, the leading near center variation of $s^2$
allows us to determine whether the SSS initially increases or
decreases away from the center, as examined in detail in
Subsection \ref{subs_peak}. Importantly, EOS families exhibiting such
peaks can still satisfy the adopted nuclear matter and astrophysical
constraints. Thus, varying $q$ reveals a new route to substantially
different sound speed behaviors, including interior peaks, while
remaining compatible with the same current constraints.

The purpose of this analysis is therefore not to eliminate EOS model
dependence completely, nor to identify one of the present nonlinear Walecka families
as the physical EOS. Rather, it is to test whether information common
to substantially different dense matter descriptions can be isolated
through the stellar structure equations. The nonlinear Walecka model
used in this work should thus be viewed as a controlled example rather
than a restriction of the strategy, which can in principle be applied
to other EOS or interaction model families. Of particular interest is
whether systematic relations and common scalings similar to those
found here persist across distinct theoretical frameworks, such as
Walecka models\,\cite{SW1986}, meta-models\,\cite{ZhangLi2018NST,XieLi2019,XieLi2021-a,ZhangLi2025Twin}, Skyrme Hartree--Fock\,\cite{Stone2003,Dutra2012,Fattoyev2012,ZhangChen2016}
and Gogny-force\,\cite{ChenCai2012,GonzalezBoquera2017,GonzalezBoquera2018,Vinas2021Gogny} descriptions, as well as microscopic or ab initio
approaches\,\cite{Marino2024,Alp2025,Zou2026PRC,
Xu2024NSTabinitio,Yang2025CPL}; establishing
such common behavior would further demonstrate the broader
applicability of the present approach and provide a stronger basis for
probing NS cores with reduced model dependence. While we focus here on
the central SSS of canonical $1.4\,M_\odot$ NSs, the same strategy can
also be extended to other stellar masses. For more massive NSs, the
larger central value of $\x$ makes higher order contributions in $\x$
to the estimator $\Theta$ increasingly important and may introduce
additional uncertainties. Systematically incorporating these
corrections, together with extending the analysis across broader EOS
frameworks and NS masses, therefore provides an important direction
for future study. The extracted values presented here remain
conditional on the EOS families considered and on the accuracy of the
IPAD-TOV approximation, but provide a concrete step toward such a more
general framework.

The paper is organized as follows. Section \ref{SEC_2} introduces the
generalized nonlinear Walecka model and derives the SSS, with
particular attention to the role of the $\omega$-meson self-interaction in determining its low and high density behavior.
Section \ref{SEC_3} constructs the corresponding EOS ensembles under
the adopted nuclear and astrophysical constraints and examines their
nuclear matter and NS properties, illustrating the model dependence
of the corresponding core properties. Section \ref{SEC_4} develops the
IPAD-TOV analysis, derives the estimator $\Theta$ for the central SSS,
and uses its common scaling across the different EOS families to
extract $s_{\rm{c}}^2$, $\x$, and the canonical NS radius with reduced
model dependence. The near center behavior of the SSS and the possible
emergence of interior sound speed peaks are also investigated.
Section \ref{SEC_5} summarizes the main results and discusses their
broader implications.
The appendices provide the detailed derivation of the analytical
expression for the SSS in $\beta$-equilibrated matter within the
nonlinear Walecka model, together with an analysis of the asymptotic
behavior of the vector mean fields.

\section{The Nonlinear Walecka Model with a Modified $\omega$-Meson Self-Interaction Term $\sim(\omega_\mu\omega^\mu)^q$}\label{SEC_2}

In this section, we establish the dense matter framework used throughout
this work and examine its implications for the SSS. In
Subsection \ref{subs_BF}, we introduce the generalized nonlinear Walecka
model and derive the relevant nuclear matter quantities, some of which are new in the literature due to the index $q$, together with
a closed analytical expression for the SSS in cold
$\beta$-equilibrated matter. We then examine the behavior of the model
in complementary density regimes. In Subsection \ref{subs_MF}, we
analyze the high density behavior of the symmetry energy, isospin
asymmetry, and vector mean fields, establishing their asymptotic
density dependence. Based on these results, in
Subsection \ref{subs_HD}, we derive the high density asymptotic limits
of $s^2$ and $\phi=P/\varepsilon$ and show how they depend on the
generalized $\omega$-meson self-interaction and its power $q$. In
Subsection \ref{subs_LD}, we consider the low density limit, where the
leading behavior is governed by the universal Fermi gas contribution
with interaction dependent corrections. Together, these results show
how the model evolves from a relatively universal low density regime
to substantially different high density behaviors, providing a
controlled set of EOS families for testing the stellar structure
scaling developed in this work.

\subsection{Basic Formalism for the Dense Matter EOS}\label{subs_BF}

The EOS of isospin asymmetric nuclear matter (ANM), defined by its binding energy per nucleon, can be expressed\,\cite{LCK08}
\begin{equation}
  E(\rho, \delta) \approx E_0(\rho) + E_{\text{sym}}(\rho)\delta^2 + \mathcal{O}(\delta^4),
\end{equation}
where $\rho = \rho_\rm n+\rho_\rm p$ is the baryon density with $\rho_\rm n$ ($\rho_\rm p$) denoting the neutron (proton) density, $\delta = (\rho_\rm n - \rho_\rm p)/(\rho_\rm n + \rho_\rm p)$ is the isospin asymmetry, $E_0(\rho) \equiv E(\rho, 0)$ is the EOS of symmetric nuclear matter (SNM), and the symmetry energy $E_{\mathrm{sym}}(\rho)$ is defined as
\begin{equation}
  E_{\mathrm{sym}}(\rho) = \left.\frac{1}{2} \frac{\partial^2 E(\rho, \delta)}{\partial \delta^2} \right|_{\delta=0}.
\end{equation}
At nuclear matter saturation density $\rho_0$, the EOS $E_0(\rho)$ of SNM can be expanded in $\chi \equiv (\rho - \rho_0)/3\rho_0$ as\,\cite{LCK08}
\begin{equation}
  E_0(\rho) \approx E_0(\rho_0) + \frac{1}{2}K_0\chi^2 + \frac{1}{6}J_0\chi^3 + \mathcal{O}(\chi^4),
\end{equation}
where $E_0(\rho_0)$ is the binding energy per nucleon in SNM at nuclear saturation $\rho_0$, $K_0\equiv 9\rho_0^2\d^2E_0(\rho)/\d\rho^2|_{\rho=\rho_0}$ is the incompressibility coefficient, and $J_0\equiv 27\rho_0^3\d^3E_0(\rho)/\d\rho^3|_{\rho=\rho_0}$ is the skewness coefficient. Similarly, the symmetry energy $E_{\mathrm{sym}}(\rho)$ can be expanded as\,\cite{LCK08}
\begin{equation}
  E_{\text{sym}}(\rho) \approx E_{\text{sym}}(\rho_0) + L\chi + \mathcal{O}(\chi^2),
\end{equation}
where $L\equiv 3\rho_0\d E_{\rm{sym}}(\rho)/\d\rho|_{\rho=\rho_0}$ is the symmetry energy density slope at nuclear matter saturation density. 

In this work, we employ the nonlinear Walecka model incorporating a modified $\omega$-meson self-interaction, with the Lagrangian given by\,\cite{Li2022}
\begin{align}
    \mathcal{L} = &\overline{\psi}\left(i\partial_\mu\gamma^\mu - M_{\rm N}\right)\psi \notag\\
    &+ g_\sigma\sigma\overline{\psi}\psi - g_\omega\omega_\mu\overline{\psi}\gamma^\mu\psi 
    - g_\rho\vec{\rho}_\mu\overline{\psi}\gamma^\mu\vec{\tau}\psi 
    + g_\delta\vec{\delta}\overline{\psi}\vec{\tau}\psi \notag\\
    &+ \frac{1}{2}\left(\partial_\mu\sigma\partial^\mu\sigma - m_\sigma^2\sigma^2\right)
    - \frac{1}{3}b_\sigma M_{\rm N}(g_\sigma\sigma)^3
    - \frac{1}{4}c_\sigma(g_\sigma\sigma)^4 \notag\\
    &- \frac{1}{4}\omega_{\mu\nu}\omega^{\mu\nu}
    + \frac{1}{2}m_\omega^2\omega_\mu\omega^\mu
    + \frac{1}{4}\frac{c_\omega}{ m_\omega^{2q-4}}\left(g_\omega^2\omega_\mu\omega^\mu\right)^q\notag \\
    &- \frac{1}{4}\vec{\rho}_{\mu\nu}\vec{\rho}^{\mu\nu}
    + \frac{1}{2}m_\rho^2\vec{\rho}_\mu\vec{\rho}^\mu
    + \frac{1}{2}\Lambda_{\rm V}g_\rho^2\vec{\rho}_\mu\vec{\rho}^\mu g_\omega^2\omega_\mu\omega^\mu\notag \\
    &+ \frac{1}{2}\left(\partial_\mu\vec{\delta}\partial^\mu\vec{\delta} 
    - m_\delta^2\vec{\delta}^2\right)
    + \frac{1}{2}C_{\delta\sigma}g_\sigma^2\sigma^2g_\delta^2\vec{\delta}^2. \label{eq:L}
\end{align}
Here, $M_{\rm N}\approx939\,\mathrm{MeV}$ is the nucleon mass, and
$m_\sigma\approx491.5\,\rm{MeV}$, $m_\omega\approx782.5\,\rm{MeV}$, $m_\rho\approx763.0\,\rm{MeV}$, and $m_\delta\approx980.0\,\rm{MeV}$ denote the corresponding
meson masses. The couplings $g_\sigma$, $g_\omega$, $g_\rho$, and
$g_\delta$ characterize the nucleon-meson interactions. The parameters
$b_\sigma$ and $c_\sigma$ govern the nonlinear self-interactions of the
$\sigma$ meson, while $\Lambda_{\rm V}$ and $C_{\delta\sigma}$ control
the $\rho$-$\omega$\,\cite{ToddRutel2005} and $\delta$-$\sigma$\,\cite{Li2022} mixing interactions,
respectively.

The new feature introduced here is the generalized $\omega$-meson
self-interaction:
\begin{equation}
  + \frac{1}{4}\frac{c_\omega}{ m_\omega^{2q-4}}\left(g_\omega^2\omega_\mu\omega^\mu\right)^q\subset\mathcal{L}.
\end{equation}
Such a term should be understood as part of an
effective description of the nuclear interaction rather than as a
fundamental one\,\cite{MullerSerot96}. In particular, higher-order interactions are
in general nonrenormalizable in the conventional power-counting sense.
Within an effective framework, however, this does not prevent them from
being used as phenomenological interaction terms over the density
regime of interest. Instead, the generalized form provides a convenient
way to control the behavior of the repulsive vector interaction at high
densities and, consequently, to explore different possible behaviors of
the EOS and the sound speed in dense matter.
We restrict ourselves to $q>1$, since for $q=1$ the additional term
becomes quadratic in the mean field of the $\omega$ meson and can be
absorbed into the ordinary quadratic mass term. Varying $q$ generates
a family of effective interactions with different high-density
behaviors. The factor $m_\omega^{2q-4}$ is included so that the coupling
coefficient $c_\omega$ remains dimensionless for arbitrary $q$.
The overall factor $1/4$ is adopted as a convenient normalization such
that the conventional quartic $\omega$-meson self-interaction is
recovered in its standard form for $q=2$.
Therefore, no particular discrete choice of $q$ is intrinsic to the
generalized interaction itself.

For the numerical analysis in this work, we select five representative
values, $q=$3/2, 2, 3, 4, 5, together with the limiting case
$c_\omega=0$, in which the $\omega$-meson self-interaction is absent.
These choices define the six nonlinear Walecka model families considered below and serve as representative realizations of the broader class of
interactions generated by varying $q$. Among them, the case $q=2$ is
of particular interest, since it recovers the conventional quartic
$\omega$-meson self-interaction widely employed in nonlinear Walecka
models. 

From the Lagrangian density in Eq.\,\eqref{eq:L}, we obtain the equations
of motion for the four meson fields. In the relativistic mean-field
(RMF) approximation, the meson fields are replaced by their expectation
values, yielding
\begin{align}
    m_\sigma^2\overline{\sigma} &= g_\sigma\bigg[\rho_\rm p^{\rm s} + \rho_\rm n^{\rm s} - b_\sigma M_{\rm N}(g_\sigma\overline{\sigma})^2 \notag\\
        &\hspace{1cm} - c_\sigma \left(g_\sigma\overline{\sigma}\right)^3 
     + C_{\delta\sigma}g_\sigma\overline{\sigma}\left(g_\delta\overline{\delta}^{(3)}\right)^2\bigg], 
        \label{eq:seqm} \\
    m_\omega^2\overline{\omega}_0 &= g_\omega\left[\rho - \frac{q}{2}\frac{c_\omega}{ m_\omega^{2q-4}}\left(g_\omega\overline{\omega}_0\right)^{2q-1} 
        - \Lambda_{\rm V}g_\omega\overline{\omega}_0\left(g_\rho\overline{\rho}_0^{(3)}\right)^2\right], 
        \label{eq:oeqm} \\
    m_\rho^2\overline{\rho}_0^{(3)} &= g_\rho\left[\rho_\rm p - \rho_\rm n
        - \Lambda_{\rm V} g_\rho\overline{\rho}_0^{(3)}\left(g_\omega\overline{\omega}_0\right)^2\right], 
        \label{eq:reqm} \\
    m_\delta^2\overline{\delta}^{(3)} &= g_\delta\left[\rho_\rm p^{\rm s} - \rho_\rm n^{\rm s} 
        + C_{\delta\sigma} g_\delta\overline{\delta}^{(3)}\left(g_\sigma\overline{\sigma}\right)^2\right].
        \label{eq:deqm}
\end{align}
Here, the overline denotes the mean-field expectation value, while the
subscript ``0'' and superscript ``3'' label the temporal component of a
four-vector and the third component in isospin space, respectively. The
scalar density $\rho_J^{\rm s}$ is given by\,\cite{Cai2017,WeiSN24,Cai2016,Long2004,Meng2006,MengZhou2015,SW1986}
\begin{align}
    \rho_J^{\rm s}
    &= \frac{2}{(2\pi)^3}
    \int_0^{k_{\rm F}^J}\d\v{k}
    \frac{M_J^*}{\sqrt{\v{k}^2+M_J^{*,2}}}
    \notag\\
    &= \frac{M_J^*}{2\pi^2}
    \left[
    k_{\rm F}^J\varepsilon_{\rm F}^{*,J}
    -M_J^{*,2}
    \ln\left(
    \frac{k_{\rm F}^J+\varepsilon_{\rm F}^{*,J}}{M_J^*}
    \right)
    \right],
    ~~ J={\rm n},{\rm p},
\end{align}
where $M_J^*$ is the nucleon Dirac effective mass\,\cite{Cai2017,Qin25},
\begin{equation}
    M_J^*
    =M_{\rm N}-g_\sigma\overline{\sigma}
    +g_\delta\overline{\delta}^{(3)}\tau_3^J,
 ~~ J={\rm n},{\rm p}.
\end{equation}
The corresponding effective Fermi energy is similarly given by
\begin{equation}
\varepsilon_{\rm F}^{*,J}
=\sqrt{k_{\rm F}^{J,2}+M_J^{*,2}},
\end{equation}
with
$k_{\rm F}^J=k_{\rm F}(1+\tau_3^J\delta)^{1/3}$ denoting the Fermi
momentum of species $J$. Here, $\tau_3^J=-1$ ($+1$) for protons
(neutrons), and $k_{\rm F}=(3\pi^2\rho/2)^{1/3}$ is the Fermi momentum of SNM at the same baryon density $\rho$.

Using the Lagrangian density introduced above, the energy density
$\varepsilon=\langle\mathcal{T}^{00}\rangle$ and pressure
$P=3^{-1}\sum_{j=1}^{3}\langle\mathcal{T}^{jj}\rangle$ of the nuclear
system can be obtained from the energy-momentum tensor
$\mathcal{T}^{\mu\nu}$\,\cite{SW1986}, given by
\begin{equation}
\mathcal{T}^{\mu\nu}
=
\sum_{\phi_a}
\frac{\partial \mathcal{L}}
{\partial(\partial_\mu \phi_a)}
\partial^\nu \phi_a
-
\mathcal{L} g^{\mu\nu}.
\end{equation}
Therefore, the expressions for energy density and pressure can be obtained as
\begin{align}
    \varepsilon =  & \varepsilon_\rm p^{\mathrm{kin}} + \varepsilon_\rm n^{\mathrm{kin}} \notag \\
    &+ \frac{1}{2} m_\sigma^2 \overline{\sigma}^2 + \frac{1}{2} m_\omega^2 \overline{\omega}_0^2 
    + \frac{1}{2} m_\rho^2 \overline{\rho}_0^{(3),2} + \frac{1}{2} m_\delta^2 \overline{\delta}^{(3),2} \notag \\
    &+ \frac{1}{3} M_{\rm N} b_\sigma \left(g_\sigma \overline{\sigma}\right)^3 + \frac{1}{4} c_\sigma \left(g_\sigma \overline{\sigma}\right)^4 + \frac{2q-1}{4} \frac{c_\omega}{m_\omega^{2q-4}} \left(g_\omega \overline{\omega}_0\right)^{2q} \notag \\
    &+ \frac{3}{2} \Lambda_{\rm V} \left(g_\rho \overline{\rho}_0^{(3)}\right)^2 \left(g_\omega \overline{\omega}_0\right)^2 - \frac{1}{2} C_{\delta\sigma} \left(g_\delta \overline{\delta}^{(3)}\right)^2 (g_\sigma \overline{\sigma})^2, \label{eq:eps}
\end{align}
as well as
\begin{align}
P =& P_\rm p^{\mathrm{kin}} + P_\rm n^{\mathrm{kin}} \notag \\
&- \frac{1}{2} m_\sigma^2 \overline{\sigma}^2 + \frac{1}{2} m_\omega^2 \overline{\omega}_0^2 + \frac{1}{2} m_\rho^2 \overline{\rho}_0^{(3),2} - \frac{1}{2} m_\delta^2 \overline{\delta}^{(3),2} \notag \\
&- \frac{1}{3} M_{\rm N} b_\sigma \left(g_\sigma \overline{\sigma}\right)^3 - \frac{1}{4} c_\sigma \left(g_\sigma \overline{\sigma}\right)^4 + \frac{1}{4}\frac{c_\omega }{ m_\omega^{2q-4}} \left(g_\omega \overline{\omega}_0\right)^{2q} \notag \\
&+ \frac{1}{2} \Lambda_{\rm V} \left(g_\rho \overline{\rho}_0^{(3)}\right)^2 \left(g_\omega \overline{\omega}_0\right)^2 + \frac{1}{2} C_{\delta\sigma} \left(g_\delta \overline{\delta}^{(3)}\right)^2 (g_\sigma \overline{\sigma})^2, \label{eq:P}
\end{align}
where $\varepsilon_J^{\mathrm{kin}}$ and $P_J^{\mathrm{kin}}$ are given by
\begin{align}
\varepsilon_J^{\mathrm{kin}} &= \frac{2}{(2\pi)^3} \int_0^{k_\rm F^J} \d \v{k}\sqrt{\v{k}^2 + M_J^{*,2}}  \notag \\
&= \frac{1}{4} \left( 3\varepsilon_\rm F^{*,J} \rho_J + M_J^* \rho_J^{\rm s} \right),~~J = \rm n,\rm p, \label{eq:29}
\end{align}
and
\begin{align}
P_J^{\mathrm{kin}} &= \frac{1}{3}\frac{2}{(2\pi)^3} \int_0^{k_\rm F^J} \d \v{k}\frac{\v{k}^2}{\sqrt{\v{k}^2 + M_J^{*,2}}}  \notag \\
&= \frac{1}{4} \left( \varepsilon_\rm F^{*,J} \rho_J - M_J^* \rho_J^{\rm s} \right), ~~ J = \rm n, \rm p. \label{eq:30}
\end{align}
The EOS of this system can be obtained through the thermodynamic relation\,\cite{Cai2017,SW1986}:
\begin{equation}
    E(\rho, \delta) = \varepsilon(\rho, \delta)/\rho - M_\rm N.
\end{equation}

Using the equations derived above, we obtain the following analytical
expression for the EOS of SNM\,\cite{Li2022,Cai2017}:
\begin{align}
    E_0(\rho)=&\varepsilon(\rho,\delta=0)/\rho-M_{\rm N}\notag\\
    =&\frac{1}{\rho}\bigg[ \frac{3}{4}\varepsilon_{\rm F}^{\ast}\rho
    +\frac{1}{4}M_0^\ast\rho^{\rm s}
    + \frac{1}{2}m_\sigma^2\overline{\sigma}^2
    + U(\overline{\sigma})\notag\\
    &+\frac{1}{2}m_\omega^2\overline{\omega}_0^2
    +\frac{1}{2}m_\rho^2\overline{\rho}_0^{(3),2}
    +\frac{1}{2}m_\delta^2\overline{\delta}^{(3),2}\notag\\
&    +\frac{2q-1}{4}\frac{c_\omega}{m_\omega^{2q-4}}
    \left(g_\omega\overline{\omega}_0\right)^{2q}\bigg]-M_{\rm N},
\end{align}
here,
$\varepsilon_{\rm F}^*=({k_{\rm F}^2+M_0^{*,2}})^{1/2}$,
$M_0^*=M_{\rm N}-g_\sigma\overline{\sigma}$ is the nucleon effective mass in SNM, and the self-interaction term for the $\sigma$ meson is defined as\,\cite{SW1986}:
\begin{equation}
U(\overline{\sigma})
=\frac{1}{3}b_\sigma M_{\rm N}\left(g_\sigma\overline{\sigma}\right)^3
+\frac{1}{4}c_\sigma\left(g_\sigma\overline{\sigma}\right)^4.
\end{equation}
The corresponding slope parameter is then given by
\begin{align}
&L_0(\rho)\equiv3\rho\frac{\mathrm{d}E_0}{\mathrm{d}\rho}
=3\left[
\frac{\varepsilon_{\rm F}^*}{4}
-\frac{M_0^*\rho^{\rm s}}{4\rho}
+g_\omega\overline{\omega}_0
\right.\notag\\
&\left.
-\frac{1}{\rho}
\left(
\frac{1}{2}m_\sigma^2\overline{\sigma}^2
+U(\overline{\sigma})
+\frac{1}{2}m_\omega^2\overline{\omega}_0^2
+\frac{(2q-1)c_\omega}{4m_\omega^{2q-4}}
\left(g_\omega\overline{\omega}_0\right)^{2q}
\right)
\right],
\label{eq:35}
\end{align}
which is directly related to the SNM pressure through
$P_0(\rho)=\rho L_0(\rho)/3$.
The incompressibility coefficient can be obtained in a similar manner as
\begin{align}
K_0(\rho)\equiv&9\rho^2\frac{\mathrm{d}^2E_0}{\mathrm{d}\rho^2}\notag\\
=&-\frac{9\rho g_\sigma^2M_0^{*,2}}
{Q_\sigma\varepsilon_{\rm F}^{*,2}}
+\frac{9\rho g_\omega^2}{Q_\omega}
+\frac{3k_{\rm F}^2}{\varepsilon_{\rm F}^*}
-6L_0(\rho),
\label{eq:33}
\end{align}
while the skewness coefficient reads\,\cite{Cai2017}
\begin{align}
J_0(\rho)
\equiv&27\rho^3\frac{\mathrm{d}^3E_0}{\mathrm{d}\rho^3}
=-\frac{3k_{\rm F}^2}{\varepsilon_{\rm F}^*}
-\frac{3k_{\rm F}^4}{\varepsilon_{\rm F}^{*,3}}
+\frac{27g_\sigma^2M_0^{*,2}\rho^2}
{Q_\sigma\varepsilon_{\rm F}^{*,3}}
\notag\\
&\times\left(
\frac{3\pi^2}{2k_{\rm F}\varepsilon_{\rm F}^*}
+\frac{2g_\sigma^2}{Q_\sigma}
-\frac{g_\sigma M_0^*\eta}{Q_\sigma^2}
-\frac{2g_\sigma^2M_0^{*,2}}
{Q_\sigma\varepsilon_{\rm F}^{*,2}}
+\frac{\varepsilon_{\rm F}^*H_\sigma}{2Q_\sigma}
\right)
\notag\\
&
-\frac{27q(q-1)(2q-1)c_\omega
g_\omega^{2q+3}\overline{\omega}_0^{2q-3}\rho^2}
{m_\omega^{2q-4}Q_\omega^3}
-9K_0(\rho).
\label{eq:34}
\end{align}

Similarly, the expressions for the symmetry energy and its slope coefficient are given by\,\cite{Li2022}
\begin{align}
E_{\mathrm{sym}}(\rho) =& \frac{k_{\rm F}^2}{6\varepsilon_{\rm F}^*}
+ \frac{\rho}{2}\left(\frac{g_\rho}{m_\rho^*}\right)^2 \notag \\
&- \frac{\rho}{2}\left(\frac{g_\delta}{m_\delta^*}\right)^2
\left(\frac{M_0^{*}}{\varepsilon_{\rm F}^{*}
}\right)^2\left[1 + \left(\frac{g_\delta}{m_\delta^*}\right)^2
A(k_{\rm F},M_0^*)\right]^{-1},
\label{eq:Esym}
\end{align}
and\,\cite{Li2022,Cai2012PLB}
\begin{align}
L(\rho)= & \frac{k_{\rm F}^2}
{6\varepsilon_{\rm F}^{*}}+\frac{k_{\rm F}^2
M_0^{*,2}}
{6\varepsilon_{\rm F}^{*,3}}
+ \frac{g_\sigma^2 M_0^{*,2} k_{\rm F}^2 \rho}
{2Q_\sigma \varepsilon_{\rm F}^{*,4}} \notag \\
&+ \frac{3g_\rho^2 \rho}{2Q_\rho}
- \frac{3g_\omega^3 g_\rho^4 \Lambda_{\rm V}
\overline{\omega}_0 \rho^2}{Q_\omega Q_\rho^2} \notag \\
& - \frac{3g_\sigma g_\delta^2 M_0^{*,2} \rho^2}
{2\varepsilon_{\rm F}^{*,3}Q_\sigma Q_\delta}
\left(
\frac{2g_\sigma M_0^{*,2}}{\varepsilon_{\rm F}^{*,2}}
+ \frac{M_0^*\Gamma}{Q_\delta}
-2g_\sigma
\right) \notag \\
&- \frac{3g_\delta^2 M_0^{*,2}\rho}
{2\varepsilon_{\rm F}^{*,2}Q_\delta}
\left(
1-\frac{2k_{\rm F}^2}{3\varepsilon_{\rm F}^{*,2}}
-\frac{H_\delta\rho}{2Q_\delta}
\right),
\label{eq:32}
\end{align}
respectively, where $m_\delta^{*,2}
=m_\delta^2-C_{\delta\sigma}g_\delta^2g_\sigma^2
\overline{\sigma}^2$. The other auxiliary quantities entering the above expressions are defined as
\begin{align}
Q_\sigma &= m_\sigma^2 + g_\sigma^2 A(k_{\rm F},M_0^*)
+2b_\sigma M_{\rm N}g_\sigma^3\overline{\sigma}
+3c_\sigma g_\sigma^4\overline{\sigma}^2,
\label{eq:36}\\
Q_\omega &= m_\omega^2
+\frac{q(2q-1)}{2}
\frac{c_\omega}{m_\omega^{2q-4}}
g_\omega^{2q}\overline{\omega}_0^{2q-2},
\label{eq:37}\\
Q_\rho &= m_\rho^{*,2}
=m_\rho^2+\Lambda_{\rm V}g_\rho^2g_\omega^2
\overline{\omega}_0^2,
\label{eq:38}\\
Q_\delta &= m_\delta^{*,2}
+g_\delta^2A(k_{\rm F},M_0^*),
\label{eq:39}
\end{align}
together with
\begin{align}
A(k_{\rm F},M_0^\ast)=& 3\left(\frac{\rho^{\rm s}}{M_0^*}
-\frac{\rho}{\varepsilon_{\rm F}^*}\right),~~
H_\sigma=\frac{2g_\sigma^2k_{\rm F}^2}
{\varepsilon_{\rm F}^{*,3}},
~~
H_\delta=\frac{2g_\delta^2k_{\rm F}^2}
{\varepsilon_{\rm F}^{*,3}},
\label{eq:40}
\end{align}
and
\begin{align}
\eta= &3g_\sigma^3\left(
\frac{2\rho^{\rm s}}{M_0^{*,2}}
-\frac{3\rho}{M_0^*\varepsilon_{\rm F}^*}
+\frac{M_0^*\rho}{\varepsilon_{\rm F}^{*,3}}
\right)\notag\\
&-2b_\sigma M_{\rm N}g_\sigma^3
-6c_\sigma g_\sigma^4\overline{\sigma},
\label{eq:41}\\
\Gamma =&3g_\sigma g_\delta^2\left(
\frac{2\rho^{\rm s}}{M_0^{*,2}}
-\frac{3\rho}{M_0^*\varepsilon_{\rm F}^*}
+\frac{M_0^*\rho}{\varepsilon_{\rm F}^{*,3}}
\right)+2C_{\delta\sigma}g_\delta^2g_\sigma^2\overline{\sigma}.
\label{eq:42}
\end{align}

It is worth noting that the compact expression for $K_0(\rho)$ in
Eq.\,\eqref{eq:33} contains no separate explicit $q$-dependent term.
The dependence on $q$ is instead encoded in $Q_\omega$ of
Eq.\,(\ref{eq:37}) and in the self-consistently determined $\omega$
mean field. In particular, the nonlinear contribution to $Q_\omega$
scales as $\overline{\omega}_0^{2q-2}$, making the vector response
increasingly sensitive to the density dependence of
$\overline{\omega}_0$ as $q$ increases once the nonlinear term becomes
important. Moreover, the inversion from the adopted nuclear matter
quantities yields $c_\omega>0$ for the EOS ensembles, so that
this nonlinear contribution to $Q_\omega$ is positive and becomes
increasingly relevant as the vector mean field grows at high densities.
Thus, changing the power of the generalized $\omega$-meson
self-interaction affects $K_0(\rho)$ through the response of the vector
mean field rather than through an additional term with an explicit
dependence on $q$. This feature is in contrast to the skewness
coefficient $J_0(\rho)$ in Eq.\,\eqref{eq:34}, where the higher-order
density response generates an explicit contribution proportional to
$q(q-1)(2q-1)$. The generalized $\omega$-meson self-interaction can
therefore have a more direct impact on the higher-order density
dependence of the SNM EOS, which becomes increasingly relevant toward
high densities.

Based on the RMF framework developed above, we further derive a closed
analytical expression for the speed of sound squared (SSS). Along the cold,
charge-neutral $\beta$-equilibrium EOS, the thermodynamic relations give
\begin{equation}
    s^2=\frac{\d P}{\d\varepsilon}
    =\frac{\rho}{\mu_{\rm n}}
    \frac{\d\mu_{\rm n}}{\d\rho}=\frac{\d\ln\mu_{\rm n}}{\d\ln\rho}.
\end{equation}
For convenience, we introduce the scaled scalar and vector mean fields
\begin{equation}
    S_\sigma=g_\sigma\overline{\sigma},~~
    S_\delta=g_\delta\overline{\delta}^{(3)},~~
    V_\omega=g_\omega\overline{\omega}_0,~~
    V_\rho=g_\rho\overline{\rho}_0^{(3)},\label{def-SV}
\end{equation}
together with
\begin{equation}
    \zeta\equiv\frac{\d\rho_{\rm p}}{\d\rho},~~
    u_J\equiv\frac{M_J^*}{\varepsilon_{\rm F}^{*,J}},~~
    F_J\equiv
    \frac{k_{\rm F}^{J,2}}
    {3\rho_J\varepsilon_{\rm F}^{*,J}},
    ~~J={\rm{n,p}}.
\end{equation}
The density responses of the mean fields can then be expressed in the
linear form $S_\sigma'=S_{\sigma,0}+S_{\sigma,1}\zeta$,
    $S_\delta'=S_{\delta,0}+S_{\delta,1}\zeta$,
    $V_\omega'=V_{\omega,0}+V_{\omega,1}\zeta$, and
    $V_\rho'=V_{\rho,0}+V_{\rho,1}\zeta$, where the prime denotes the total derivative with respect to $\rho$
along the $\beta$-equilibrium sequence. The subscripts 0
and 1 label the coefficients of the composition-independent and
$\zeta$-dependent parts of the corresponding field responses,
respectively. We then obtain
\begin{align}
    s^2=\frac{\rho}{\mu_{\rm n}}
    \bigg[
    &F_{\rm n}
    -u_{\rm n}S_{\sigma,0}
    +u_{\rm n}S_{\delta,0}
    +V_{\omega,0}-V_{\rho,0}\notag\\
    &+\left(
    -F_{\rm n}
    -u_{\rm n}S_{\sigma,1}
    +u_{\rm n}S_{\delta,1}
    +V_{\omega,1}-V_{\rho,1}
    \right)\zeta
    \bigg].
\label{eq:rmf_s2_closed}
\end{align}
The explicit expressions for the response coefficients and $\zeta$,
together with the detailed derivation of Eq.\,\eqref{eq:rmf_s2_closed}, are given in Appendix \ref{app:sound_speed}.

\subsection{High Density Behavior of Symmetry Energy and Meson Fields}
\label{subs_MF}

The generalized $\omega$-meson self-interaction introduced above can
substantially modify the high density behavior of the EOS and,
consequently, the density dependence of the SSS. To understand the
origin of these effects, we first examine how the power $q$ affects
the symmetry energy, isospin asymmetry, and meson mean fields at high
densities. This analysis establishes the leading density dependence
of the vector fields and provides the basis for deriving the
asymptotic limits of $s^2$ and $\phi=P/\varepsilon$ in the next
subsection.

At asymptotically high densities, the scalar fields become saturated
and their source terms grow much more slowly than the baryon density.
The leading mesonic density dependence is therefore controlled
primarily by the vector mean fields. Using
$V_\omega=g_\omega\overline{\omega}_0$ and
$V_\rho=g_\rho\overline{\rho}_0^{(3)}$ of (\ref{def-SV}), their field equations can be
written as
\begin{align}
\rho&=
\left(
\frac{m_\omega^2}{g_\omega^2}
+\Lambda_{\rm V}V_\rho^2
\right)V_\omega
+\frac{q}{2}\frac{c_\omega}{m_\omega^{2q-4}}
V_\omega^{2q-1},
\label{eq:omega_high}\\
\rho\delta&=
-\left(
\frac{m_\rho^2}{g_\rho^2}
+\Lambda_{\rm V}V_\omega^2
\right)V_\rho.
\label{eq:rho_high}
\end{align}
The two vector fields are coupled through the mixed
$\rho$-$\omega$ interaction, while the source of the $\rho$ field
depends explicitly on the isospin asymmetry $\delta$. It is therefore
necessary to first determine the asymptotic behavior of $\delta$.

In $\beta$-equilibrium, $
    \mu_{\rm n}-\mu_{\rm p}
    \approx4E_{\rm{sym}}(\rho)\delta
    \approx\mu_{\rm e}$,
where, together with charge neutrality, the ultra-relativistic lepton
chemical potential scales as
$\mu_{\rm e}\sim[(1-\delta)\rho]^{1/3}$, up to a composition-dependent
numerical factor. The asymptotic behavior of $\delta$ is therefore
directly connected to that of the symmetry energy.
The kinetic contribution satisfies
\begin{equation}
    E_{\rm{sym}}^{\rm{kin}}(\rho)=k_{\rm F}^2/6\varepsilon_{\rm F}^{\ast}
    \sim\rho^{1/3}.
\end{equation}
The $\delta$ meson contribution is suppressed at high densities. In particular, for $k_{\rm F}\gg M_0^*$, the function $A$ behaves as
\begin{equation}
A(k_{\rm F},M_0^*)=3\left(\frac{\rho^{\rm s}}{M_0^*}-\frac{\rho}{\varepsilon_{\rm F}^*}\right)
\approx\frac{k_{\rm F}^2}{\pi^2}\sim\rho^{2/3}.
\end{equation}
Since $m_\delta^{*,2}$ approaches a finite constant while $g_\delta^2 A\sim\rho^{2/3}$ increases with density, the response function $A$ dominates the denominator in the high-density limit. Consequently, the $\delta$-meson contribution is suppressed by both the relativistic factor $(M_0^*/\varepsilon_{\rm F}^*)^2$ and the increasing response function, yielding
\begin{equation}
E_{\rm{sym}}^\delta(\rho)\approx
-\frac{\rho}{2A}\left(\frac{M_0^*}{k_{\rm F}}\right)^2
=-\frac{M_0^{*,2}}{3k_{\rm F}}
\sim-M_0^{*,2}\rho^{-1/3}.
\end{equation}
This contribution becomes even more strongly suppressed if $M_0^*$ decreases with increasing density. The only
interaction contribution that can modify the leading asymptotic
behavior is therefore the $\rho$ meson term\,\cite{ToddRutel2005},
\begin{equation}
E_{\rm{sym}}^{\rho}(\rho)
=
\frac{1}{2}
\frac{\rho}
{m_\rho^2/g_\rho^2+\Lambda_{\rm V}V_\omega^2}.
\label{eq:Esym_rho_high}
\end{equation}

To determine its density dependence self consistently, we write
$V_\omega\sim\rho^\nu$, with $\nu>0$.
At sufficiently high densities,
$\Lambda_{\rm V}V_\omega^2$ dominates over $m_\rho^2/g_\rho^2$, so
that
$E_{\rm{sym}}^\rho(\rho)\sim\rho^{1-2\nu}$. Combining this with the kinetic
contribution gives
\begin{equation}
    E_{\rm{sym}}(\rho)\sim\rho^y,
    ~~
    y=\max\left(3^{-1},1-2\nu\right).
\label{eq:Esym_scaling_general}
\end{equation}
The $\beta$-equilibrium condition then yields
\begin{equation}\label{delta-case}
\delta(\rho
)\sim
\begin{cases}
\rho^{2\nu-2/3}, & \nu<1/3,\\
\rho^0, & \nu\geq1/3.
\end{cases}
\end{equation}
Thus, the isospin asymmetry decreases asymptotically for
$\nu<1/3$, whereas it approaches a finite value for
$\nu\geq1/3$.

We can now return to the vector field equations. Since
$\Lambda_{\rm V}V_\omega^2$ dominates the denominator of the $\rho$
field equation at high densities, $
    V_\rho\sim
    -{\delta\rho}/{\Lambda_{\rm V}V_\omega^2}$,
which gives
\begin{equation}
V_\rho\sim
\begin{cases}
\rho^{1/3}, & \nu<1/3,\\
\rho^{1-2\nu}, & \nu\geq1/3.
\end{cases}
\label{eq:R_general_scaling}
\end{equation}
For $\nu\neq1/3$, the mixed vector term
$V_\rho^2V_\omega$ grows more slowly than the source term $\rho$ and
is therefore asymptotically sub-leading in the $\omega$ field equation.
For $c_\omega\neq0$, the nonlinear term
$V_\omega^{2q-1}$ eventually dominates over the linear term, giving
the leading balance $
    \rho\sim V_\omega^{2q-1}$,
and hence $
    V_\omega\sim\rho^{1/(2q-1)}$.
Thus, the power $q$ directly controls the asymptotic growth of the
isoscalar vector field.

The boundary case $q=2$ corresponds to $\nu=1/3$ and requires special
attention. In this case,
$V_\rho\sim V_\omega\sim\rho^{1/3}$, so that both
$V_\omega^{2q-1}$ and $V_\rho^2V_\omega$ scale as $\rho$. The mixed
vector term therefore contributes at the same leading order as the
$\omega$ self-interaction, but does not change the asymptotic scaling
$V_\omega\sim\rho^{1/3}$. Combining the different regimes, for
$c_\omega\neq0$ we obtain
\begin{equation}
V_\omega\sim\rho^{1/(2q-1)},~~
V_\rho\sim
\begin{cases}
\rho^{(2q-3)/(2q-1)}, & 1<q<2,\\
\rho^{1/3}, & q\geq2.
\end{cases}
\label{eq:R_q_scaling}
\end{equation}
The change at $q=2$ therefore separates two distinct asymptotic
regimes of the vector fields.

Finally, for $c_\omega=0$, the nonlinear $\omega$ self-interaction is
absent and the linear term in Eq.\,\eqref{eq:omega_high} must balance
the source term. This gives $
    V_\omega\sim\rho$ and $
    V_\rho\sim\rho^{-1}$,
with the mixed vector term asymptotically sub-leading. 
Eqs.\,(\ref{eq:omega_high}), (\ref{eq:rho_high}), and (\ref{delta-case})
also admit formal asymptotic branches other than
$V_\omega\sim\rho$ and $V_\rho\sim\rho^{-1}$.
The branch branch
$V_\omega\sim V_\rho\sim\rho^{1/3}$ requires
$\Lambda_{\rm V}\geq128/3\pi^2\approx4.32$.
Since the values of $\Lambda_{\rm V}$ obtained from the empirical nuclear matter
sampling ranges are typically $\lesssim\mathcal{O}(0.1)$, as shown in
TAB.\,\ref{tab:sampling_ranges}, neither alternative branch is realized
in the present ensembles. A detailed derivation is given in
Appendix \ref{app:vector_fields}. Thus, the isoscalar vector field grows linearly with density for $c_\omega=0$, but only sub-linearly for $c_\omega\neq0$. These qualitatively
different vector field scalings determine the dominant contributions
to the high density EOS and, consequently, the asymptotic limits of
the EOS-parameter and SSS discussed in the next subsection.

\subsection{High Density Asymptotic Limits of the SSS $s^2$ and the EOS- parameter
$\phi=P/\varepsilon$}\label{subs_HD}

Having established the asymptotic behavior of the vector mean fields,
we now determine the dominant contributions to the energy density and
pressure and thereby the limiting values of $s^2$ and
$\phi=P/\varepsilon$. Since the scalar fields become saturated
asymptotically, their contributions are sub-leading. The potentially
leading contributions are the ultra-relativistic kinetic terms\,\cite{CaiLi2022PRC},
\begin{equation}
    \varepsilon_{\rm{kin}}(\rho)
    \sim
    P_{\rm{kin}}(\rho)
    \sim\rho^{4/3},
\end{equation}
and the vector field terms proportional to
$V_\omega^2$, $V_\omega^{2q}$, $V_\rho^2$, and
$V_\omega^2V_\rho^2$. Their relative importance depends on $q$ and on
whether the generalized $\omega$-meson self-interaction is present.

For $c_\omega=0$, the asymptotic behavior
$V_\omega\sim\rho$ gives $V_\omega^2\sim\rho^2$, which grows faster
than the kinetic contribution $\rho^{4/3}$ and the remaining mesonic
terms. The quadratic $\overline{\omega}_0$ contribution therefore
dominates both the energy density and pressure. Since it enters
$\varepsilon$ and $P$ with the same coefficient:
\begin{equation}\label{ps-1}
\mbox{case 1: }    \phi
    \rightarrow
    s^2
    \rightarrow1,
    ~~\mbox{when } c_\omega=0.
\end{equation}
Thus, without the $\omega$ self-interaction, the EOS asymptotically
approaches the maximally stiff causal behavior with a
linear relation between pressure and energy density,
$P\sim\varepsilon\sim\rho^2$. This is the classical Zel'dovich model\,\cite{Zeldovich1961}.

For $1<q<2$, we have
$\nu=1/(2q-1)>1/3$. The $\overline{\omega}_0$ self-interaction scales
as $
    V_\omega^{2q}
    \sim\rho^{2q/(2q-1)}
    =\rho^{1+\nu}$,
which grows faster than the kinetic contribution $\rho^{4/3}$.
Meanwhile, using $V_\rho\sim\rho^{1-2\nu}$, the mixed vector term
behaves as
$V_\omega^2V_\rho^2\sim\rho^{2-2\nu}$ and is sub-leading because
$2-2\nu<1+\nu$. The high density EOS is therefore dominated by the
$\omega$ self-interaction. Its leading contributions to the energy
density and pressure satisfy
\begin{equation}
    \varepsilon_{\omega,{\rm{self}}}
    \approx
    \frac{2q-1}{4}V_\omega^{2q},
    ~~
    P_{\omega,{\rm{self}}}
    \approx
    \frac{1}{4}V_\omega^{2q},
\end{equation}
which gives
\begin{equation}\label{ps-2}
\mbox{case 2: }    \phi
    \rightarrow
    s^2
    \rightarrow
    \frac{1}{2q-1},
    ~~\mbox{when }1<q<2.
\end{equation}
The asymptotic stiffness in this regime therefore retains an explicit
dependence on the power $q$.

For $q>2$, in contrast, $\nu<1/3$ and
$V_\rho\sim\rho^{1/3}$. The $\omega$ self-interaction and mixed vector
terms scale as $
    V_\omega^{2q}\sim\rho^{1+\nu}$ and $
    V_\omega^2V_\rho^2\sim\rho^{2\nu+2/3}$,
respectively. Since both exponents are smaller than $4/3$, these
terms, together with $V_\omega^2$ and $V_\rho^2$, are asymptotically
sub-leading. The ultra-relativistic kinetic contribution
$\sim\rho^{4/3}$ therefore dominates, yielding\,\cite{CollinsPerry1975,Bjorken1983}
\begin{equation}\label{ps-3}
\mbox{case 3: }    \phi
    \rightarrow
    s^2
    \rightarrow
    \frac{1}{3},
    ~~\mbox{when }q>2.
\end{equation}
Thus, although the underlying RMF interaction remains non-conformal,
the leading high density EOS approaches the ultra-relativistic value
$P/\varepsilon=s^2=1/3$.

The boundary $q=2$ is particularly instructive. In this case,
$V_\omega\sim V_\rho\sim\rho^{1/3}$, so that
$
    V_\omega^4
    \sim
    V_\omega^2V_\rho^2
    \sim\rho^{4/3}$,
of the same order as the ultra-relativistic kinetic contribution.
Thus, unlike the $q>2$ regime, the kinetic, $\omega$ self-interaction,
and mixed vector contributions all coexist at leading order.
Nevertheless, they give the same asymptotic EOS-parameter, and hence
\begin{equation}\label{ps-4}
\mbox{case 4: }    \phi
    \rightarrow
    s^2
    \rightarrow
    \frac{1}{3},
    ~~\mbox{when }q=2.
\end{equation}

The value $q=2$ therefore marks a change in the physical origin of the
asymptotic limit, even though the limit itself remains continuous.
The corresponding high-density scalings of the relevant mean fields,
composition, and thermodynamic quantities are summarized in
TAB.\,\ref{tab:high_density_scalings}. In particular, the distinct
scalings of $V_\omega$, $V_\rho$, and $\delta$ lead consistently to
the asymptotic limits of $\phi$ and $s^2$ discussed above.
The symmetry energy also exhibits distinct high-density behaviors, with $E_{\rm{sym}}(\rho)\sim\rho^{1/3}$ for $c_\omega=0$ and $1<q\leq2$, but $E_{\rm{sym}}(\rho)\sim\rho^{(2q-3)/(2q-1)}$ for $q>2$, corresponding, for example, to $E_{\rm{sym}}(\rho)\sim\rho^{3/5}$ and $\rho^{5/7}$ for $q=3$ and $q=4$, respectively. Notably, although the generalized $\omega$ self-interaction is isoscalar in nature and does not directly distinguish between neutrons and protons, it can substantially alter the density dependence of the symmetry energy, even changing its asymptotic power-law behavior.

\renewcommand*\tablename{\small TAB.}
\begin{table}[h!]
    \centering
    \small
    \renewcommand{\arraystretch}{1.25}
    \setlength{\tabcolsep}{3pt}
    \begin{tabular}{ccccc}
        \hline\hline
        Quantity
        & $c_\omega=0$
        & $1<q<2$
        & $q=2$
        & $q>2$ \\
        \hline

        $V_\omega$
        & $1$
        & ${1}/{(2q-1)}$
        & $1/3$
        & ${1}/{(2q-1)}$ \\

        $V_\rho$
        & $-1$
        & ${(2q-3)}/{(2q-1)}$
        & $1/3$
        & $1/3$ \\

        $\delta$
        & $0$
        & $0$
        & $0$
        & ${2(2-q)}/{3(2q-1)}$ \\

        $E_{\rm{sym}}(\rho)$
        & $1/3$
        & $1/3$
        & $1/3$
        & ${(2q-3)}/{(2q-1)}$ \\

        \hline

        $V_\omega^{2q}$
        & /
        & ${2q}/{(2q-1)}$
        & $4/3$
        & ${2q}/{(2q-1)}$ \\

        $\varepsilon,\,P$
        & $2$
        & ${2q}/{(2q-1)}$
        & $4/3$
        & $4/3$ \\

        \hline

        $\phi,\,s^2$
        & $1$
        & ${1}/{(2q-1)}$
        & $1/3$
        & $1/3$ \\

        \hline\hline
    \end{tabular}
    \caption{High density asymptotic behavior of the relevant RMF
quantities for different $\omega$ self-interaction regimes.
For the quantities above the last horizontal line, the entries
denote the power law exponent $\vartheta$ defined by $\mathcal{O}\sim\rho^\vartheta$ for quantity $\mathcal{O}$;
the last row gives the asymptotic values of $\phi$ and $s^2$.
The kinetic contributions universally scale as
$\varepsilon_{\rm{kin}},P_{\rm{kin}}\sim\rho^{4/3}$ and are therefore
not listed separately; moreover, lepton chemical potentials scale as $\mu_{\rm e}=\mu_{\mu}\sim\rho^{1/3}$, $V_\omega=g_\omega\overline{\omega}_0$ and $ V_\rho=g_\rho\overline{\rho}_0^{(3)}$.}
    \label{tab:high_density_scalings}
\end{table}

\renewcommand*\figurename{\small FIG.}
\begin{figure}[h!]
    \centering
    \includegraphics[width=0.95\columnwidth]{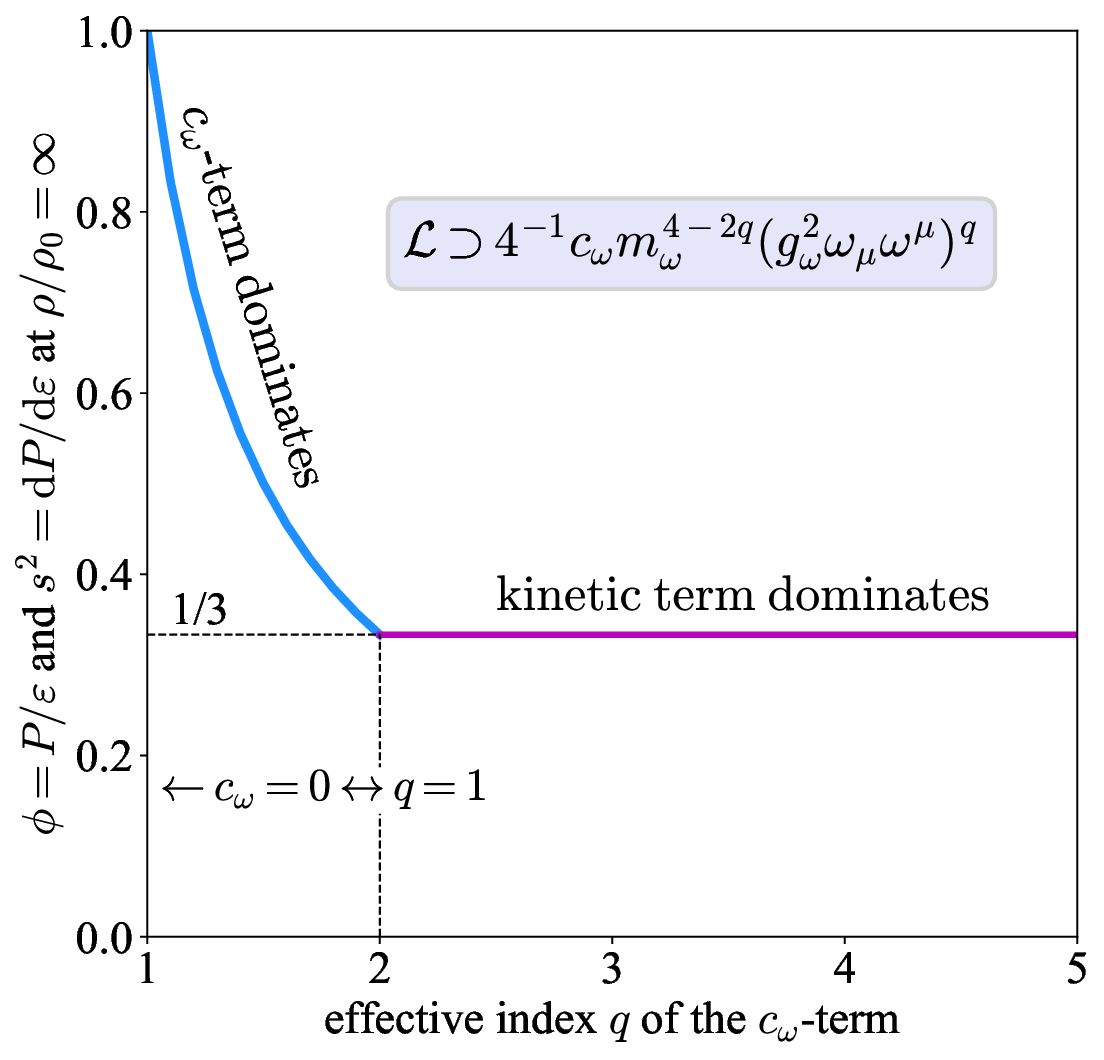}
    \caption{(Color online). Asymptotic limits of the EOS-parameter
$\phi$ and the SSS $s^2$ as $\rho/\rho_0\rightarrow\infty$.
For $c_\omega=0$, the quadratic $\omega$ meson contribution dominates,
giving $\phi\rightarrow s^2\rightarrow1$. For $1<q<2$, the generalized
$\omega$-meson self-interaction dominates and gives
$\phi\rightarrow s^2\rightarrow1/(2q-1)$, as shown by the blue curve.
At $q=2$, the $\omega$ self-interaction, mixed vector interaction, and
ultra-relativistic kinetic contribution all scale as $\rho^{4/3}$ and
give the common limit $1/3$. For $q>2$, the kinetic contribution
dominates and the same limit $\phi\rightarrow s^2\rightarrow1/3$ is
obtained, as indicated by the horizontal magenta line. The $q=1$
endpoint represents only the asymptotic continuation of the $q>1$
result and coincides with the $c_\omega=0$ limit.}
    \label{fig:phi-large-rho}
\end{figure}

Combining the above four cases of (\ref{ps-1}), (\ref{ps-2}), (\ref{ps-3}) and (\ref{ps-4}), the asymptotic limits for the EOS-parameter and the $s^2$ can be summarized:
\begin{equation}
\frac{\rho}{\rho_0}\to\infty:\;
    \phi=\frac{P}{\varepsilon}
    \rightarrow
    s^2
    \rightarrow
    \begin{cases}
    1,
    & c_\omega=0,
    \\
    1/(2q-1),
    & 1<q\leq2,
    \\
    1/3,
    & q>2.
    \end{cases}
\label{eq:cs2_asymptotic_limit}
\end{equation}
as shown in FIG.\,\ref{fig:phi-large-rho}. The two branches for
$c_\omega\neq0$ connect continuously at $q=2$, but their physical
origins are different: the $\omega$ self-interaction dominates for
$1<q<2$, whereas the ultra-relativistic kinetic contribution dominates
for $q>2$. At $q=2$, the kinetic, $\omega$ self-interaction, and mixed
vector contributions coexist at leading order and give the same limit
$1/3$. We also note that the $q=1$ endpoint represents only the
asymptotic continuation of the $q>1$ result. Although
$1/(2q-1)\rightarrow1$ as $q\rightarrow1^+$, this limit should not in
general be identified with the $c_\omega=0$ case at finite densities.

These distinct asymptotic behaviors provide a controlled way to
generate substantially different high density EOS evolutions within
the same underlying RMF framework. More importantly, NS interiors
probe finite densities rather than the strict limit
$\rho\rightarrow\infty$. The different asymptotic tendencies can
nevertheless already influence $s^2$ and $\phi$ over the density
range reached in NS cores, leading to appreciably different core
behaviors among the EOS families. It is this finite density
manifestation, rather than the asymptotic limits themselves, that is
directly relevant to the central SSS extraction considered in this
work.
The resulting diversity provides a useful testing ground for the
stellar structure analysis developed below. If common information on
the central SSS can be identified across EOS families with such
different high density behaviors, its extraction can become less
sensitive to the particular dense matter model used to describe the
NS core.

\subsection{Low Density Behavior of the SSS and the Factor $f_0\equiv s^2/\phi$}\label{subs_LD}

It is also instructive to examine the SSS in the opposite,
low-density regime. Expanding the closed analytical expression for
$s^2$ in Eq.\,\eqref{eq:rmf_s2_closed} for $\rho\rightarrow0$, we
obtain straightforwardly,
\begin{equation}
    s^2=
    \frac{(3\pi^2)^{2/3}}{3M_{\rm N}^2}\rho^{2/3}
    +\frac{\mathcal{C}}{M_{\rm N}}\rho
    +\mathcal O(\rho^{4/3}),
    \label{eq:app_s2_low_density}
\end{equation}
where the coefficient $\mathcal C$ is constructed by meson fields:
\begin{equation}
    \mathcal C \equiv
    \left({g_\omega^2}/{m_\omega^2}
    +{g_\rho^2}/{m_\rho^2}\right)
    -\left({g_\sigma^2}/{m_\sigma^2}
    +{g_\delta^2}/{m_\delta^2}\right).
    \label{eq:app_Cn}
\end{equation}
The leading term of $s^2$, proportional to $\rho^{2/3}$, represents the ideal
Fermi gas contribution, while the term proportional to $\rho$ gives
the leading interaction correction. The latter is governed by the
competition between the repulsive vector interactions and the
attractive scalar interactions. If the combined vector repulsion
exceeds the scalar attraction, $\mathcal C>0$, the interaction
correction increases the SSS and stiffens the EOS relative to the
ideal Fermi gas. Conversely, for $\mathcal C<0$, the scalar attraction
dominates this correction, reducing the SSS and softening the EOS.

The low-density behavior can also be characterized through the ratio
between the SSS and the EOS-parameter
$\phi= P/\varepsilon$. 
Similar to the
SSS, $\phi$ admits the following low-density expansion:
\begin{equation}
\phi\approx
\frac{(3\pi^2)^{2/3}}{5M_{\rm N}^{2}}\rho^{2/3}
+\frac{\mathcal C}{2M_{\rm N}}\rho
+\mathcal O(\rho^{4/3}).
\end{equation}Defining $f_0\equiv s^2/\phi$\,\cite{CaiLi2025Review}, its
low-density expansion is
\begin{equation}
    f_0\approx\frac{5}{3}
    +\frac{5}{6}\frac{M_{\rm N}\mathcal C}
    {(3\pi^2)^{2/3}}\rho^{1/3}
    +\mathcal O(\rho^{2/3})
    \label{eq:app_f0_low_density}
\end{equation}
Equivalently, by using the neutron Fermi momentum in PNM
$k_{\rm F}^{\rm n}=(3\pi^2\rho)^{1/3}$,
Eq.\,\eqref{eq:app_f0_low_density} can be expressed as
\begin{equation}
    f_0\approx\frac{5}{3}
    +\frac{5M_{\rm N}\mathcal C}
    {18\pi^2}k_{\rm F}^{\rm n}
    +\mathcal O\left(k_{\rm F}^{{\rm n},2}\right)
    =\frac{5}{3}\left(1+\frac{M_{\rm N}\mathcal{C}k_{\rm{F}}^{\rm{n}}}{6\pi^2}\right)+\mathcal{O}\left(k_{\rm{F}}^{\rm{n},2}\right).
    \label{eq:app_f0_low_density_kF}
\end{equation}

Therefore, $f_0$ approaches the ideal Fermi-gas value $5/3$ in the
strict low-density limit $\rho\rightarrow0$. At finite but low
densities, however, the leading interaction correction scales as
$\mathcal C\rho^{1/3}$ and determines whether $f_0$ is shifted upward
or downward from $5/3$. A positive $\mathcal C$, corresponding to a
dominant net vector repulsion, drives $f_0$ above $5/3$, whereas a
negative $\mathcal C$, associated with a dominant net scalar
attraction, drives it below $5/3$. The relatively slow
$\rho^{1/3}$ dependence of this correction also implies that
noticeable deviations from the ideal Fermi-gas value may already
develop at finite low densities.

These considerations suggest that $f_0$ should remain a quantity of
order unity in the low-density regime, with values broadly around
$f_0\approx1$-$2$ depending on the sign and magnitude of $\mathcal C$.
The ideal Fermi-gas value $5/3$ naturally lies within this range, while
the leading interaction correction shifts $f_0$ toward its lower or
upper side according to the relative importance of scalar attraction
and vector repulsion. Interestingly, this characteristic range is
also consistent with recent chiral perturbation theory (ChPT)
calculations of low-density nuclear matter\,\cite{Tews2018,Essick2021}, providing an
independent microscopic indication that $f_0$ remains of order unity
in this density regime.

The low-density behavior provides a useful complement to the
high-density analysis presented above. At sufficiently low densities,
the leading behavior of the SSS is controlled by the universal
Fermi-gas contribution, with the interaction dependence entering only
through sub-leading corrections such as $\mathcal C$. By contrast, as
the density increases, the nonlinear interactions become increasingly
important and generate the substantially different behaviors of
$s^2$ and $\phi$ among the different model families. The evolution
from the relatively universal low-density behavior to the strongly
model-dependent high-density behavior is particularly relevant for
the stellar-structure scaling studied below, which aims to extract the
central SSS while reducing its sensitivity to the underlying
dense-matter EOS.

\section{Nuclear Matter and NS Properties of the Nonlinear RMF Model Ensembles}\label{SEC_3}

This section examines the nuclear matter and NS properties of the six
RMF model ensembles. In Subsection \ref{subs_EOSCon}, we construct the
EOS ensembles and impose the adopted nuclear and astrophysical
constraints. In Subsection \ref{subs_NM}, we examine their nuclear
matter properties and compare them with available empirical and
microscopic constraints. In Subsection \ref{subs_NS}, we investigate
the resulting NS mass-radius (M-R) relations. In
Subsection \ref{subs_phis2}, we further analyze the profiles of
$\phi=P/\varepsilon$ and $s^2$ inside NSs, with particular attention
to their dependence on the high density behavior of the different
RMF families. These results provide the motivation for the IPAD-TOV
analysis in the next section.

\subsection{EOS Construction and Constrained Model Sampling}\label{subs_EOSCon}

We construct the EOS of cold, charge-neutral matter in
$\beta$-equilibrium using the RMF framework introduced above. In the
NS core, charge neutrality and $\beta$-equilibrium require
$\rho_{\rm{p}}=\rho_{\rm{e}}+\rho_\mu$ and
$\mu_{\rm{n}}-\mu_{\rm{p}}=\mu_{\rm{e}}=\mu_\mu$ when muons are
present. Electrons and muons are treated as ideal Fermi gases, and
their contributions are added to the RMF energy density and
pressure. The core-crust transition density
$\rho_{\rm{t}}$ is determined using the thermodynamic
method\,\cite{Xu2009,Kubis2007,Cai2012}. In the inner crust,
$\rho_{\rm{out}}<\rho<\rho_{\rm{t}}$, we adopt\,\cite{Xu2009,Kubis2007,Cai2012}
\begin{equation}
    P(\varepsilon)=a_{\rm{crust}}+b_{\rm{crust}}\varepsilon^{4/3},
\end{equation}
where the coefficients $a_{\rm{crust}}$ and $b_{\rm{crust}}$ are determined by
matching continuously to the core and outer-crust
EOSs\,\cite{Iida1997},
giving
\begin{equation}
a_{\rm{crust}}=
\frac{P_{\rm{out}}\varepsilon_{\rm{t}}^{4/3}
-P_{\rm{t}}\varepsilon_{\rm{out}}^{4/3}}
{\varepsilon_{\rm{t}}^{4/3}-\varepsilon_{\rm{out}}^{4/3}},~~
b_{\rm{crust}}=
\frac{P_{\rm{t}}-P_{\rm{out}}}
{\varepsilon_{\rm{t}}^{4/3}-\varepsilon_{\rm{out}}^{4/3}},
\end{equation}
where the subscripts ``${\rm t}$'' and ``${\rm{out}}$'' refer to the core-crust and inner-outer crust boundaries, respectively.
For densities below the outer boundary
$\rho_{\rm{out}}\approx2.46\times10^{-4}\,{\rm{fm}}^{-3}$, we use
the Baym--Pethick--Sutherland (BPS) EOS as well as the Feynman--Metropolis--Teller
(FMT) EOS\,\cite{Baym1971}.

To construct broad ensembles for the six RMF model families, we sample
the characteristic nuclear matter parameters over the ranges summarized
in TAB.\,\ref{tab:sampling_ranges}. Specifically, we take
$0.14\,{\rm{fm}}^{-3}\lesssim\rho_0\lesssim0.16\,{\rm{fm}}^{-3}$
for the saturation density\,\cite{Dutra2014RMF},
$-17\,{\rm{MeV}}\lesssim E_0(\rho_0)\lesssim-15\,{\rm{MeV}}$
for the binding energy per nucleon\,\cite{Dutra2014RMF},
$220\,{\rm{MeV}}\lesssim K_0\lesssim260\,{\rm{MeV}}$
for the incompressibility\,\cite{Xu2021K0,Khan2012,KhanMargueron2013,Blaizot1980,Colo2004,GargColo2018,XieLi2021,WangDeng2025}, and
$-600\,{\rm{MeV}}\lesssim J_0\lesssim0\,{\rm{MeV}}$
for the skewness coefficient\,\cite{Cai2017}. The nucleon effective
mass in SNM is sampled within
$0.5\lesssim M_0^*(\rho_0)/M_{\rm{N}}\lesssim0.7$\,\cite{Chen2007,Jaminon1989,Fuchs1995,Li2018EffectiveMass}.
For the isovector sector, we adopt
$29\,{\rm{MeV}}\lesssim E_{\rm{sym}}(\rho_0)\lesssim35\,{\rm{MeV}}$
for the symmetry energy and
$30\,{\rm{MeV}}\lesssim L\lesssim80\,{\rm{MeV}}$
for its slope parameter\,\cite{LiHan2013,LiCaiXieZhang2021}.
These ranges are broadly consistent with current terrestrial
nuclear matter constraints.

\renewcommand*\tablename{\small TAB.}
\begin{table}[h!]
    \centering
    \renewcommand{\arraystretch}{1.2}
    \setlength{\tabcolsep}{6pt}
    \begin{tabular}{cc}
        \hline\hline
        Quantity & Range or constraint \\
        \hline
        $\rho_0$
        & $0.14\,\rm{fm}^{-3}\lesssim\rho_0\lesssim0.16\,\rm{fm}^{-3}$ \\

        $E_0(\rho_0)$
        & $-17\,\rm{MeV}\lesssim E_0(\rho_0)\lesssim-15\,\rm{MeV}$ \\

        $K_0$
        & $220\,\rm{MeV}\lesssim K_0\lesssim260\,\rm{MeV}$ \\

        $J_0$
        & $-600\,\rm{MeV}\lesssim J_0\lesssim0\,\rm{MeV}$ \\

        $M_0^*(\rho_0)/M_{\rm{N}}$
        & $0.5\lesssim M_0^*(\rho_0)/M_{\rm{N}}\lesssim0.7$ \\

        $E_{\rm{sym}}(\rho_0)$
        & $29\,\rm{MeV}\lesssim E_{\rm{sym}}(\rho_0)\lesssim35\,\rm{MeV}$ \\

        $L$
        & $30\,\rm{MeV}\lesssim L\lesssim80\,\rm{MeV}$ \\

        \hline
        $\Lambda_{1.4}$
        & $\Lambda_{1.4}\lesssim580$ \\

        $M_{\rm{TOV}}$
        & $M_{\rm{TOV}}\gtrsim2.01M_\odot$ \\
        \hline\hline
    \end{tabular}

    \caption{Sampling ranges of the nuclear matter parameters and
    astrophysical constraints imposed on the resulting NS EOSs.}
    \label{tab:sampling_ranges}
\end{table}

For each of the six RMF model families, the nuclear matter parameters
are sampled uniformly within the specified ranges, and the corresponding
coupling constants are obtained by solving the analytical relations
between the model couplings and the nuclear matter parameters, while
$g_\delta$ and $C_{\delta\sigma}$ are fixed to their values in the
FSU-$\delta$6.7 model of Ref.\,\cite{Li2022}.
In particular, a positive $c_\omega$ should be obtained via this procedure.
For the
$c_\omega=0$ family, fixing $c_\omega=0$ reduces the number of
independent coupling constants by one, so that $J_0$ is not required
as an input parameter.
In addition to the nuclear matter constraints, we impose the two
astrophysical requirements
$\Lambda_{1.4}\lesssim580$\,\cite{Abbott2017,Abbott2018} and
$M_{\rm{TOV}}\gtrsim2.01M_\odot$\,\cite{Fonseca2021}.
Any sampled parameter set that violates either requirement is
discarded. The sampling and filtering procedure is repeated until
$1000$ accepted parameter sets are obtained for each RMF family,
yielding a total ensemble of $6000$ physically constrained
NS EOSs.

It is important to emphasize that the EOS ensembles constructed in this
way are intentionally broad. The sampled nuclear matter parameters span
wide ranges allowed by current empirical knowledge, while the six RMF
families exhibit substantially different behaviors at high densities,
despite satisfying the same basic nuclear and astrophysical constraints.
Such a broad ensemble is essential for the extraction of NS core
properties considered below. In particular, an estimate of the central
SSS obtained from a narrowly restricted class of EOSs could largely
reflect the assumptions built into that specific model class. By
allowing substantial variations in both the nuclear parameters and the
high density behavior of the EOS, the present ensembles provide a
meaningful basis for examining whether the central SSS can be extracted
with reduced sensitivity to the underlying dense matter model.

\renewcommand*\figurename{\small FIG.}
\begin{figure}[h!]
    \centering
    \includegraphics[width=0.98\columnwidth]{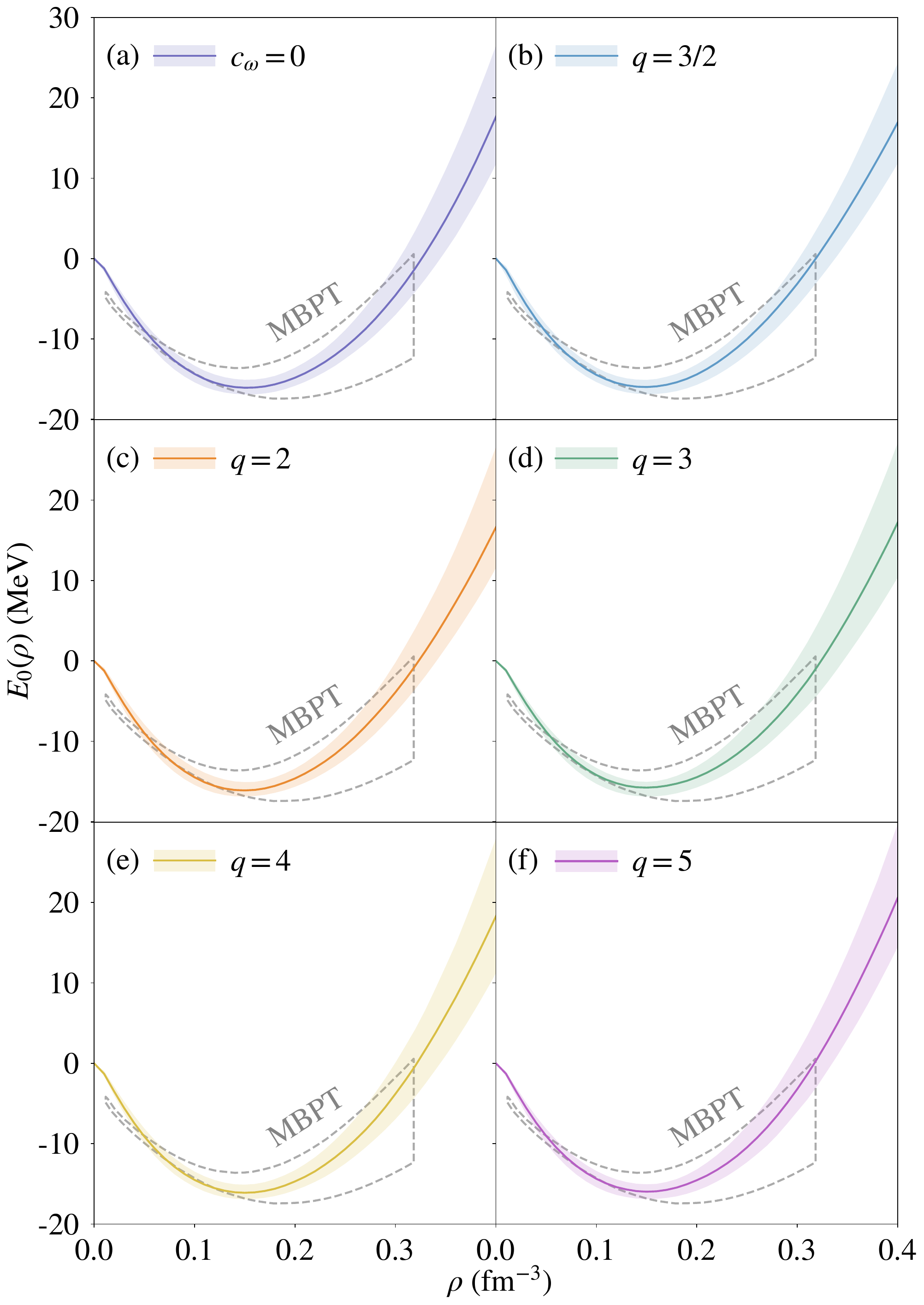}
    \caption{(Color online). EOS $E_0(\rho)$ of SNM
    as a function of baryon density $\rho$ for
    the six RMF families: (a) $c_\omega=0$ and
    (b)-(f) $q=3/2$, 2, 3, 4, 5, respectively.
    Solid curves denote the medians and colored bands the
    $95\%$ intervals of the RMF ensembles. The gray dashed
    boundaries indicate the constraint from many-body
    perturbation theory (MBPT) calculations with chiral effective interaction\,\cite{Drischler2026}.}
    \label{fig:E0}
\end{figure}

\renewcommand*\figurename{\small FIG.}
\begin{figure}[h!]
    \centering
    \includegraphics[width=\columnwidth]{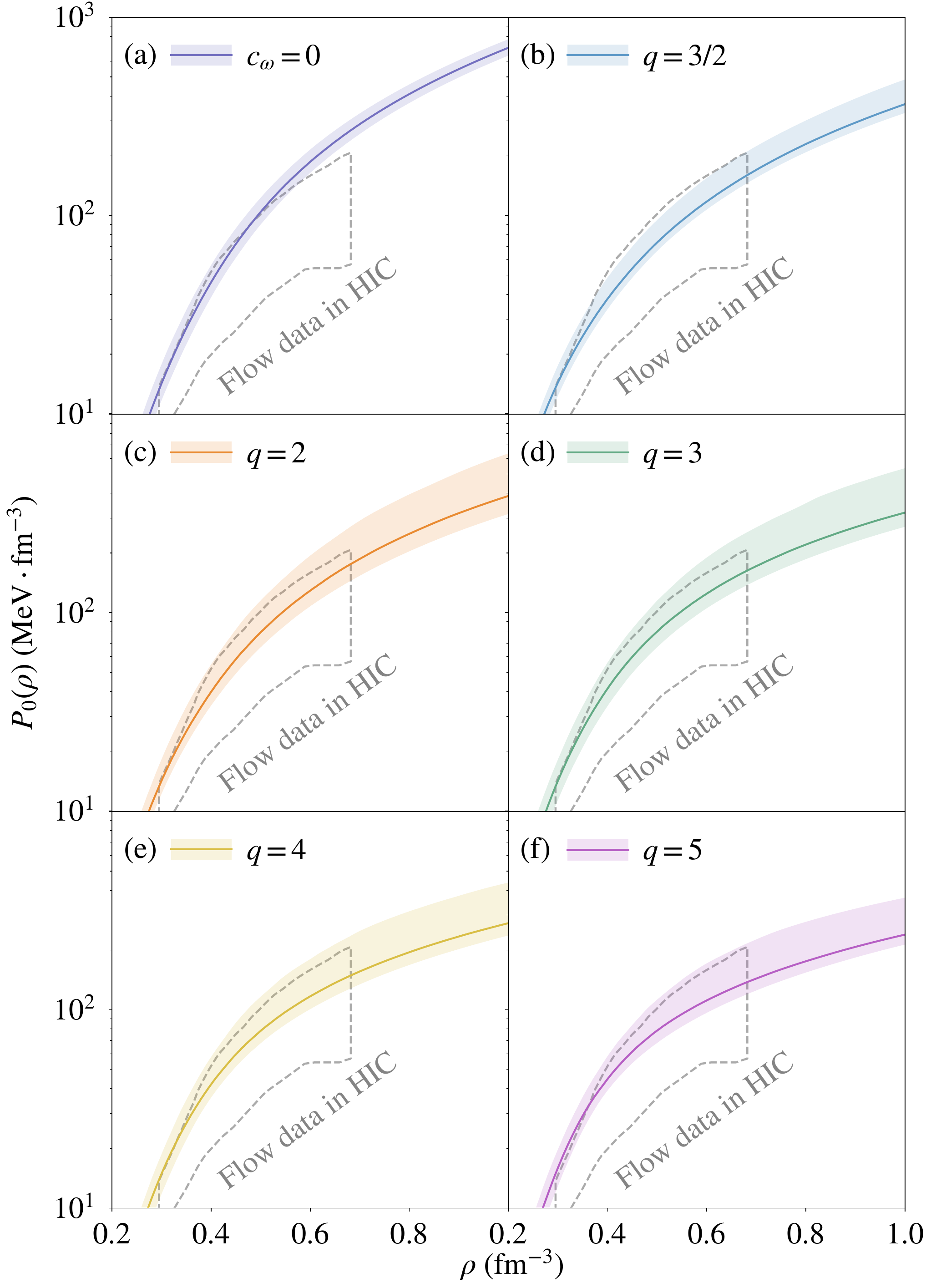}
    \caption{(Color online). Pressure of SNM,
    $P_0(\rho)$, as a function of baryon density $\rho$ for
    the six RMF families similar as FIG.\,\ref{fig:E0}. The gray dashed
    boundaries indicate the constraint inferred from flow
    data in heavy ion collisions (HICs)\,\cite{Danielewicz2002}.}
    \label{fig:P_SNM}
\end{figure}

\renewcommand*\figurename{\small FIG.}
\begin{figure}[h!]
    \centering
    \includegraphics[width=0.98\columnwidth]{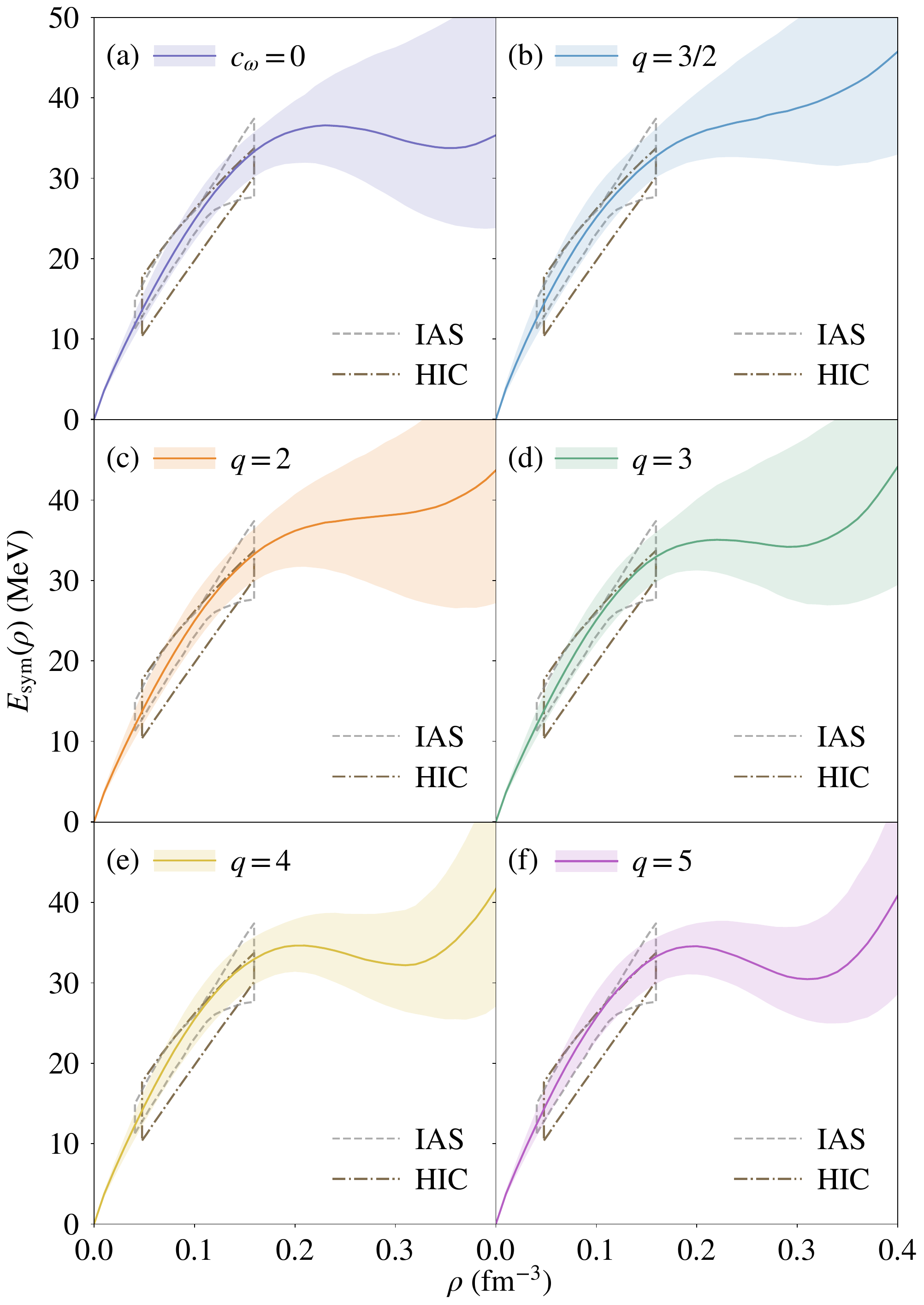}
    \caption{(Color online). Symmetry energy 
    $E_{\rm{sym}}(\rho)$ as a function of baryon density
    $\rho$ for the six RMF families similar as FIG.\,\ref{fig:E0}. The gray dashed and brown dash-dotted boundaries indicate the constraints
    from isobaric analog states (IAS)\,\cite{Danielewicz2014}
    and HIC\,\cite{Tsang2009}, respectively.}
    \label{fig:E_sym}
\end{figure}

\renewcommand*\figurename{\small FIG.}
\begin{figure}[h!]
    \centering
    \includegraphics[width=0.98\columnwidth]{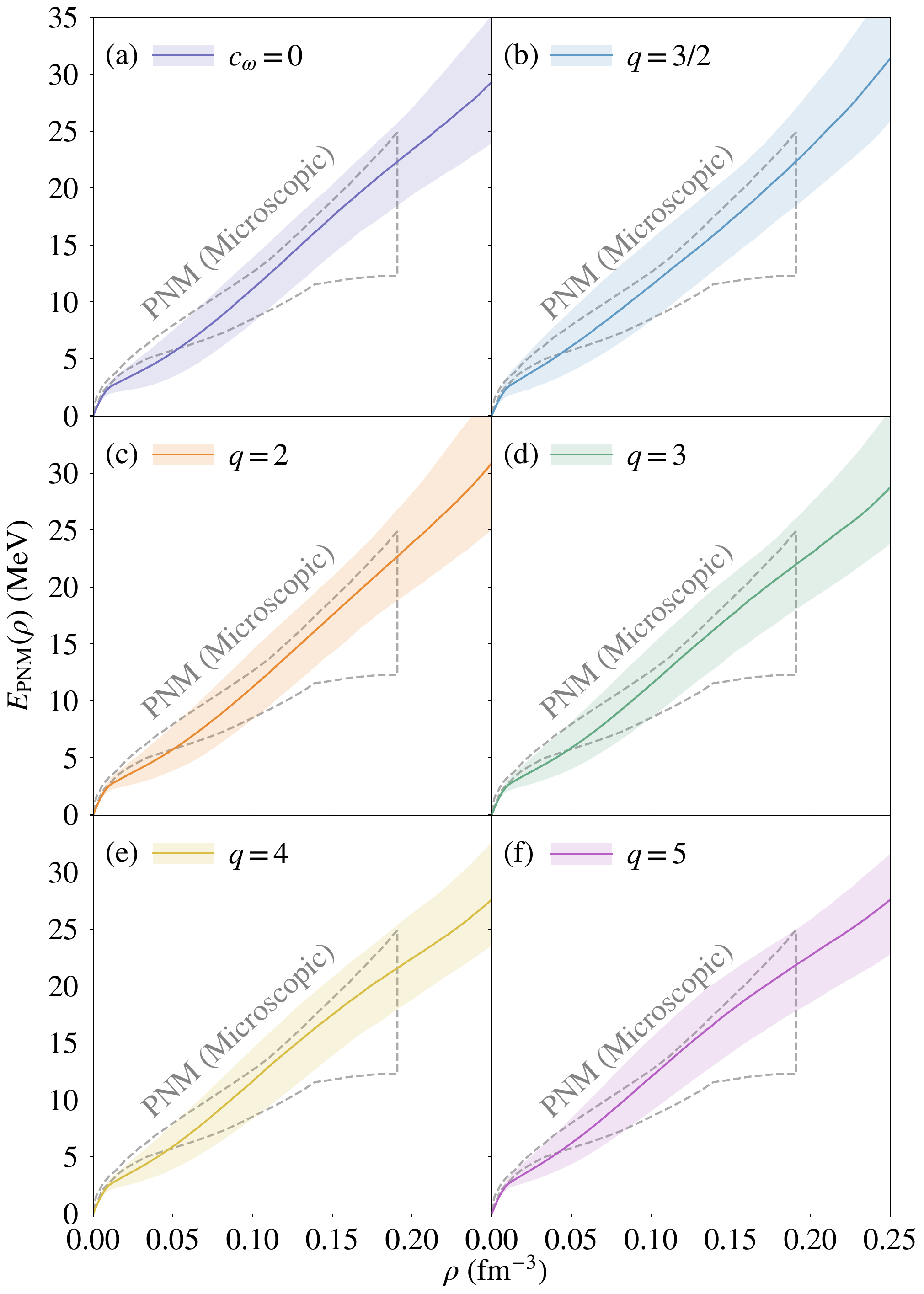}
    \caption{(Color online). EOS $E_{\rm{PNM}}(\rho)$ of PNM as a function of baryon density
    $\rho$ for the six RMF families similar as FIG.\,\ref{fig:E0}.
     The gray dashed boundaries indicate the constraint from microscopic
    neutron matter calculations\,\cite{WangChen2025,ZhangChen2023,Huth2021}.}
    \label{fig:E_PNM}
\end{figure}

\subsection{Nuclear Matter Properties of the RMF Ensembles: EOS of SNM, Symmetry Energy and EOS of PNM}\label{subs_NM}

After constructing the six constrained RMF ensembles, we first examine
their nuclear matter properties and compare them with available
empirical and microscopic constraints. In particular, we consider the
energy per nucleon (EOS) and pressure of SNM, the nuclear symmetry energy,
and the EOS of PNM. For each quantity, we construct the
median and the $95\%$ interval of the accepted EOSs within each RMF
family. The resulting comparisons are shown in
FIGs.\,\ref{fig:E0}-\ref{fig:E_PNM}.

FIG.\,\ref{fig:E0} shows the EOS of SNM,
$E_0(\rho)$, at densities between $0\lesssim\rho\lesssim0.4\,\rm{fm}^{-3}$. The results are compared with the
constraint inferred from the recent many-body perturbation theory (MBPT)
calculations using the chiral effective interaction\,\cite{Drischler2026}. The six RMF families exhibit similar behavior over the density range shown, and their $95\%$ intervals broadly overlap with the MBPT constraint. This consistency is expected since all six ensembles are constructed using the same empirical ranges of the nuclear matter parameters around the saturation density. The pressure of SNM, $P_0(\rho)$, is shown in
FIG.\,\ref{fig:P_SNM} over a broader density range and is compared
with the constraint inferred from flow measurements in heavy ion
collisions (HICs)\,\cite{Danielewicz2002}. Differences among the six families
become more visible with increasing density, reflecting the growing
influence of their different $\omega$-meson self-interactions.
Nevertheless, the $95\%$ intervals remain broadly compatible with the
flow constraint. The main exception is the $c_\omega=0$ family, whose
band extends moderately above the constraint region over part of the
density range. This behavior indicates the comparatively stiff EOS of
this family and anticipates the more pronounced differences that emerge
at higher densities.

The isovector sector is examined in FIG.\,\ref{fig:E_sym}, where the
symmetry energy $E_{\rm{sym}}(\rho)$ is compared with constraints from
isobaric analog states (IAS) and HICs\,\cite{Danielewicz2014,Tsang2009}.
The $95\%$ intervals of all six families broadly overlap with the
empirical constraint regions, and the differences among them remain
relatively limited over the densities where these constraints are
available, despite their distinct behaviors at higher densities. This
is expected because the $c_\omega$ term represents an isoscalar
interaction, and its influence on $E_{\rm{sym}}(\rho)$ and the PNM EOS
is transmitted through their coupled analytical expressions, in which
the isoscalar mean fields enter together with the isovector
contributions. We also observe a flattening of $E_{\rm{sym}}(\rho)$
over an intermediate density range, which is mainly associated with
the $\delta$-$\sigma$ mixing interaction governed by
$C_{\delta\sigma}$. As a further test, FIG.\,\ref{fig:E_PNM} shows
$E_{\rm{PNM}}(\rho)$ together with constraints from microscopic neutron
matter calculations\,\cite{WangChen2025,ZhangChen2023,Huth2021}. The
six families again give broadly overlapping $95\%$ intervals and
remain consistent with the microscopic constraints over the density
range shown. The simultaneous agreement of $E_{\rm{sym}}(\rho)$ and
$E_{\rm{PNM}}(\rho)$ with their respective constraints provides an
additional check of the isovector properties of the constructed
ensembles.

Taken together, FIGs.\,\ref{fig:E0}-\ref{fig:E_PNM} demonstrate an
important feature of the RMF ensembles constructed in this work. The
six families are broadly consistent with available nuclear matter
information over the density regions where empirical or microscopic
constraints are available, while they are not required to exhibit the
same behavior at higher densities. Thus, the substantial differences
among these families at densities relevant to NS interiors cannot be
simply attributed to a failure to reproduce the basic nuclear matter
constraints considered here. Instead, they reflect the remaining
freedom in extrapolating the EOS beyond the density region that is
relatively well constrained. This provides a broad and physically
meaningful set of EOS ensembles for examining NS properties and, in
particular, for assessing how robustly the central SSS can be extracted
in the presence of different high density behaviors.

\subsection{NS Mass-Radius Relations}\label{subs_NS}

\begin{figure}[h!]
    \centering
    \includegraphics[width=0.95\columnwidth]{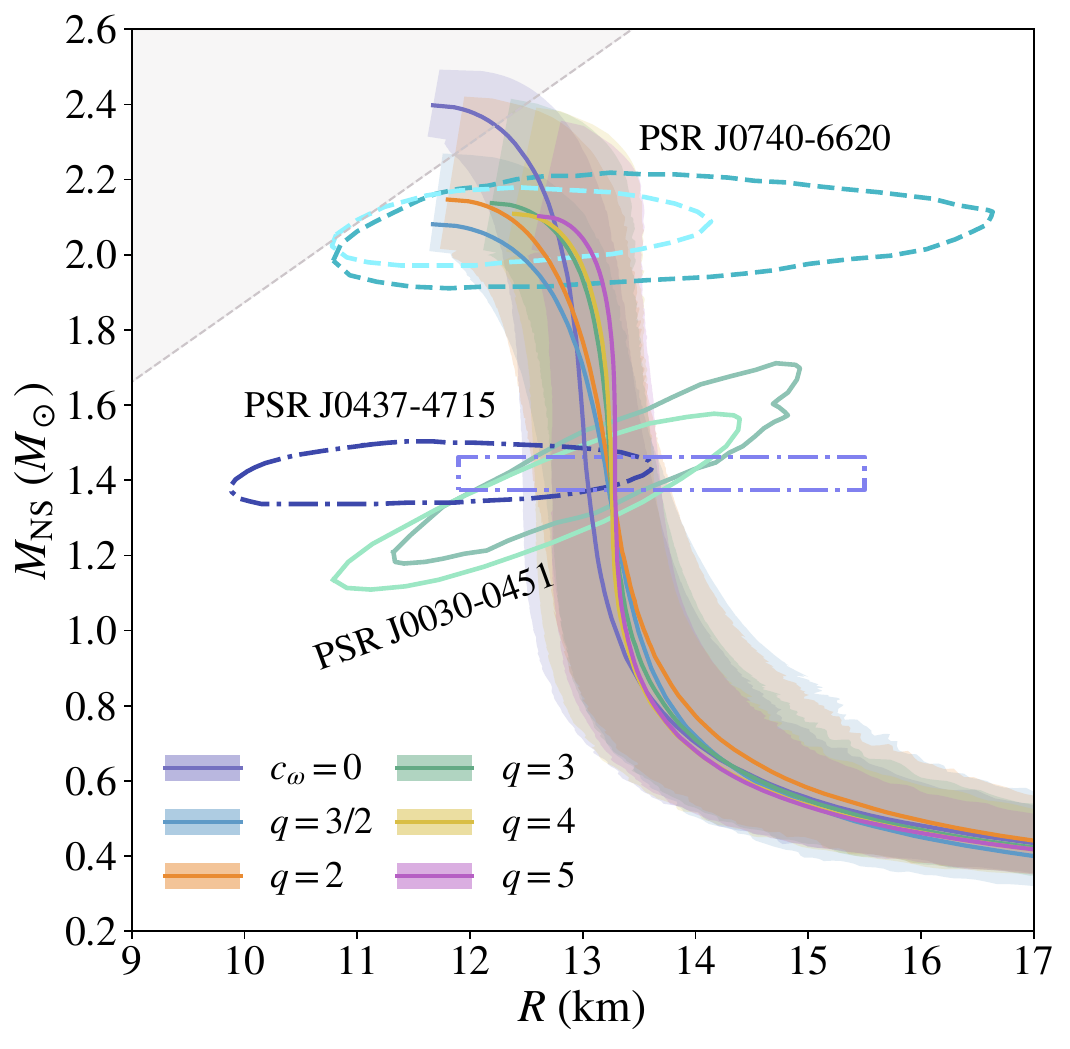}
    \caption{(Color online). NS M-R relations for the six RMF
    families. Solid curves denote the medians and colored bands the
    $95\%$ intervals of the RMF ensembles. Observational constraints
    for PSR J0030+0451\,\cite{Miller19,Riley19},
    PSR J0740+6620\,\cite{Miller21,Riley21}, and
    PSR J0437-4715\,\cite{Choud24,Miller26} are shown for comparison.
    The gray dashed line denotes the effective causality boundary
    $R/\rm{km}\gtrsim4.73\,M_{\rm{NS}}/M_\odot+1.14$
    from Ref.\,\cite{CaiLiZhang2023PRD} in the IPAD-TOV framework, with the upper left gray region
    disfavored by this relation.}
    \label{fig:MR}
\end{figure}

We next examine the M-R relations predicted by the six RMF
ensembles constructed in Subsection \ref{subs_EOSCon}. In
FIG.\,\ref{fig:MR}, we show them,
together with the recent NICER observational constraints for
PSR J0030+0451\,\cite{Miller19,Riley19},
PSR J0740+6620\,\cite{Miller21,Riley21}, as well as
PSR J0437-4715\,\cite{Choud24,Miller26}. The solid curves denote the
median relations, while the colored bands represent the corresponding
$95\%$ intervals. All six families are broadly consistent with the
observational constraints shown in the figure. At low and intermediate
masses, the substantial overlap among the bands reflects their common
calibration to empirical nuclear matter properties around saturation
density. For canonical NSs, the predicted radii span approximately
$12$-$14\,\mathrm{km}$.

\renewcommand*\figurename{\small FIG.}

The differences among the families become more pronounced toward large
NS masses, where the stellar structure increasingly probes the high
density EOS and is therefore sensitive to the form of the
$\omega$-meson self-interaction. In particular, the $c_\omega=0$
family generally supports larger maximum masses, consistent with its
stiffer asymptotic behavior, whereas the families with finite
$c_\omega$ exhibit distinct high mass trajectories and termination
points. Nevertheless, the broad overlap of their $95\%$ intervals and
the present observational contours precludes a clear discrimination
among the different families. This diversity is useful for the present
analysis, as it provides EOS ensembles spanning substantially different
high density behaviors while remaining compatible with current
astrophysical constraints. We also show an effective M-R boundary
obtained previously within the IPAD-TOV framework\,\cite{CaiLiZhang2023PRD},
namely ${R}/{\mathrm{km}}\gtrsim
4.73{M_{\rm{NS}}}/{M_\odot}+1.14$. Among the six families, the
$c_\omega=0$ family predicts the most compact maximum-mass
configurations and therefore approaches, or may partially cross, this
boundary, whereas the median relations of the finite-$c_\omega$
families remain broadly consistent with it.

\subsection{Radial Profiles of $\phi=P/\varepsilon$ and $s^2$ in NSs}\label{subs_phis2}

\begin{figure}[h!]
    \centering
    \includegraphics[width=0.95\columnwidth]{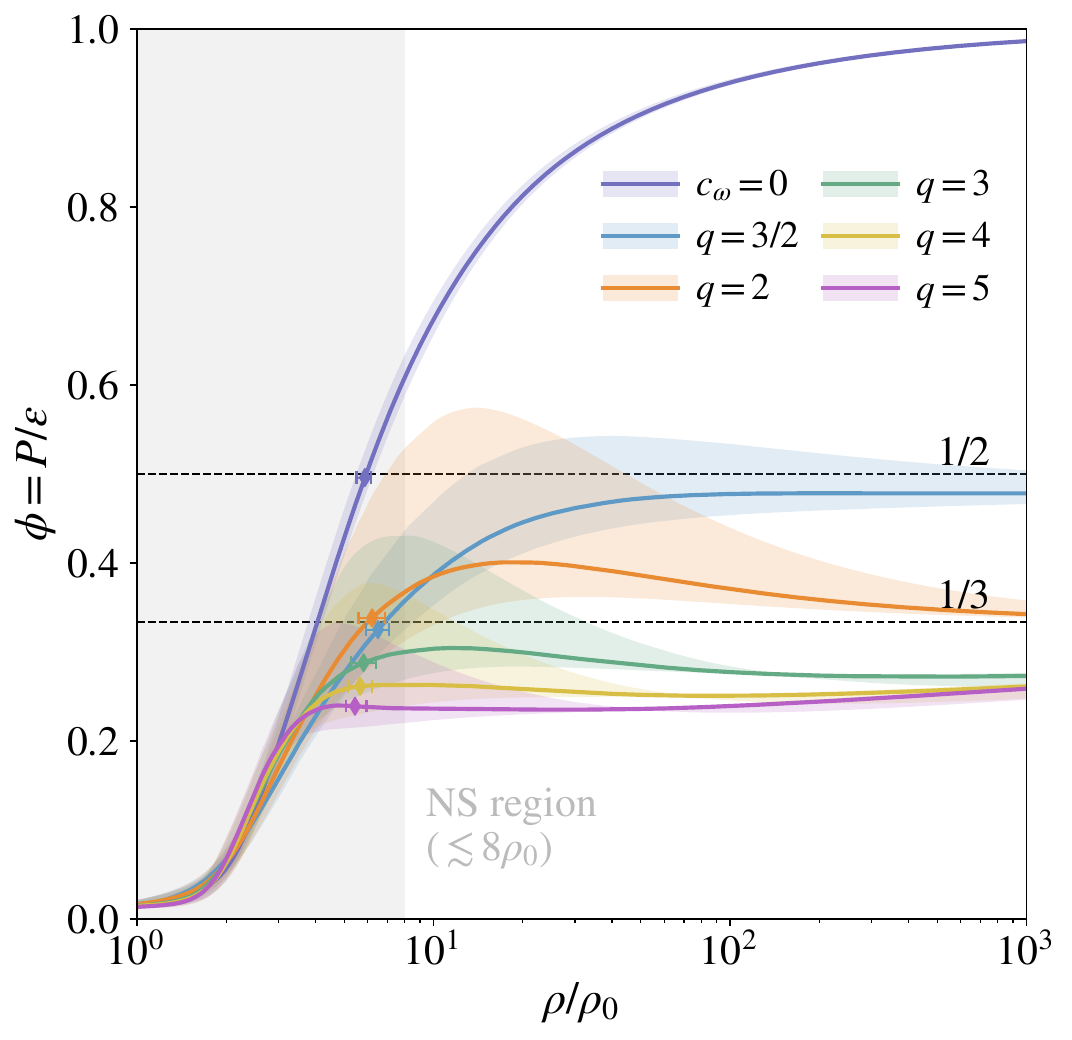}
    \caption{(Color online). EOS-parameter
    $\phi\equiv P/\varepsilon$ as a function of normalized baryon
    density $\rho/\rho_0$ for cold, charge neutral matter in
    $\beta$-equilibrium. Different colors denote the six RMF families,
    with solid curves showing the medians and colored bands their
    $95\%$ intervals. The gray region indicates the approximate
    density range relevant to NS interiors,
    $\rho\lesssim8\rho_0$. The circular marker and short horizontal
    segment for each family denote, respectively, the median and
    $95\%$ interval of the central density of the corresponding
    TOV NSs.}
    \label{fig:P/eps}
\end{figure}

\begin{figure}[h!]
    \centering
    \includegraphics[width=0.95\columnwidth]{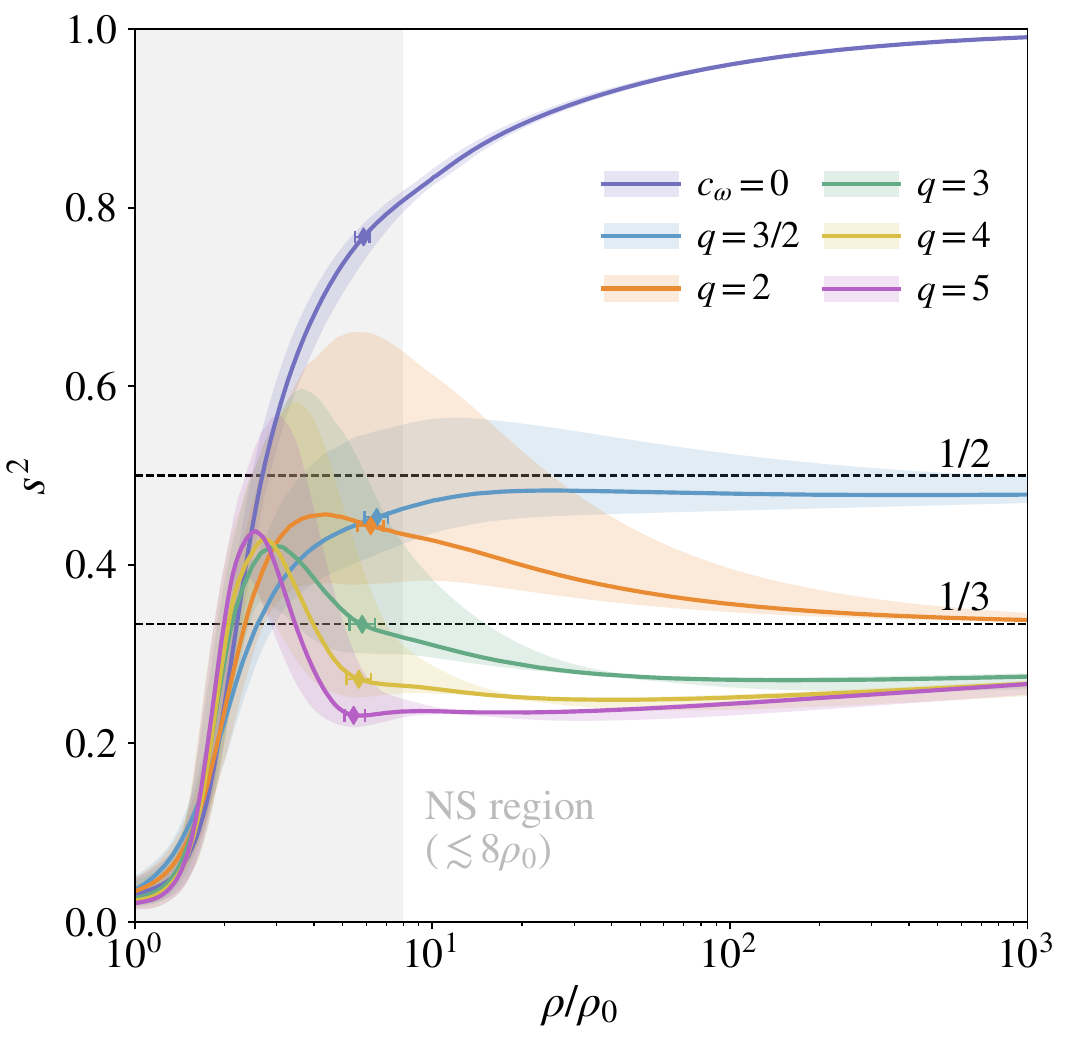}
    \caption{(Color online). The same as FIG.\,\ref{fig:P/eps} but for the SSS profile $s^2\equiv\d P/\d\varepsilon$ for different EOS families.}
    \label{fig:sss_band}
\end{figure}

Although the six families give broadly similar macroscopic stellar
properties within current observational uncertainties, their EOSs can
behave quite differently in the stellar interior. To illustrate this
point, we calculate the EOS-parameter
$\phi$ and the SSS
$s^2$ for cold, charge neutral matter in
$\beta$-equilibrium. For each family, the median and $95\%$ interval
are shown in FIGs.\,\ref{fig:P/eps} and \ref{fig:sss_band},
respectively; the $s^2$ profile is obtained via Eq.\,(\ref{eq:rmf_s2_closed}). We extend the calculation to $10^3\rho_0$ only to expose the
asymptotic behavior clearly; at such extreme densities, a purely
nucleonic description is no longer expected to be realistic and
quark degrees of freedom should eventually become relevant. The gray
region, $\rho\lesssim8\rho_0$, approximately indicates the density
range relevant to NSs. Importantly, as will be discussed below, the
distinct asymptotic behaviors already leave pronounced imprints on
the EOS within this NS-relevant density range.
The results in FIGs.\,\ref{fig:P/eps} and \ref{fig:sss_band}
demonstrate that EOS ensembles satisfying broadly the same nuclear and
astrophysical constraints can nevertheless exhibit markedly different
behavior at high densities. They also confirm the analytical
asymptotic limits derived in Subsection \ref{subs_HD}. For $c_\omega\neq0$,
both $\phi$ and $s^2$ approach $1/2$ for $q=3/2$ and $1/3$ for
$q\geq2$, whereas the $c_\omega=0$ family approaches the causal limit
of unity. For $q>2$, the convergence toward $1/3$ is relatively slow
over the density range displayed, but the upturn of both quantities at
the highest densities is consistent with the predicted asymptotic
behavior. 
Moreover, both $\phi$ and $s^2$ exhibit rather similar behaviors among the six families at densities $\rho\lesssim3\rho_0$, primarily because they are constrained by the same empirical nuclear matter properties around saturation density.
Related asymptotic properties of nonlinear RMF models were also
analyzed in Ref.\,\cite{MullerSerot96}, which provides a detailed
discussion of the high density behavior of different nonlinear
interactions.

\renewcommand*\figurename{\small FIG.}

More importantly for NSs, substantial differences among the six
families already emerge at finite densities well before the asymptotic
regime is reached. This behavior can be understood from the competition
between the linear and nonlinear terms in the $\omega$ field equation.
At relatively low densities, $V_\omega$ is small and the linear term
$m_\omega^2V_\omega/g_\omega^2$ dominates over the nonlinear
$\omega$ self-interaction. The resulting rapid growth of $V_\omega$
enhances the vector repulsion and drives $s^2$ upward. As the density
increases, the nonlinear term
$q c_\omega m_\omega^{4-2q}V_\omega^{2q-1}/2$
becomes increasingly important and suppresses the growth of the vector
field, which eventually follows the asymptotic scaling
$V_\omega\sim\rho^{1/(2q-1)}$.
Since the exponent $1/(2q-1)$ decreases with increasing $q$, the vector
field grows more slowly for larger $q$ once the nonlinear interaction
becomes important. Consequently, the growth of the repulsive vector
contribution is suppressed more efficiently, leading to a stronger
softening of the EOS.

\renewcommand*\figurename{\small FIG.}

\begin{table*}
    \centering
    \renewcommand{\arraystretch}{1.25}
    \setlength{\tabcolsep}{6pt}
    \begin{tabular}{cccccccc}
        \hline\hline
        Configuration&Quantity
        & $c_\omega=0$
        & $q=3/2$
        & $q=2$
        & $q=3$
        & $q=4$
        & $q=5$ \\
        \hline

        Canonical&$\rho_{\rm c}/\rho_0$
        & $2.66^{+0.20}_{-0.20}$
        & $2.76^{+0.33}_{-0.25}$
        & $2.71^{+0.33}_{-0.25}$
        & $2.58^{+0.31}_{-0.24}$
        & $2.56^{+0.32}_{-0.25}$
        & $2.49^{+0.29}_{-0.19}$ \\

        &$\x$
        & $0.146^{+0.009}_{-0.009}$
        & $0.136^{+0.010}_{-0.005}$
        & $0.140^{+0.009}_{-0.007}$
        & $0.140^{+0.008}_{-0.005}$
        & $0.141^{+0.011}_{-0.005}$
        & $0.141^{+0.009}_{-0.004}$ \\

        &$s_{\rm{c}}^2$
        & $0.500^{+0.035}_{-0.057}$
        & $0.356^{+0.063}_{-0.032}$
        & $0.401^{+0.090}_{-0.059}$
        & $0.413^{+0.090}_{-0.051}$
        & $0.433^{+0.077}_{-0.060}$
        & $0.443^{+0.077}_{-0.055}$ \\

        &$s_{\rm{c}}^2/\x$
        & $3.425^{+0.319}_{-0.444}$
        & $2.618^{+0.473}_{-0.304}$
        & $2.864^{+0.659}_{-0.460}$
        & $2.950^{+0.651}_{-0.401}$
        & $3.071^{+0.557}_{-0.488}$
        & $3.142^{+0.553}_{-0.439}$ \\
        \hline

    Maximum-mass&$\rho_{\rm{c}}/\rho_0$
        & $5.87^{+0.28}_{-0.37}$
        & $6.50^{+0.58}_{-0.58}$
        & $6.21^{+0.64}_{-0.62}$
        & $5.81^{+0.60}_{-0.54}$
        & $5.66^{+0.55}_{-0.51}$
        & $5.44^{+0.51}_{-0.38}$ \\

        &$\x$
        & $0.499^{+0.012}_{-0.025}$
        & $0.328^{+0.059}_{-0.028}$
        & $0.339^{+0.117}_{-0.048}$
        & $0.289^{+0.124}_{-0.036}$
        & $0.261^{+0.116}_{-0.031}$
        & $0.239^{+0.093}_{-0.024}$ \\

        &$s_{\rm{c}}^2$
        & $0.768^{+0.012}_{-0.030}$
        & $0.453^{+0.094}_{-0.041}$
        & $0.443^{+0.216}_{-0.065}$
        & $0.331^{+0.184}_{-0.031}$
        & $0.270^{+0.130}_{-0.019}$
        & $0.230^{+0.073}_{-0.012}$ \\

        &$s_{\rm{c}}^2/\x$
        & $1.539^{+0.081}_{-0.071}$
        & $1.381^{+0.310}_{-0.278}$
        & $1.307^{+0.663}_{-0.490}$
        & $1.145^{+0.652}_{-0.503}$
        & $1.034^{+0.513}_{-0.465}$
        & $0.962^{+0.320}_{-0.378}$ \\

        \hline
        
       Peak Quantities &$\rho_{\rm{pk}}/\rho_0$
        & $/$
        & $19.37^{+24.75}_{-11.41}$
        & $4.44^{+4.75}_{-0.77}$
        & $2.91^{+0.90}_{-0.41}$
        & $2.65^{+0.81}_{-0.39}$
        & $2.50^{+0.59}_{-0.27}$ \\

        &$s_{\rm{pk}}^2\equiv s^2(\rho_{\rm{pk}})$
        & $/$
        & $0.493^{+0.075}_{-0.022}$
        & $0.457^{+0.205}_{-0.074}$
        & $0.423^{+0.174}_{-0.061}$
        & $0.435^{+0.150}_{-0.059}$
        & $0.444^{+0.126}_{-0.050}$ \\

        \hline\hline
    \end{tabular}

    \caption{The median values and $95\%$ intervals of the characteristic
    quantities for the six RMF families. The central density, EOS parameter
    $\x=P_{\rm c}/\varepsilon_{\rm c}$, SSS $s_{\rm c}^2$, and the ratio
    $s_{\rm c}^2/\x$ are evaluated for both canonical and maximum-mass configuration NSs. The last two rows highlight the peak
    structure of the SSS, with
    $s_{\rm{pk}}^2\equiv s^2(\rho_{\rm{pk}})$ denoting its maximum value
    and $\rho_{\rm{pk}}$ the density at which the maximum is reached.
    The comparison demonstrates that both the central properties and,
    in particular, the magnitude and location of the SSS peak vary
    substantially among EOS families satisfying the adopted nuclear
    and astrophysical constraints.}

    \label{tab:family_intervals}
\end{table*}

The competition between these two regimes naturally produces the peak structure of the SSS seen in FIG.\,\ref{fig:sss_band}. The initial increase of the vector repulsion drives $s^2$ upward, whereas the subsequent suppression of the growth of $V_\omega$ causes $s^2$ to turn over and decrease. Moreover, the nonlinear contribution depends on $V_\omega^{2q-1}$ and therefore grows more rapidly relative to the linear term for larger $q$. To estimate the density scale at which the nonlinear contribution becomes comparable to the linear one, we consider the $\omega$-meson field equation in SNM, where the isovector mean field vanishes. Writing the baryon density as $\rho=m_\omega^2 V_\omega/g_\omega^2+q c_\omega V_\omega^{2q-1}/2m_\omega^{2q-4}$ and equating the two contributions, we obtain the characteristic density
\begin{equation}
\rho_{\rm{turn}}\approx
\frac{2m_\omega^3}{g_\omega^2}
\left(
\frac{2}
{q c_\omega g_\omega^2}
\right)^{1/(2q-2)}.
\end{equation}
Although the fitted couplings, particularly $c_\omega$ and $g_\omega$, affect the precise density at which the nonlinear interaction becomes important, the above estimate shows that $\rho_{\rm{turn}}$ approaches the characteristic density scale $2m_\omega^3/g_\omega^2$ as $q\to\infty$ for fixed couplings. Taking $m_\omega=782.5\,{\rm{MeV}}$ and $g_\omega\approx10$-$20$, this corresponds to roughly $2$-$8\rho_0$, indicating that the onset of nonlinear $\omega$-field suppression remains associated with densities relevant to NS interiors. In the present ensembles, increasing $q$ tends to shift the turnover toward lower densities, bringing the peak of $s^2$ further into the density range accessible inside NSs.
The generalized $\omega$ self-interaction therefore provides a possible mechanism for generating an interior peak of the SSS within a purely nucleonic EOS description. For $q\gtrsim2$, it is probably that there would be a peak in $s^2$ profile within NS densities.
As quantified in TAB.\,\ref{tab:family_intervals}, the peak magnitude
remains typically around $s_{\rm{pk}}^2\sim0.4$-$0.5$, whereas its
location changes substantially with $q$. For $q=3/2$, the SSS develops
only a shallow peak at a relatively high density, with a median
$\rho_{\rm{pk}}/\rho_0\approx19$. As $q$ increases, the peak shifts
rapidly toward lower densities, with $\rho_{\rm{pk}}/\rho_0\approx4.4$
for $q=2$ and about $2.5$-$2.9$ for $q=3$-$5$, placing it well
within the density range characteristic of NS cores for larger $q$. A similar peak can also develop in $\phi=P/\varepsilon$ as shown in FIG.\,\ref{fig:P/eps}, typically at a
higher density than that of $s^2$. This ordering can be understood
from the trace-anomaly decomposition of the SSS\,\cite{Fujimoto2022},
\begin{equation}
s^2(\varepsilon)=\frac{1}{3}-\Delta-\varepsilon\frac{\d\Delta}{\d\varepsilon},
\end{equation} where $\Delta=1/3-\phi$. The derivative term directly reflects the density dependence of $\phi$ (or equivalently $\Delta$), allowing $s^2$ to reach its maximum before $\phi$ does\,\cite{CaiLi2025Review}.

An interesting example is provided by the $q=3$ family. As seen in
FIG.\,\ref{fig:sss_band}, $s^2$ happens to remain close to $1/3$ over
a substantial part of the density range accessible inside NSs, while
FIG.\,\ref{fig:P/eps} shows that the EOS-parameter $\phi=P/\varepsilon$ is also close
to $1/3$ in approximately the same region. Since
$s^2\to\phi\to1/3$ represents a characteristic thermodynamic behavior of
conformal matter\,\cite{BedaqueSteiner2015,Annala2020,Fujimoto2022,Marczenko2023}, an analysis restricted to this particular model
family might appear to suggest the emergence of approximate
conformality inside NSs. In the present case, however, this behavior
is coincidental: It results from the specific competition among the
RMF interaction terms at finite density rather than from the
restoration of conformal symmetry. Indeed, the different model
families satisfying the same adopted constraints exhibit markedly
different values and density dependences of both $s^2$ and $\phi$ in
the same NS density region. This example therefore illustrates a more
general caution in model-dependent inference: When an assumed EOS
model possesses a particular limiting or characteristic behavior,
its apparent realization within the density range probed by NSs need
not imply that the underlying matter has reached the corresponding
physical regime.

For the $q\geq4$ families, $s^2$ develops a shallow minimum at higher
densities within the density range relevant to NSs, as shown in
FIG.\,\ref{fig:sss_band}. After the nonlinear $\omega$ interaction
drives $s^2$ downward from its peak, the ultra-relativistic kinetic
contribution becomes increasingly important. As discussed in
Subsection \ref{subs_HD}, this contribution eventually dominates for
$q>2$ and drives $s^2$ back toward $1/3$. The competition between the
suppression of the vector contribution and the increasing importance
of the kinetic contribution therefore naturally produces the shallow
minimum before the asymptotic regime is reached.

The differences among the six families can be quantified more directly
through their central properties. TAB.\,\ref{tab:family_intervals}
summarizes the medians and $95\%$ intervals of the central density,
the central EOS-parameter $\x=\phi_{\rm{c}}$, the central SSS
$s_{\rm{c}}^2$, and their ratio $s_{\rm{c}}^2/\x$ for both TOV NSs
and canonical $1.4\,M_\odot$ NSs. For canonical NSs, the central
densities and $\x$ remain relatively similar among the six
families, with median values
$\rho_{\rm c}/\rho_0\approx2.5$-$2.8$ and
$\x\approx0.14$-$0.15$. In contrast,
$s_{\rm{c}}^2$ shows a more pronounced family dependence, with
median values ranging from about $0.36$ for $q=3/2$ to $0.50$ for
$c_\omega=0$, while $s_{\rm{c}}^2/\x$ remains of order
$3$. The differences become substantially larger for TOV NSs, whose
median central densities reach approximately $5.4$-$6.5\rho_0$.
In this case, $\x$ varies from about $0.24$ to $0.50$,
while $s_{\rm{c}}^2$ spans a considerably broader range, from
about $0.23$ for $q=5$ to $0.77$ for $c_\omega=0$. Correspondingly,
$s_{\rm{c}}^2/\x$ decreases systematically from about
$1.54$ for $c_\omega=0$ to $0.96$ for $q=5$. This behavior shows that
the relation between the local stiffness $s_{\rm c}^2$ and the
accumulated stiffness $\x=P_{\rm c}/\varepsilon_{\rm c}$ itself
becomes increasingly model dependent at the high central densities
reached by TOV NSs. The growing spread from canonical to TOV
configurations therefore reflects the increasing sensitivity of the
NS center to the high density EOS.

The ratio $s_{\rm c}^2/\x$ has a particularly transparent physical
interpretation, since it corresponds to the local
polytropic index,
\begin{equation}
\frac{\d\ln P}{\d\ln\varepsilon}=\frac{s^2}{\phi}.\end{equation} For canonical NSs,
$s_{\rm c}^2/\x$ remains well above unity for all six families, with
median values of about $2.6$-$3.4$, reflecting that their centers
probe a relatively moderate density regime where the high density
differences among the model families are not yet fully developed.
For the TOV configurations, in contrast, this ratio decreases
systematically with increasing $q$ and reaches values close to, or
even below, unity for the large-$q$ families. This trend is not
surprising from the construction of the present models: Increasing
$q$ enhances the nonlinear suppression of the $\omega$ field and
softens the EOS at high densities, naturally driving
$s_{\rm c}^2/\x$ toward smaller values. In particular, the $q=5$
family can reach $s_{\rm c}^2/\x\lesssim1$ as a consequence of this
built-in high density behavior. Similar to the conformal-like
behavior of the $q=3$ family discussed above, the significance here
lies not in interpreting this feature of a particular family as
evidence for conformal symmetry, but in the fact that EOS families
with markedly different behaviors of $s_{\rm c}^2/\x$ can all satisfy
the same adopted nuclear and astrophysical constraints.

Against the conformal-like behavior found for some of the large-$q$
families within the NS density range, the $c_\omega=0$ family provides
a useful opposite limiting reference. For this family,
$s_{\rm{c}}^2/\x\approx1.54$, showing that the central
EOS of the TOV configurations remains well away from the asymptotic
relation $s^2/\phi\to1$. Nevertheless, without the nonlinear $\omega$
self-interaction, the vector field retains the asymptotic behavior
$V_\omega\sim\rho$, so that the quadratic vector contribution scales
as $\rho^2$ and eventually dominates the EOS. At sufficiently high
densities, both $\phi=P/\varepsilon$ and $s^2$ therefore approach
unity, and hence $s^2/\phi\to1$, as seen in
FIGs.\,\ref{fig:P/eps} and \ref{fig:sss_band}. This asymptotic
behavior should thus be distinguished from the finite-density
properties actually reached inside the TOV configurations. The
exceptionally stiff tendency of this family is also reflected in the
SNM pressure band, which partially exceeds the HIC flow constraint in
FIG.\,\ref{fig:P_SNM}. The $c_\omega=0$ family is therefore less
favored as a realistic description of dense
matter\,\cite{Hornick2018,ZhuLiLiu2023}, while providing a useful
contrast for illustrating the softening induced by the nonlinear
$\omega$ self-interaction. A related indication can be seen in the
M-R relations in FIG.\,\ref{fig:MR}, where the $c_\omega=0$ family
approaches the effective causality boundary obtained from the
IPAD-TOV analysis\,\cite{CaiLiZhang2023PRD} and shows some tension
with it, consistent with its exceptionally stiff high density EOS.

This observation is also relevant to the potential effective upper
bound on the central EOS-parameter inferred from the IPAD-TOV framework. Our previous analysis suggested
$\x\lesssim0.385$ for stable NSs\,\cite{CaiLiMa2026X}. The TOV configurations
of all five families with $c_\omega>0$ are broadly consistent with
this bound, whereas the $c_\omega=0$ family reaches
$\x_{\rm{TOV}}\approx0.50$. Interestingly, the latter is also the
family that shows tension with the independent HIC flow constraint\,\cite{Danielewicz2002}.
Thus, although substantially larger values of $\x$ can be generated
within the RMF framework, the very stiff EOSs required to support such
values are independently disfavored by nuclear matter constraints.
The consistency between the HIC flow constraint and the effective
upper bound on $\x$ obtained from intrinsic NS stellar structure
therefore provides a complementary indication of the physical
relevance of this bound. 

These results expose a fundamental ambiguity in extracting NS core
properties from the input EOS. Except for the particularly stiff
$c_\omega=0$ family, the different RMF families considered here are
broadly consistent with both current nuclear matter constraints and
astrophysical observations, yet they predict markedly different core
properties, including $s_{\rm{c}}^2$ and $\x$. Given the still limited
observational constraints on the high density EOS, agreement with
existing data alone is therefore insufficient to determine which of
these distinct core behaviors is physically realized.

{\it \textcolor{black}{Can information intrinsic to NS stellar
structure itself be used to reduce the ambiguity arising from EOS
model dependence and provide a more robust extraction of the core
properties such as the central SSS?}} This question points to an important limitation of
conventional inferences of the dense matter EOS from astrophysical
observations. Once a particular EOS model or parametrization is
adopted, the observational data primarily constrain the parameters
within that assumed model space. The inferred central pressure, energy
density, and SSS then inherit the high density behavior encoded in the
underlying EOS model. Consequently, even when different EOS models
reproduce the same macroscopic observables, their inferred core
properties need not agree, as explicitly demonstrated by the present
RMF families. In this sense, the resulting constraints on the NS core
are generally conditional on the adopted EOS description rather than
being determined by observations alone. In the next section, we
address this question using the IPAD-TOV framework and extract the
central SSS with reduced sensitivity to the underlying dense matter
model.

\section{Probing the Central Sound Speed through the IPAD-TOV Framework}\label{SEC_4}

The model dependence illustrated above motivates an approach that
probes NS central properties with minimal reliance on a specific
nuclear EOS. The IPAD-TOV framework provides such an approach by
deriving approximate scaling relations directly from the stellar
structure equations. Previous studies have established scalings
connecting NS mass, radius, compactness, and tidal deformability to
the central pressure and energy density, and used them to constrain
the central EOS\,\cite{CaiLiZhang2023PRD,CaiLi2025Review,CaiLi2025Trace}.
Here, we extend this framework to investigate the central SSS and its
behavior inside NSs. In Subsection \ref{subs_CSSS}, we derive a
an improved approximation for the central SSS from the IPAD-TOV analysis; its physical interpretation is given in Subsection \ref{subs_PhysTheta}.
In Subsection \ref{subs_SA}, we examine its scaling behavior across
the different EOS families and use their common behavior to extract
the central SSS and related properties of canonical NSs with reduced
model dependence. Finally, in Subsection \ref{subs_peak}, we examine
the near center behavior of the SSS and the EOS-parameter
$\phi=P/\varepsilon$, with particular attention to the possible
emergence of interior sound speed peaks.

\subsection{$\Theta(\x,\Psi)$: Estimator for the Central SSS from IPAD-TOV}\label{subs_CSSS}

The IPAD-TOV framework is based on intrinsic and perturbative analyses
of the dimensionless TOV equations\,\cite{CaiLi2025Review}. By expanding
the stellar structure equations about the center, it connects global NS
properties with the central EOS without specifying a particular nuclear
EOS. Introducing the dimensionless radius $\widehat{r}=r/Q$ with $Q=1/\sqrt{4\pi \varepsilon_{\rm{c}}}$, the pressure
and energy density near the stellar center can be expanded as\,\cite{CaiLiZhang2023ApJ}
\begin{align}
    \widehat{P}(\widehat{r})
    \approx&\x+b_2\widehat{r}^{2}+b_4\hr^4+\mathcal{O}(\widehat{r}^{6}),\label{P-exp}\\
    \widehat{\varepsilon}(\widehat{r})
    \approx&1+a_2\widehat{r}^{2}+a_4\hr^4+\mathcal{O}(\widehat{r}^{6}),\label{eps-exp}
\end{align}
where $\widehat{P}=P/\varepsilon_{\rm{c}}$,
$\widehat{\varepsilon}=\varepsilon/\varepsilon_{\rm{c}}$, and
$\x=P_{\rm{c}}/\varepsilon_{\rm{c}}$. Here, $b_2$ and $a_2$
characterize the leading radial variations of the pressure and energy
density away from the stellar center, respectively. The dimensionless
TOV equations\,\cite{CaiLiZhang2023ApJ}
\begin{equation}
\frac{\d\hP}{\d\hr}=- \frac{\heps\widehat{M} (1 + \hP/\heps) (1 + \hr^3 \hP/\widehat{M})}{\hr^2 (1 - 2\widehat{M}/\hr)};\frac{\d\widehat{M}}{\d\hr} = \hr^2 \heps,\label{eq:TOV1}
\end{equation}
give the coefficients $
    b_2=-({1+4\x+3\x^2})/{6}$ and $
    a_2={b_2}/{s_{\rm{c}}^2}$,
with $s_{\rm{c}}^2$ the central SSS; here $\widehat{M}=M/Q$.
The latter relation shows explicitly how the central SSS controls the
relative radial variations of the pressure and energy density. Moreover, $a_4$ and $b_4$ are higher-order coefficients\,\cite{CaiLi2025Review}.

Integrating the energy density then gives the near-center expansion
of the dimensionless NS mass\,\cite{CaiLiZhang2023ApJ}:
\begin{equation}
    \widehat{M}_{\rm{NS}}\equiv
    \widehat{M}(\widehat{R})=
    \int_0^{\widehat{R}}\d\hr\hr^2\heps(\hr)
    \approx\frac{1}{3}\widehat{R}^{3}
    +\frac{1}{5}a_2\widehat{R}^{5}+\cdots.
\end{equation}
Truncating the pressure profile at quadratic order and imposing
$\widehat{P}(\widehat{R})\approx0$ gives\,\cite{CaiLiZhang2023ApJ}
\begin{equation}
    \widehat{R}\approx
    \left(\frac{-\x}{b_2}\right)^{1/2}
    =
    \left(\frac{6\x}{1+4\x+3\x^2}\right)^{1/2}.
\end{equation}
Keeping only the leading mass term then yields the
scaling\,\cite{CaiLiZhang2023ApJ}
\begin{equation}
    M_{\rm{NS}}
    \sim\frac{\widehat{M}_{\rm{NS}}}
    {\sqrt{\varepsilon_{\rm{c}}}}
    \sim\frac{1}{\sqrt{\varepsilon_{\rm{c}}}}
    \left(\frac{\x}{1+4\x+3\x^2}\right)^{3/2}.
    \label{eq:IPAD_mass_leading}
\end{equation}
This relation provides the key connection between the global stellar
mass and the central EOS quantities in the leading-order IPAD-TOV
analysis.

Along an equilibrium sequence for a fixed EOS, taking the logarithmic
derivative of Eq.\,\eqref{eq:IPAD_mass_leading} with respect to
$\varepsilon_{\rm{c}}$ relates the central SSS directly to the
variation of the stellar mass\,\cite{CaiLiZhang2023PRD},
\begin{equation}
    s_{\rm{c}}^2
    \approx
    \x\left(
        1+\frac{1+\Psi}{3}
        \frac{1+3\x^2+4\x}{1-3\x^2}
    \right)\equiv \overline{s_{\rm{c}}^2},
    \label{eq:sc2_IPAD_leading}
\end{equation}
where $\Psi$ is the logarithmic slope of the NS mass with respect to the central
energy density, defined in Eq.\,(\ref{def-Psi}). Thus, at this order, $s_{\rm{c}}^2$ is determined by
the central EOS-parameter $\x$ and the response $\Psi$ of the stellar
mass along the equilibrium sequence. In this sense, $\Psi$ carries
information on how the global stellar structure responds to a change
of the central EOS.

To improve this leading-order approximation, we retain the sub-leading
mass term $a_2\widehat{R}^{5}/5$, which contains explicit information
on the central SSS through $a_2=b_2/s_{\rm{c}}^2$. This modifies
Eq.\,\eqref{eq:IPAD_mass_leading} to
\begin{equation}
    M_{\rm{NS}}\sim
    \frac{\lambda}{\sqrt{\varepsilon_{\rm{c}}}}
    \left(\frac{\x}{1+3\x^2+4\x}\right)^{3/2}
    ,~~\mbox{with }
    \lambda\equiv
    1-\frac{3\x}{5s_{\rm{c}}^2}.
    \label{eq:IPAD_mass_corrected}
\end{equation}
Using $
    {\d\x}/{\d\ln\varepsilon_{\rm{c}}}
    =s_{\rm{c}}^2-\x$,
the logarithmic slope of
Eq.\,\eqref{eq:IPAD_mass_corrected} gives
\begin{equation}
    \frac{1+\Psi}{2}\approx
    (s_{\rm{c}}^2-\x)
    \frac{\d}{\d\x}
    \ln\left[\lambda
        \left(\frac{\x}{1+3\x^2+4\x}\right)^{3/2}
    \right].
    \label{eq:IPAD_matching}
\end{equation}
This relation contains $s_{\rm{c}}^2$ implicitly through both the
derivative of $\x$ and the correction factor $\lambda$. To obtain an
explicit approximation, we expand locally in $\x$ at fixed
$\Psi$, \begin{equation}
    s_{\rm{c}}^2
    \approx A_\Psi\x+B_\Psi\x^2+\mathcal{O}(\x^3).\end{equation}
Correspondingly,
\begin{equation}
    \lambda=
    \left(1-\frac{3}{5A_\Psi}\right)
    \left[
        1+\frac{B_\Psi}
        {A_\Psi}\frac{3}{5A_\Psi-3}\x
        +\mathcal{O}(\x^2)
    \right].
\end{equation}
Matching the constant and linear terms in $\x$ in
Eq.\,\eqref{eq:IPAD_matching} yields
\begin{equation}
    A_\Psi=\frac{4+\Psi}{3},
~~
    B_\Psi=\frac{4}{3}
    \frac{(1+\Psi)(4+\Psi)(11+5\Psi)}
    {50+37\Psi+5\Psi^2}.
\end{equation}

\renewcommand*\figurename{\small FIG.}
\begin{figure}[h!]
    \centering
    \includegraphics[width=0.9\columnwidth]{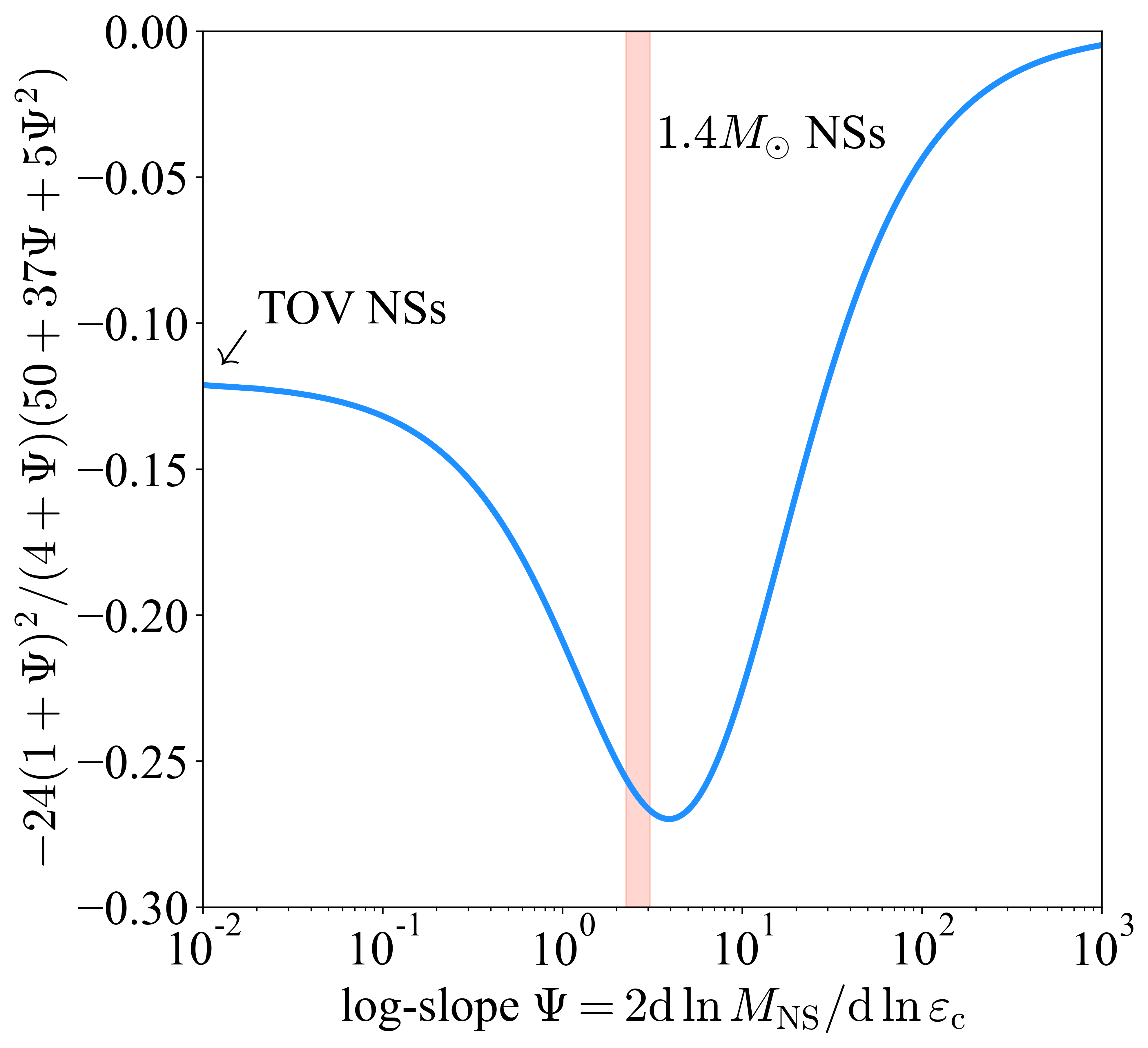}
    \caption{(Color online). The correction term
    ``$-24(1+\Psi)^2/(4+\Psi)(50+37\Psi+5\Psi^2)$'' of the central SSS in Eq.\,(\ref{eq:sc2_IPAD_corrected})
    as a function of the log-slope
    $\Psi=2\d\ln M_{\rm{NS}}/\d\ln\varepsilon_{\rm c}$.
    The shaded band indicates $\Psi\approx2.66\pm0.40$\,\cite{Shi2026} for canonical
     NSs, while the maximum-mass (TOV) configuration corresponds to
    $\Psi=0$.}
    \label{fig:fPsi}
\end{figure}

We next incorporate the resulting $\mathcal{O}(\x^2)$ correction
through a multiplicative factor $1+\kappa\x$ applied to
Eq.\,\eqref{eq:sc2_IPAD_leading}. Matching its expansion to
$A_\Psi\x+B_\Psi\x^2+\mathcal{O}(\x^3)$ gives
\begin{align}
    s_{\rm{c}}^2\approx&
   \overline{s_{\rm{c}}^2}\times
    \left[
        1-\frac{24(1+\Psi)^2}
        {(4+\Psi)(50+37\Psi+5\Psi^2)}\x
    \right].
    \label{eq:sc2_IPAD_corrected}
\end{align}
The factor in the square bracket provides a relative $\mathcal{O}(\x)$
correction to the leading-order result $\overline{s_{\rm{c}}^2}$ and
therefore modifies $s_{\rm{c}}^2$ at $\mathcal{O}(\x^2)$. This
correction originates from the first sub-leading term in the
near-center mass expansion and thus incorporates additional information
on the central EOS while retaining the EOS-independent structure of
the IPAD-TOV analysis. 
As shown in FIG.\,\ref{fig:fPsi}, the correction coefficient ``$-24(1+\Psi)^2/(4+\Psi)(50+37\Psi+5\Psi^2)$'' in Eq.\,\eqref{eq:sc2_IPAD_corrected} is negative and bounded for $\Psi\geq0$, with its magnitude remaining well below unity over the entire domain. Its magnitude reaches the maximum at about $\Psi\approx3.92$, where the coefficient is approximately $-0.270$. This minimum lies near the region relevant for canonical NSs, for which $\Psi\approx2.66\pm0.40$\,\cite{Shi2026}, and explains why the correction is most pronounced for such stars. Moreover, if one uses the effective bound $\x\lesssim0.385$\,\cite{CaiLiMa2026X}, the magnitude of the correction is at most approximately $10\%$, supporting the perturbative treatment adopted in the IPAD-TOV analysis. At larger $\Psi$, its magnitude decreases and vanishes asymptotically. At the maximum-mass configuration, $\Psi=0$, for which $A_0=4/3$, $B_0=88/75$, and the correction coefficient becomes $-3/25$. Consequently, Eq.\,\eqref{eq:sc2_IPAD_corrected} reduces to the TOV approximation obtained previously in Ref.\,\cite{CaiLi2025Trace}, namely $s_{\rm c}^2\approx\overline{s_{\rm c}^2}\cdot(1-3\x/25)$, with the leading-order expression $\overline{s_{\rm c}^2}$ defined in Eq.\,(\ref{eq:sc2_IPAD_leading}).

\subsection{Physical Interpretation of the Central SSS Estimator}\label{subs_PhysTheta}

In the following, we denote the approximate expression for the central SSS in
Eq.\,\eqref{eq:sc2_IPAD_corrected} by $\Theta(\mathrm{X},\Psi)$, which acts as an effective estimator for $s_{\rm c}^2$.
It is worth emphasizing that $\Theta(\x,\Psi)$ is not an empirical
scaling variable, but follows directly from the dimensionless TOV
equations as an approximate physical expression for
$s_{\rm{c}}^2$. The auxiliary scales and prefactors in $M_{\rm{NS}}$ cancel in the
derivation, as done for Eq.\,(\ref{eq:sc2_IPAD_leading}), leaving $\Theta$ determined only by $\x$ and $\Psi$.
In particular, imposing the causal condition $\Theta\leq1$ yields an
upper bound on $\x$ highly consistent with independent theoretical
estimates\,\cite{Cai2024Front}, supporting this physical interpretation. At the same time,
$\Theta$ remains approximate, since higher order terms and the
associated nonlinear dependence on $\x$ are not fully included.

\begin{table}[h!]
    \centering
    \renewcommand{\arraystretch}{1.1}
    \setlength{\tabcolsep}{3pt}
    \begin{tabular}{ccc}
        \hline\hline
        Property
        & $s_{\rm{c}}^2$ (Numerical)
        & $\Theta(\x,\Psi)$ (IPAD-TOV) \\
        \hline
        Meaning
        & Physical central SSS
        & Analytical estimator \\
        Origin
        & Model EOS
        & TOV equations \\
        Model dependence
        & EOS-dependent
        & EOS-independent\\
        Accessible range
        & Model-dependent
        & Causal domain $[0,1]$ \\
        Accuracy
        & Exact for given EOS
        & Approximate \\
        Higher orders
        & Included
        & Partially included \\
        \hline\hline
    \end{tabular}
    \caption{Comparison between the physical central SSS
    $s_{\rm{c}}^2$ and its analytical estimator
    $\Theta(\x,\Psi)$. The accessible range of
    $s_{\rm{c}}^2$ is determined by the adopted EOS model,
    whereas $\Theta$ follows from the intrinsic structure
    of the dimensionless TOV equations and admits the
    causal domain $[0,1]$ at the level of the approximation.
    The values of $\Theta$ sampled by a particular EOS
    family may nevertheless occupy only part of this domain.}
    \label{tab:sc2_Theta_comparison}
\end{table}

This distinction is important when interpreting the numerical
relations between $\Theta$ and $s_{\rm{c}}^2$. Physically, $\Theta$ is
an analytical estimator of the central SSS. If all higher order
corrections were included, it would coincide with the physical
$s_{\rm{c}}^2$. The present expression for $\Theta$, however, retains
only the leading corrections and is therefore not exact. The fitted
relation $
s_{\rm c}^2\approx\alpha\Theta+\beta$ provides an effective calibration of this finite-order approximation
for each EOS family. Deviations of $\alpha$ and $\beta$ from unity and
zero may reflect omitted higher-order corrections, the restricted
domain sampled by the family, or residual model dependence; if they
are sufficiently large and systematic, they may also signal tension
between the model predictions and the physical relation represented
by $\Theta$. Since each EOS family samples only a limited range of
$\Theta$ and $s_{\rm c}^2$, rather than the full causal interval from
zero to unity, a strong correlation between these quantities may still
be present even when the fitted slope and intercept differ appreciably
from unity and zero. For the same reason, extrapolating a
family-dependent fit to $s_{\rm c}^2=1$ may extend it far beyond the
sampled domain and could produce an artificial constraint on $\x$.
The causal bound should instead be imposed directly on the underlying
physical estimator, $\Theta\leq1$\,\cite{CaiLiMa2026X}.
TAB.\,\ref{tab:sc2_Theta_comparison} gives the comparison and connection between a model-based $s_{\rm c}^2$ and the estimator $\Theta(\x,\Psi)$.
Since $\Theta$ estimates the same physical quantity as the numerically
calculated $s_{\rm c}^2$, a strong correlation between them is expected
within the relevant domain. This correlation is examined quantitatively
in the following subsection.

\subsection{Scaling Analysis and Common Extraction of the Central SSS}\label{subs_SA}

\begin{figure}[h!]
    \centering
    \includegraphics[width=0.95\columnwidth]{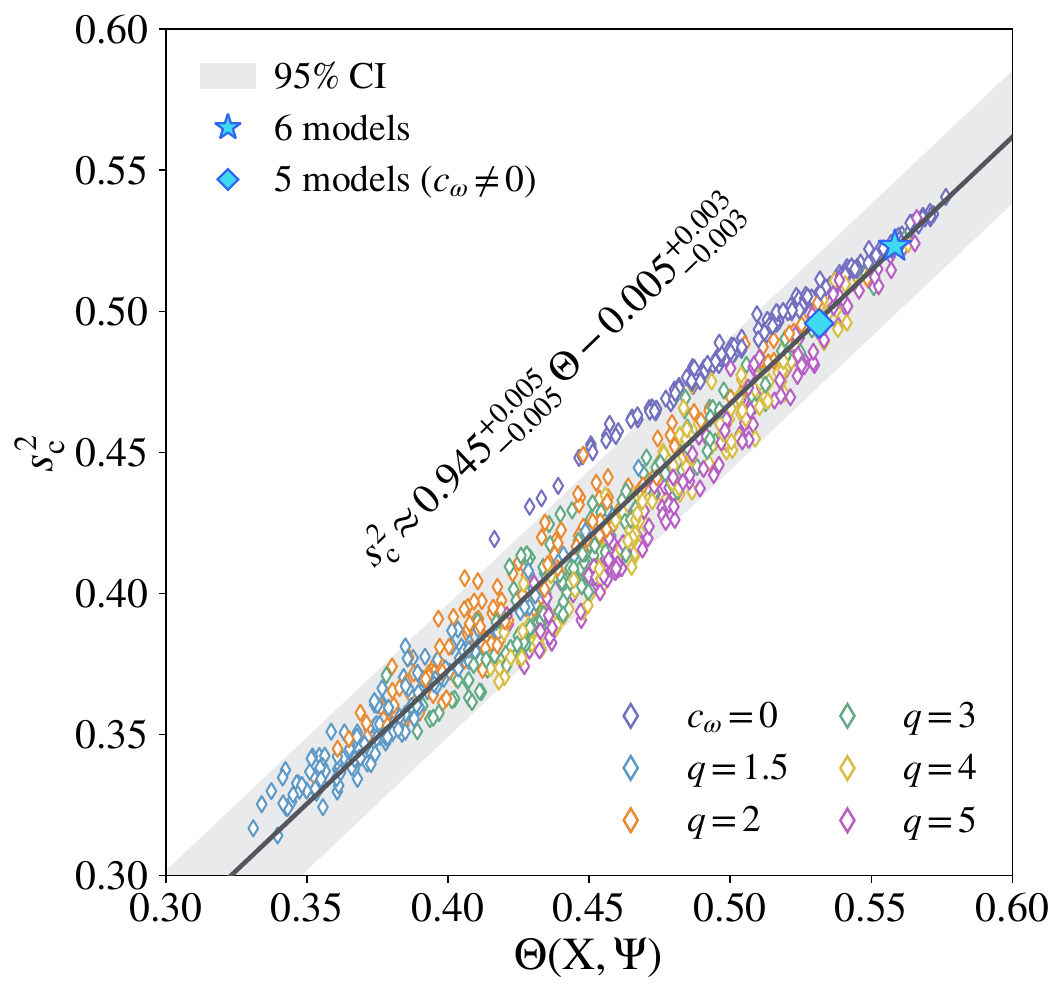}
    \caption{(Color online). Correlation between central SSS
    $s_{\mathrm{c}}^2$ and the analytical estimator $\Theta(\mathrm{X},\Psi)$
    for canonical $1.4\,M_\odot$ NSs.
    The open symbols show 150 randomly selected configurations
    from each of the six RMF families.
    The solid black line shows the linear fit to the combined
    ensemble, and the gray band indicates the $95\%$
    CI. The fitted relation is displayed
    in the figure.
    The filled star and diamond mark the common estimates.}
    \label{fig:sss_scaling}
\end{figure}

To assess the improved expression of Eq.\,(\ref{eq:sc2_IPAD_corrected}), we use the six EOS families
constructed above, each containing 1000 parameter sets. For each EOS, we identify the $1.4\,M_\odot$ configuration
along its equilibrium sequence, extract the central EOS
parameter $\x$ and the sequence derivative $\Psi$, and
calculate $s_{\mathrm{c}}^2$ directly from the EOS at the
corresponding central density. 
FIG.\,\ref{fig:sss_scaling} compares the $\Theta(\x,\Psi)$ with
the exact $s_{\mathrm{c}}^2$, displaying 150 randomly selected
cases from each family for clarity. The fits use all samples rather than only the displayed points.

\renewcommand*\figurename{\small FIG.}

Despite differences in their asymptotic limits and
finite density behavior, all six families exhibit
tight linear correlations between $\Theta$ and
$s_{\mathrm{c}}^2$. We characterize each family by 
\begin{equation}
s_{\mathrm{c}}^2\approx\alpha_j\Theta(\x,\Psi)+\beta_j,
\end{equation}
where $j$ labels the EOS family. The fitted coefficients and
coefficients of determination are summarized in
TAB.\,\ref{tab:sc2_family_scaling} with $\rm R^2$ values ranging from approximately
$0.94$ to $0.99$. Thus, the $\Theta(\x,\Psi)$ of Eq.\,\eqref{eq:sc2_IPAD_corrected} captures the main behavior of the central speed of sound within each family.
The differences in the fitted slopes and intercepts
quantify the residual family dependence.

\begin{table}[h!]
    \centering
    \renewcommand{\arraystretch}{1.1}
    \setlength{\tabcolsep}{6pt}
    \begin{tabular}{cccc}
        \hline\hline
        EOS family & $\alpha_j$ & $\beta_j$ & $\rm R^2$ \\
        \hline
        $c_\omega=0$ & $0.700\pm0.004$ & $0.135\pm0.002$ & 0.99 \\
        $q=3/2$      & $0.878\pm0.014$ & $0.024\pm0.005$ & 0.94 \\
        $q=2$        & $0.941\pm0.012$ & $0.003\pm0.005$ & 0.96 \\
        $q=3$        & $0.999\pm0.015$ & $-0.033\pm0.007$ & 0.94 \\
        $q=4$        & $1.072\pm0.014$ & $-0.074\pm0.007$ & 0.96 \\
        $q=5$        & $1.098\pm0.012$ & $-0.091\pm0.006$ & 0.97 \\
        \hline
        All families & $0.945\pm0.005$ & $-0.005\pm0.003$ & 0.95 \\
        \hline\hline
    \end{tabular}
    \caption{Linear scalings $s_{\mathrm{c}}^2\approx
    \alpha_j\Theta+\beta_j$ for canonical NSs. The last row refers
    to a fit to the combined samples of all six families. The quoted
    uncertainties in $\alpha_j$ and $\beta_j$ represent 95\%
    confidence intervals.}
    \label{tab:sc2_family_scaling}
\end{table}

These separate scaling relations allow us to examine whether the
different EOS families can provide a common estimate of the central
SSS. Since the values of $\Theta$ sampled by different EOS families
are generally different, there is no reason to expect the fitted
coefficients $\alpha_j$ and $\beta_j$ to be approximately the same
across the families, as seen in TAB.\,\ref{tab:sc2_family_scaling}.
Nevertheless, $s_{\rm{c}}^2$ is a physical property of a given NS
and should not be strongly sensitive to which EOS family is used to
infer it. This motivates us to search for a region in which the
different family relations,
$s_{\rm{c}}^2\approx\alpha_j\Theta+\beta_j$, give similar values of
$s_{\rm{c}}^2$. Schematically, this condition can be physically expressed as
\begin{equation}\label{def-MM}
    \underbrace{\alpha_j\Theta^{(j)}+\beta_j}_{\rm{family}\;j}
    \approx s_{\rm{c}}^2 \approx\underbrace{\alpha_{j'}\Theta^{(j')}+\beta_{j'}}_{\rm{family}\;j'}\mbox{ for a given }M_{\rm{NS}}.
\end{equation}
where the superscripts $(j)$ and $(j')$ are introduced on $\Theta$ to label the corresponding EOS families and make this relation more transparent. Such a convergence would indicate a central SSS that is less
sensitive to the particular form of the $\omega$ self-interaction.

A natural way to implement the condition in
relation \eqref{def-MM} is to minimize the dispersion among the family
relations. Specifically, we determine
the common estimate through a least squares minimization,
\begin{equation}
    \left(\Theta^{\ast},s_{\rm c}^{2,\ast}\right)
    =\underset{\Theta,s_{\rm c}^2}{\operatorname{arg\,min}}
    \frac{1}{N_{\mathrm{f}}}
    \sum_{j=1}^{N_{\mathrm{f}}}
    \left(\alpha_j\Theta+\beta_j-s_{\rm c}^2\right)^2,
    \label{eq:sc2_common_point}
\end{equation}
where $N_{\mathrm{f}}$ is the number of included families and equal
weight is assigned to each family. The resulting
$(\Theta^{\ast},s_{\rm c}^{2,\ast})$ corresponds to the point where the family
predictions are closest to one another, with
$s_{\rm c}^{2,\ast}$ giving the common estimate of the
central SSS. Using $\Psi_{1.4}\approx2.66\pm0.40$ from our previous analysis
in Ref.\,\cite{Shi2026}, the corresponding $\x^{\ast}$ is obtained by
solving $\Theta^{\ast}=\Theta(\x^\ast,\Psi_{1.4})$.

The resulting estimates and their $95\%$ confidence intervals are
listed in TAB.\,\ref{tab:sc2_results}, with the six family and five
family results marked by a star and a diamond, respectively, in
FIG.\,\ref{fig:sss_scaling}. The root mean square residuals among the
family predictions are approximately $0.0044$ and $0.0041$, indicating
a close convergence of the family relations near the extracted points.
Excluding the $c_\omega=0$ family lowers the extracted
$s_{\rm{c}}^2$ by about $5\%$ and $\x$ by about $3\%$, showing a
relatively weak sensitivity to the choice of EOS families. Both
extractions give $s_{\rm{c}}^2$ close to $0.5$, consistent with our
previous result\,\cite{Shi2026}, and above the conformal value $1/3$.
It is worth emphasizing that individual EOS families can predict
central SSS values either below or above $1/3$, as shown in
FIG.\,\ref{fig:sss_band}. The present result instead follows from the
convergence of the family scaling relations through the common
IPAD-TOV variable $\Theta(\x,\Psi)$. Accordingly, the extracted point
characterizes the agreement among these relations rather than an
average or the most densely populated region of the individual EOS
samples. The inferred values remain conditional on the EOS families
considered and the accuracy of the corresponding scaling relations.

\begin{table}[h!]
    \centering
    \renewcommand{\arraystretch}{1.35}  
    \setlength{\tabcolsep}{6pt}      
    \begin{tabular}{ccccc}
        \hline\hline
        $N_{\mathrm{f}}$
        & $\Theta^\ast$
        & $s_{\mathrm{c}}^{2,\ast}$
        & $\x^{\ast}$
        & $R^\ast_{1.4}\,(\mathrm{km})$ \\
        \hline
        6
        & $0.558^{+0.046}_{-0.046}$
        & $0.523^{+0.044}_{-0.044}$
        & $0.173^{+0.020}_{-0.018}$ 
        & $11.73^{+1.08}_{-0.92}$ \\
        5 $(c_\omega\neq0)$
        & $0.531^{+0.092}_{-0.092}$
        & $0.496^{+0.092}_{-0.092}$
        & $0.167^{+0.030}_{-0.029}$ 
        & $11.92^{+1.76}_{-1.28}$ \\
        \hline\hline
    \end{tabular}
    \caption{Extracted values of $\Theta^{\ast}$, $s_{\mathrm{c}}^{2,\ast}$,
 $\x^{\ast}$ and $R_{1.4}^\ast\,(\rm{km})$ for canonical NSs. The quoted uncertainties correspond to the $95\%$ confidence intervals. $N_{\mathrm{f}}$ is the number of included families for the optimization.}
    \label{tab:sc2_results}
\end{table}

We also note that the values of $\x$ directly sampled by the six EOS
families tend to lie below the extracted $\x^\ast$, as shown in
TAB.\,\ref{tab:family_intervals}. This is not inconsistent with the
present analysis, since $\x^\ast$ is not determined by the distribution
of $\x$ among the EOS samples, but is inferred from the common behavior
of the family scaling relations through $\Theta(\x,\Psi)$. It therefore
provides another illustration of how the IPAD-TOV scaling can extract
central EOS information that is not directly apparent from the sampled
distributions.
The extracted $\x^\ast$ can be further mapped to the radius of a
canonical NS using the $\x$-$\xi$ relation obtained within the IPAD-TOV
framework\,\cite{Shi2026}. The resulting estimates are
$R_{1.4}^\ast\approx11.73^{+1.08}_{-0.92}\,\rm{km}$ or
$11.92^{+1.76}_{-1.28}\,\rm{km}$ for the six or five family
analyses, respectively, as listed in TAB.\,\ref{tab:sc2_results}.
The former is slightly smaller than the latter because including
the comparatively stiff $c_\omega=0$ family shifts the convergence
of the family relations toward a smaller inferred radius.
In addition, these radii are somewhat smaller than those more directly populated
by the original M-R ensembles, see FIG.\,\ref{fig:MR}. Thus, the radius estimates provide
another manifestation of the same feature: the common IPAD-TOV
scaling can infer stellar properties from the convergence of the
family relations rather than simply reflecting the distributions of
the input EOS samples.

The combined samples of all six EOS families also exhibit a relatively
tight linear relation between $\Theta$ and $s_{\rm{c}}^2$, with
$\rm R^2\approx0.95$, as shown in FIG.\,\ref{fig:sss_scaling}.
This is particularly noteworthy because the combined fit does not
distinguish among the individual EOS families. Moreover, the common
estimates obtained from both the six family and five family analyses
lie close to this combined relation. The agreement demonstrates the
internal consistency of the extraction: The central SSS inferred from
the convergence of the individual family relations is also consistent
with the scaling followed by the full ensemble of EOS samples.
The combined scaling provides a complementary way to infer
$s_{\rm{c}}^2$, with the important difference that $\x$ is now an
input rather than being jointly extracted from the minimum dispersion
among the family relations. For an NS of a specified mass, given $\x$
and the corresponding $\Psi$, one can evaluate $\Theta(\x,\Psi)$ and
directly infer $s_{\rm{c}}^2$ from the combined fit. For canonical
$1.4\,M_\odot$ NSs, using $\Psi\approx2.66\pm0.40$ together with, e.g.,
$\x\approx0.173$ and $0.167$ gives
$s_{\rm{c}}^2\approx0.52\pm0.06$ and $0.50\pm0.06$, respectively.
These values agree closely with the corresponding common estimates,
providing a further demonstration of the consistency between the two
extraction procedures.
Adopting a smaller value of $\x\approx0.16$ gives $s_{\rm{c}}^2\approx0.47\pm0.06$, while a further reduction to $\x\approx0.14$, as suggested by the model calculations summarized in TAB.\,\ref{tab:family_intervals}, yields $s_{\rm{c}}^2\approx0.39\pm0.05$. 
Interestingly, a central SSS of order $0.5$ at densities characteristic of
canonical NS cores is broadly compatible with the agnostic-EOS and Bayesian
analyses of Refs.\,\cite{Altiparmak2022,EckerRezzolla2022,Pang2024}. Although these
studies employ different EOS constructions and physical assumptions, they
also favor a substantial stiffening of dense matter over the density range
relevant to NS interiors. In particular, Ref.\,\cite{Pang2024} demonstrated
that such behavior can arise naturally from a crossover to quarkyonic
matter. This comparison should nevertheless be regarded as qualitative,
since the inferred SSS profile and the location of its maximum depend on
the adopted model and constraints.

As summarized in TAB.\,\ref{tab:sc2_family_scaling}, the fitted
coefficients for the families with $c_\omega>0$ remain broadly close
to the ideal values $\alpha_j\approx1$ and $\beta_j\approx0$. This
supports the physical interpretation of $\Theta$ as an analytical
estimator of $s_{\rm c}^2$, although the present expression for
$\Theta$ is approximate and each family samples only a limited range
of the central quantities. The combined fit is also close to the ideal
relation, with $\alpha\approx0.945$ and
$\beta\approx-0.005$, indicating that much of the
family-dependent variation averages out when the full ensemble is
considered. The main exception is the $c_\omega=0$ family, for which
$\alpha\approx0.700$ and $\beta\approx0.135$ differ
appreciably from unity and zero, respectively. Correspondingly,
FIG.\,\ref{fig:sss_scaling} shows that this family is systematically
displaced from the other families. As discussed above, in the absence
of the $\omega$ self-interaction, the $c_\omega=0$ family produces a
particularly stiff SNM EOS, whose pressure band partially exceeds the
HIC flow constraint shown in FIG.\,\ref{fig:P_SNM}. Its larger
departure from the ideal coefficients is therefore consistent with a
more strongly modified mapping between the approximate $\Theta$ and
the numerically calculated $s_{\rm c}^2$. To examine how this distinct
family affects the common estimate, we perform the extraction both
with all six families and with the five families having $c_\omega>0$.
The comparison provides a useful measure of the sensitivity of the
extracted central quantities to the choice of EOS families.

The extracted $s_{\rm c}^2$ for canonical NSs is also consistent with
the recent analysis based on the $\Lambda$-$D(\x,\Psi)$ correlation,
where $D(\x,\Psi)$ is the intrinsic tidal response function\,\cite{Shi2026,Stroud2026}. Adopting
$190\lesssim\Lambda_{1.4}\lesssim580$, the values of $\x$ being about $0.133\lesssim\x\lesssim0.185$ inferred
from the $\Lambda$-$D(\x,\Psi)$ correlation, when directly inserted
into the estimator $\Theta(\x,\Psi)$, give approximately
$0.39\lesssim s_{\rm c}^2\equiv \Theta(\x,\Psi)\lesssim0.62$. Here, $\Theta(\x,\Psi)$ is defined as the approximate physical
expression for $s_{\rm c}^2$ derived from IPAD-TOV, namely
Eq.\,(\ref{eq:sc2_IPAD_corrected}), hence the notation ``$\equiv$''. The agreement provides an
independent consistency check on the present extraction and further
supports the use of $\Theta(\x,\Psi)$ as a direct probe of the central
SSS. Indeed, as shown in TAB.\,\ref{tab:sc2_family_scaling}, the
overall values of $\alpha$ and $\beta$ obtained using all EOS families
are very close to unity and zero, respectively, demonstrating that
$s_{\rm c}^2$ is well approximated directly by $\Theta(\x,\Psi)$.

The present analysis focuses on canonical $1.4\,M_\odot$ NSs, but the
same approach can be extended to estimate the central SSS
of NSs with other masses. For more massive NSs, the central EOS
parameter $\x$ generally increases, making higher order terms in $\x$
in the approximation to $\Theta(\x,\Psi)$ of Eq.\,(\ref{eq:sc2_IPAD_corrected}) potentially more important.
A systematic extension of $\Theta$ to higher orders in $\x$ would
therefore allow the accuracy of the scaling to be improved for more
massive NSs and its applicability to be examined over a broader mass
range. We leave such an extension for future study.

\subsection{Near Center Behavior and Interior Sound Speed Peaks}
\label{subs_peak}

\renewcommand*\figurename{\small FIG.}
\begin{figure}[h!]
    \centering
    \includegraphics[width=0.84\columnwidth]{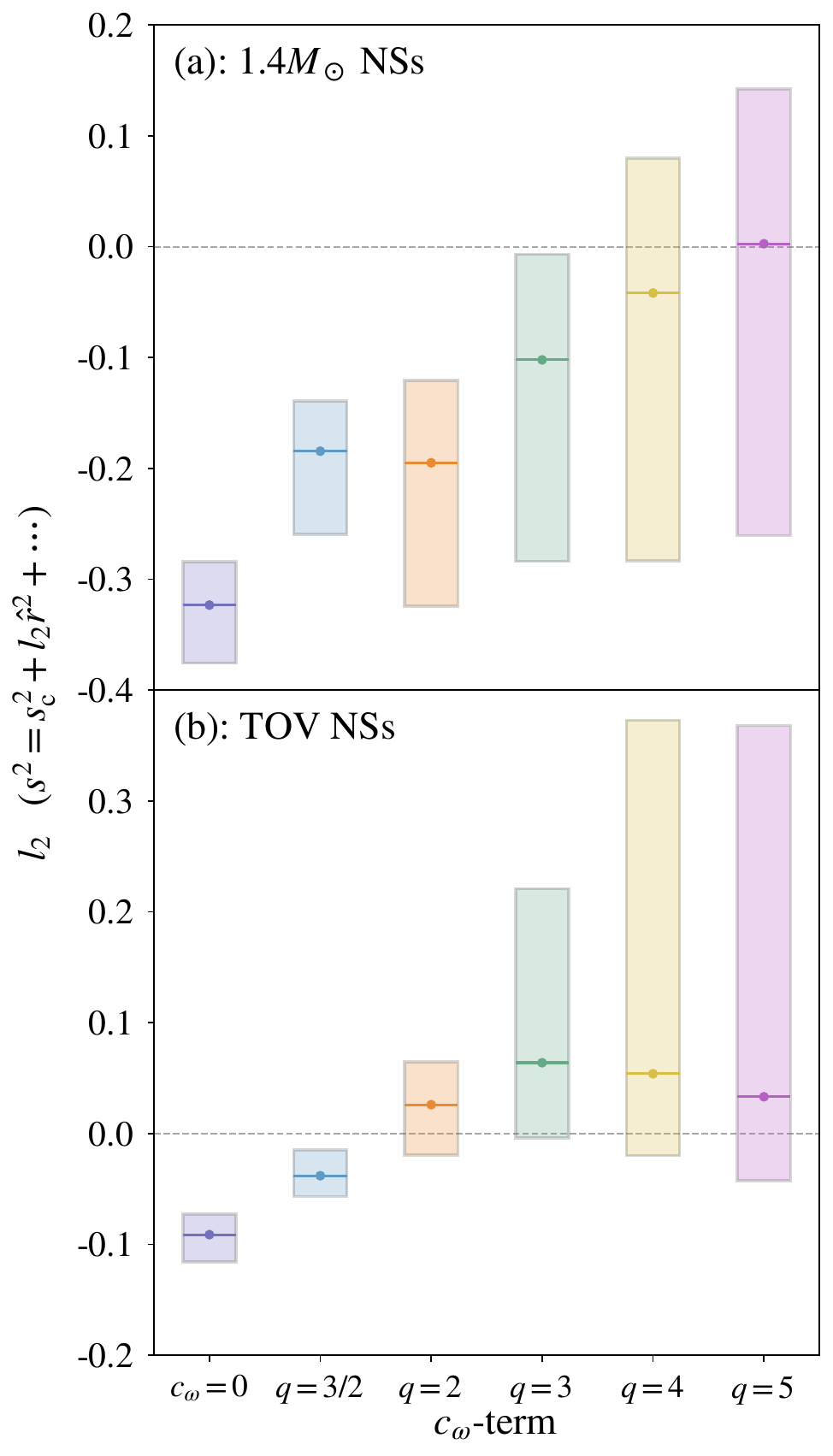}
    \caption{(Color online). Quadratic coefficient $l_2$ in the
    near center expansion
    $s^2(\hr)\approx s_{\rm{c}}^2+l_2\widehat{r}^{2}
    +\mathcal{O}(\widehat{r}^{4})$,
    where $\widehat{r}$ is the dimensionless radius.
    Colored rectangles span the $95\%$ intervals within
    each RMF family, and short horizontal solid lines
    mark the medians.}
    \label{fig:l2}
\end{figure}

\renewcommand*\figurename{\small FIG.}
\begin{figure}[h!]
    \centering
    \includegraphics[width=0.84\columnwidth]{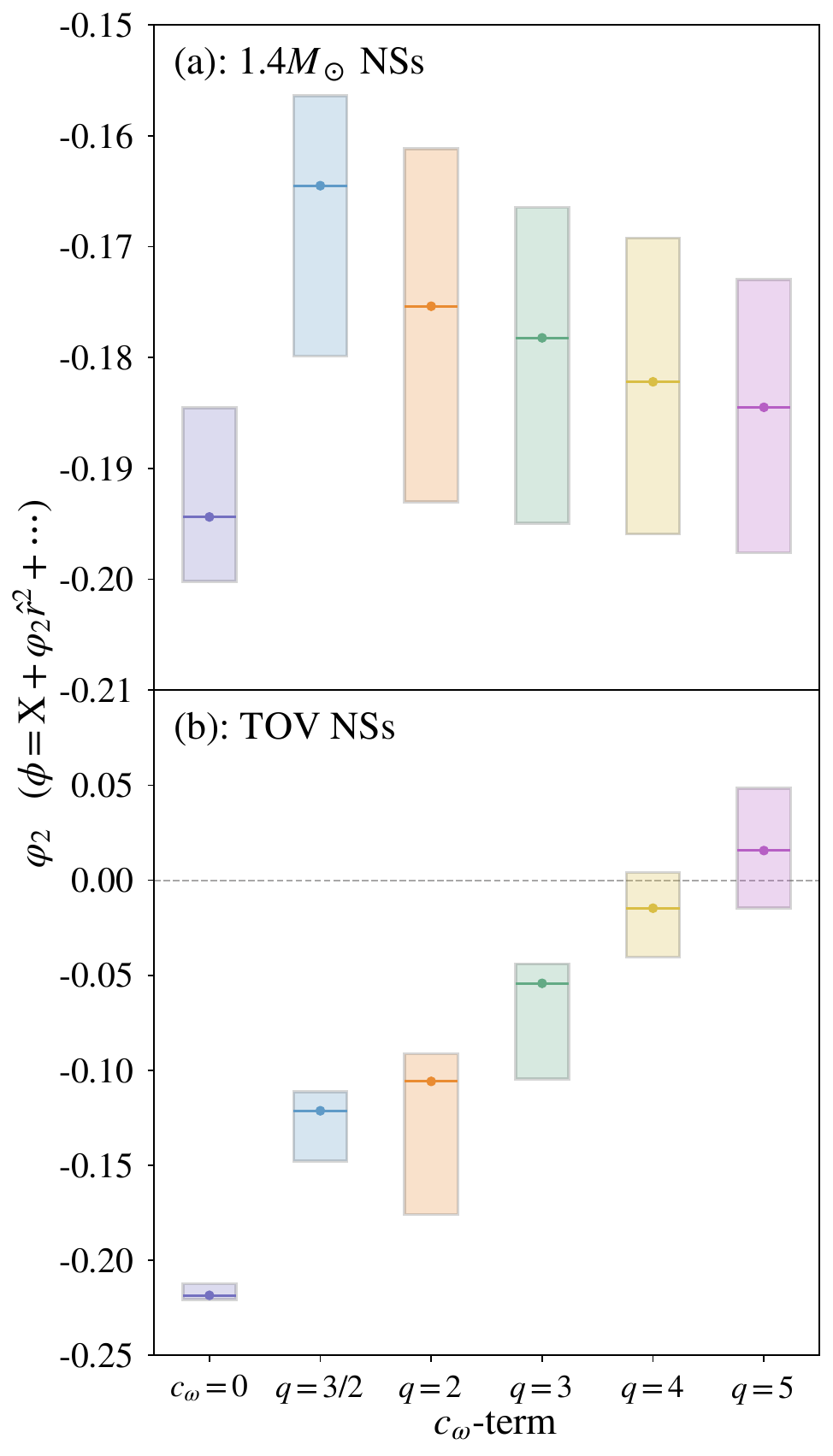}
    \caption{(Color online). The same as FIG.\,\ref{fig:l2} but for the quadratic coefficient $\varphi_2$
    in the near center expansion
    $\phi(\hr)\approx \x+\varphi_2\widehat{r}^{2}
    +\mathcal{O}(\widehat{r}^{4})$.}
    \label{fig:phi2}
\end{figure}

The near center expansion underlying the IPAD-TOV analysis also
provides a direct way to characterize how the EOS properties evolve
away from the stellar center. In particular, the SSS and the EOS
parameter $\phi\equiv P/\varepsilon$ can be expanded in terms of the
dimensionless radius $\widehat{r}$ as\,\cite{CaiLi2024SSS}
\begin{equation}
    s^2(\hr)\approx s_{\rm{c}}^2+l_2\widehat{r}^{2}
    +\mathcal{O}(\widehat{r}^{4}),
\end{equation}
and\,\cite{CaiLiMa2026phi}
\begin{equation}
    \phi(\hr)\approx\x+\varphi_2\widehat{r}^{2}
    +\mathcal{O}(\widehat{r}^{4}),
\end{equation}
respectively. Here, $l_2=2s_{\rm c}^2/b_2\cdot(b_4-s_{\rm c}^2a_4)$\,\cite{CaiLi2024SSS} and $\varphi_2=b_2(1-\x/s_{\rm c}^2)$\,\cite{CaiLiMa2026phi} obtained from Eq.\,(\ref{P-exp}) and Eq.\,(\ref{eps-exp}) characterize the leading
radial variations of $s^2$ and $\phi$ away from the stellar center,
while odd order terms vanish owing to center symmetry\,\cite{CaiLi2025Review}. A positive quadratic coefficient therefore
indicates an initial outward increase of the corresponding quantity.
For the SSS, an initial increase (corresponding to $l_2>0\leftrightarrow a_4>b_4/s_{\rm{c}}^2$) followed by its eventual decrease
toward the stellar surface implies the presence of an interior sound
speed peak\,\cite{CaiLi2024SSS}.

The distributions of $l_2$ for canonical and TOV NSs are shown in
FIG.\,\ref{fig:l2}. For canonical NSs, the $95\%$ intervals of $l_2$
extend into the positive region only for the $q=4$ and $q=5$
families, as shown in panel (a). Thus, within the present ensembles,
an interior sound-speed peak becomes possible for canonical NSs only
at relatively large $q$, roughly for $q\gtrsim4$. This is consistent with the behavior found
in Subsection \ref{subs_NS}, where increasing $q$ shifts the
sound-speed peak toward lower densities and hence into the density
range probed by canonical NSs, as shown in
FIG.\,\ref{fig:sss_band}.
For example, for the $q=5$ EOS family, the median peak position is approximately $\rho_{\rm{pk}}/\rho_0\approx2.50$ which is very near its central energy density (see TAB.\,\ref{tab:family_intervals}). For TOV NSs, whose centers probe substantially higher densities, the
$95\%$ intervals of $l_2$ extend into the positive region already for
$q\geq2$, as shown in panel (b). Thus, the higher densities reached
in these massive stars allow the condition for an interior
sound-speed peak to be realized over a broader range of the RMF
families. This mass dependence is qualitatively consistent with the
scale-independent analysis of Ref.\,\cite{EckerRezzolla2022}, in
which canonical NSs tend to exhibit no interior sound-speed peak,
whereas such a peak becomes more prevalent as the NS mass approaches
$M_{\rm{TOV}}$.

Beyond diagnosing the presence of an interior peak, the coefficient
$l_2$ also helps characterize the difference
$\Delta s^2\equiv s_{\rm{pk}}^2-s_{\rm{c}}^2$ between
the peak and central SSSs for TOV NSs. Using the quartic expansion
$s^2(\hr)\approx s_{\rm{c}}^2+l_2\hr^2
+l_4\hr^4+\cdots$ in 
$\hr$\,\cite{CaiLi2024SSS}, the peak condition gives
$\hr_{\rm{pk}}^2\approx-l_2/2l_4$, and hence
\begin{equation}
\Delta s^2\approx\frac{1}{2}l_2\hr_{\rm{pk}}^2.
\end{equation}
The central density
expansion further implies approximately
$\hr_{\rm{pk}}^2\sim
1-\rho_{\rm{pk}}/\rho_{\rm{c}}$\,\cite{CaiLiZhang2023PRD}.
Since the $q=2$-$5$ families have comparable positive median values
of $l_2$, they exhibit similar initial growth of the SSS away from
the center, whereas the total increase accumulated before reaching
the peak is mainly governed by its effective separation from the
center. In particular, as shown in TAB.\,\ref{tab:family_intervals},
$\rho_{\rm{pk}}/\rho_{\rm{c}}$ decreases systematically
with increasing $q$, indicating an increasing separation in density
between the stellar center and the SSS peak. 
This provides an intuitive explanation for the
corresponding growth of $\Delta s^2$ with $q$ shown in
FIG.\,\ref{fig:sss_band}.
For example, the relatively large
$\rho_{\rm{pk}}/\rho_{\rm{c}}$ for the $q=2$ family
places the peak closer to the central density and consequently leads
to a comparatively shallow peak.
Thus, $l_2$ measures the initial radial increase of the SSS near the
stellar center, while the relative peak density determines the
effective distance over which this increase can accumulate before the
peak is reached. Together, these two quantities control the prominence
of the interior SSS peak: A larger positive $l_2$ produces a more rapid
initial increase, whereas a smaller
$\rho_{\rm{pk}}/\rho_{\rm{c}}$ places the peak farther
from the center and allows this increase to persist over a larger
portion of the stellar core. Although this quartic-order (in $\hr$) argument captures the leading trend,
the precise location and magnitude of the peak may also receive
non-negligible contributions from $l_6$ and still higher order
coefficients, particularly for the larger-$q$ families.

Similarly, the corresponding behavior of $\phi=P/\varepsilon$ provides
additional information on the near center evolution of the EOS.
As shown in panel (a) of FIG.\,\ref{fig:phi2}, $\varphi_2$ remains
negative throughout the displayed intervals for canonical NSs,
indicating that $\phi$ initially decreases away from the center.
For TOV NSs, however, the distributions extend into the positive
region for the $q=4$ and $q=5$ families, as shown in panel (b) of
FIG.\,\ref{fig:phi2}. The present RMF calculations therefore allow
$\phi$ to increase initially away from the center for sufficiently
large $q$, followed by a decrease at larger radii. It is worth noting,
however, that the intrinsic IPAD-TOV analysis indicates that $\phi$
should decrease monotonically from the stellar center\,\cite{CaiLiMa2026phi} due to the basic property $\x/s_{\rm{c}}^2<1$ (as shown by Eq.\,(\ref{eq:sc2_IPAD_leading}) or Eq.\,(\ref{eq:sc2_IPAD_corrected})) and $b_2<0$. The positive values of $\varphi_2$ found for some large-$q$ EOSs
should therefore be interpreted as signaling incompatibility with the
intrinsic properties of the TOV equations, rather than as a generic feature of stellar
structure, and consequently disfavor these models.

These results provide a stellar manifestation of the density
dependence discussed in Subsection \ref{subs_NS}. As $q$ increases,
the nonlinear $\omega$ self-interaction becomes important at lower
densities and increasingly suppresses the growth of the vector field,
allowing the SSS to turn over within the density range probed by NS
interiors. Consequently, an interior sound speed peak can occur in
lower mass NSs for larger $q$, whereas for smaller $q$ such behavior,
when present, is reached only in more massive configurations. The
near center coefficients $l_2$ and $\varphi_2$ thus provide a useful
characterization of how the density dependence of the EOS manifests
itself within NS interiors, with $l_2$ directly connecting the
underlying high density behavior to the possible emergence of an
interior sound speed peak.

\section{Summary and Outlook}\label{SEC_5}

In this work, we constructed six families of generalized RMF models
with different $\omega$-meson self-interactions and generated 1000
EOSs for each family under the same nuclear and astrophysical
constraints. We find that the nonlinear $\omega$ interaction plays a
key role in controlling the high density vector repulsion and the
resulting SSS. At asymptotically high densities, both the SSS $s^2$ and the EOS-parameter $\phi=P/\varepsilon$ approach unity for $c_\omega=0$,
$1/(2q-1)$ for $1<q<2$, and $1/3$ for $q\geq2$, demonstrating
qualitatively different limiting behaviors among the EOS families.
At finite densities relevant to NS interiors, increasing $q$
suppresses the growth of the vector field at progressively lower
densities and favors the emergence of an interior sound speed peak.
This provides a purely nucleonic mechanism for generating
nonmonotonic sound speed behavior inside NSs, without requiring a
change of the underlying degrees of freedom. The central expansions
further show that the occurrence and radial structure of this peak
depend on both $q$ and the stellar mass: For larger $q$, the turnover
can already occur in lower-mass NSs, whereas for smaller $q$ it is
shifted toward more massive configurations. These results illustrate
how qualitatively different sound speed profiles can emerge from the
density dependence of the nuclear interaction even among EOSs
satisfying the same nuclear and astrophysical constraints.

Despite satisfying the same adopted constraints, the six EOS families
predict appreciably different central properties, particularly
$s_{\rm{c}}^2$ and $\x$. This demonstrates that present nuclear and
astrophysical constraints alone do not uniquely determine the high
density EOS or NS core properties. To reduce this model dependence,
we used the IPAD-TOV framework to derive the improved estimator
$\Theta(\x,\Psi)$ for the central SSS directly from the stellar
structure equations, without specifying a particular EOS. For
canonical $1.4\,M_\odot$ NSs, $\Theta$ exhibits tight linear
correlations with the numerical $s_{\rm{c}}^2$ within all six
families, with $\rm R^2\approx0.94$-$0.99$. Remarkably, these
correlations persist despite the substantially different high density
behaviors and sound speed profiles of the underlying EOS families.
The residual differences among the family scaling relations encode
their remaining model dependence, while their convergence identifies
a region in which the inferred central SSS becomes considerably less
sensitive to the particular EOS family. This turns the model diversity
itself into useful information and provides the basis for extracting
common properties of the NS core from their shared stellar structure
behavior.

Minimizing the dispersion among the family scaling relations gives
$s_{\rm{c}}^{2,\ast}\approx0.523$ and $\x^\ast\approx0.173$ using
all six families, while the corresponding five-family analysis gives
$s_{\rm{c}}^{2,\ast}\approx0.496$ and $\x^\ast\approx0.167$.
Through the relation between $\x$ and the stellar radius, these
estimates further correspond to
$R_{1.4}^\ast\approx11.73\,\rm{km}$ and
$11.92\,\rm{km}$, respectively. Remarkably, despite their distinctly
different high density asymptotic behaviors, the EOS families converge
on a rather common picture of the canonical NS core, with
$s_{\rm{c}}^2\sim0.5$, $\x\sim0.17$, and
$R_{1.4}\sim12\,\rm{km}$. The inferred central SSS is appreciably
above the conformal value $1/3$, suggesting a relatively stiff inner
core at densities characteristic of canonical NSs. The combined
samples also follow a relatively tight common
$\Theta$-$s_{\rm{c}}^2$ scaling with $\rm R^2\approx0.95$,
providing an independent consistency check of the extraction.

The inferred values remain conditional on the EOS families considered
and the accuracy of the scaling approximation. Nevertheless, the
convergence among EOS families with very different asymptotic behaviors
raises the interesting possibility that canonical NSs probe a density
regime in which the central SSS is less sensitive to the ultimate
high density limit of the EOS. Extending the analysis beyond RMF
models, particularly to microscopic EOSs and EOSs with phase
transitions or new degrees of freedom, will provide an important test
of this possibility. Applications to NSs over a broader mass range,
together with higher order corrections in $\x$, may further reveal
how this common behavior evolves and eventually breaks down as
progressively higher densities are probed.

\section*{Acknowledgment} 
This work was supported in part by the National Natural Science Foundation of China under contract Nos. 12688101 and 12547102, and the STCSM under Grant Nos.~23590780100 and 23JC1400200. 

\appendix
\renewcommand{\theequation}{\Alph{section}.\arabic{equation}}
\setcounter{equation}{0}

\section{Analytical Expression for the SSS in $\beta$-equilibrium NS Matter}
\label{app:sound_speed}

We derive the analytical expression for the speed of sound squared
along the cold, charge neutral $\beta$-equilibrium EOS. Throughout
this appendix, $P$ and $\varepsilon$ include the lepton contributions,
and primes denote total derivatives with respect to $\rho$ along the
equilibrium sequence. We first establish the thermodynamic relation
that connects the SSS to the density response of the neutron chemical
potential. Charge neutrality,
$\rho_\rm p=\rho_\rm e+\rho_\mu$, together with chemical equilibrium,
$\mu_\rm n=\mu_\rm p+\mu_\rm e$ and $\mu_\mu=\mu_\rm e$ when muons
are present, gives
\begin{align}
\d\varepsilon
&=\sum_i\mu_i\d\rho_i\notag\\
&=\mu_{\rm{n}}\d\rho_{\rm{n}}
+\mu_{\rm{p}}\d\rho_{\rm{p}}
+\mu_{\rm{e}}\d\rho_{\rm{e}}
+\mu_\mu\d\rho_\mu \notag\\
&=\mu_{\rm{n}}(\d\rho_{\rm{n}}+\d\rho_{\rm{p}})
=\mu_{\rm{n}}\d\rho,
\end{align}
while the pressure can be written as
\begin{align}
P
&=\sum_i\rho_i\mu_i-\varepsilon\notag\\
&=\rho_{\rm{n}}\mu_{\rm{n}}
+\rho_{\rm{p}}\mu_{\rm{p}}
+\rho_{\rm{e}}\mu_{\rm{e}}
+\rho_\mu\mu_\mu-\varepsilon \notag\\
&=(\rho_{\rm{n}}+\rho_{\rm{p}})\mu_{\rm{n}}
-\varepsilon
=\rho\mu_{\rm{n}}-\varepsilon.
\end{align}
It then follows that
\begin{equation}
    \d P=\rho\d\mu_{\rm n}+\mu_{\rm n}\d\rho-\d\varepsilon
    =\rho\d\mu_{\rm n},
\end{equation}
and consequently
\begin{equation}
    s^2=\frac{\d P}{\d\varepsilon}
    =\frac{\rho}{\mu_{\rm n}}
    \frac{\d\mu_{\rm n}}{\d\rho}=\frac{\d\ln\mu_{\rm n}}{\d\ln\rho}.
    \label{eq:app_s2_thermo}
\end{equation}
Thus, the problem reduces to determining the density response
$\mu_{\rm n}'$ along the $\beta$-equilibrium sequence.

For this purpose, we use the scaled mean fields
\begin{align}
    S_\sigma=&g_\sigma\overline{\sigma},\\
    S_\delta=&g_\delta\overline{\delta}^{(3)},\\
    V_\omega=&g_\omega\overline{\omega}_0,\\
    V_\rho=&g_\rho\overline{\rho}_0^{(3)}.
\end{align}
The nucleon chemical potentials are then
\begin{equation}
    \mu_J=\varepsilon_{\rm F}^{*,J}
    +V_\omega-\tau_3^J V_\rho,
    ~~ J={\rm{n,p}}.
\end{equation}
To separate the explicit density dependence from that induced by
changes in the proton fraction, we define
\begin{equation}
    \zeta\equiv\frac{\d\rho_{\rm p}}{\d\rho},
    ~~
    u_J\equiv\frac{M_J^*}{\varepsilon_{\rm F}^{*,J}},
    ~~
    F_J\equiv
    \frac{k_{\rm F}^{J,2}}
    {3\rho_J\varepsilon_{\rm F}^{*,J}}.
\end{equation}
Using $(k_{\rm F}^J)'=k_{\rm F}^J\rho_J'/3\rho_J$, we obtain
\begin{equation}
    \left(\varepsilon_{\rm F}^{*,J}\right)'
    =\frac{k_{\rm F}^J(k_{\rm F}^J)'
    +M_J^*(M_J^*)'}{\varepsilon_{\rm F}^{*,J}}
    =F_J\rho_J'
    -u_J\left(S_\sigma'-\tau_3^J S_\delta'\right),
\end{equation}
where $M_J^*=M_{\rm N}-S_\sigma+\tau_3^J S_\delta$,
with $\tau_3^{\rm n}=+1$ and $\tau_3^{\rm p}=-1$.
Since $\rho_{\rm n}'=1-\zeta$ and $\rho_{\rm p}'=\zeta$,
the corresponding derivatives of the chemical potentials become
\begin{align}
    \mu_{\rm n}'&=
    F_{\rm n}(1-\zeta)
    -u_{\rm n}S_\sigma'
    +u_{\rm n}S_\delta'
    +V_\omega'-V_\rho',
    \label{eq:app_mun_prime}\\
    \mu_{\rm p}'&=
    F_{\rm p}\zeta
    -u_{\rm p}S_\sigma'
    -u_{\rm p}S_\delta'
    +V_\omega'+V_\rho'.
    \label{eq:app_mup_prime}
\end{align}
Eq.\,\eqref{eq:app_mun_prime} shows that evaluating the SSS
requires the density responses of the four meson mean fields, as well
as the composition response $\zeta$. We determine the field responses
first.

The scalar density satisfies the differential relation
\begin{equation}
    \d\rho_J^{\rm s}
    =u_J\d\rho_J+A_J\d M_J^*,
    ~~
    A_J=3\left(
    \frac{\rho_J^{\rm s}}{M_J^*}
    -\frac{\rho_J}{\varepsilon_{\rm F}^{*,J}}
    \right).
    \label{eq:app_scalar_response}
\end{equation}
Using Eq.\,\eqref{eq:app_scalar_response}, the derivatives of the
scalar source terms are
\begin{align}
    (\rho_{\rm n}^{\rm s}+\rho_{\rm p}^{\rm s})'
    =&
    u_{\rm n}(1-\zeta)+u_{\rm p}\zeta\notag\\
&    -(A_{\rm n}+A_{\rm p})S_\sigma'
    +(A_{\rm n}-A_{\rm p})S_\delta',
    \\
    (\rho_{\rm p}^{\rm s}-\rho_{\rm n}^{\rm s})'
 =   &
    u_{\rm p}\zeta-u_{\rm n}(1-\zeta)\notag\\
&    +(A_{\rm n}-A_{\rm p})S_\sigma'
    -(A_{\rm n}+A_{\rm p})S_\delta'.
\end{align}
The corresponding vector source derivatives are simply
\begin{equation}
    (\rho_{\rm n}+\rho_{\rm p})'=1,
    ~~
    (\rho_{\rm p}-\rho_{\rm n})'=2\zeta-1.
\end{equation}
Differentiating Eqs.\,\eqref{eq:seqm}-\eqref{eq:deqm}, substituting
the above source derivatives, and collecting the field responses on
the left-hand side lead to two coupled linear systems,
\begin{align}
    \begin{pmatrix}
        \mathcal Z_\sigma & \mathcal Z_{\sigma\delta}\\
        \mathcal Z_{\sigma\delta} & \mathcal Z_\delta
    \end{pmatrix}
    \begin{pmatrix}
        S_\sigma'\\S_\delta'
    \end{pmatrix}
    &=
    \begin{pmatrix}
        u_{\rm n}+(u_{\rm p}-u_{\rm n})\zeta\\
        -u_{\rm n}+(u_{\rm p}+u_{\rm n})\zeta
    \end{pmatrix},
    \label{eq:app_scalar_matrix}\\
    \begin{pmatrix}
        \mathcal Z_\omega & \mathcal Z_{\omega\rho}\\
        \mathcal Z_{\omega\rho} & \mathcal Z_\rho
    \end{pmatrix}
    \begin{pmatrix}
        V_\omega'\\V_\rho'
    \end{pmatrix}
    &=
    \begin{pmatrix}
        1\\2\zeta-1
    \end{pmatrix}.
    \label{eq:app_vector_matrix}
\end{align}
Here, the response coefficients are
\begin{align}
    \mathcal Z_\sigma
    &=\frac{m_\sigma^2}{g_\sigma^2}
    +2b_\sigma M_{\rm N}S_\sigma
    +3c_\sigma S_\sigma^2
    -C_{\delta\sigma}S_\delta^2
    +A_{\rm p}+A_{\rm n},\\
    \mathcal Z_\delta
    &=\frac{m_\delta^2}{g_\delta^2}
    -C_{\delta\sigma}S_\sigma^2
    +A_{\rm p}+A_{\rm n},\\
    \mathcal Z_{\sigma\delta}
    &=A_{\rm p}-A_{\rm n}
    -2C_{\delta\sigma}S_\sigma S_\delta,\\
    \mathcal Z_\omega
    &=\frac{m_\omega^2}{g_\omega^2}
    +\frac{q(2q-1)c_\omega}
    {2m_\omega^{2q-4}}V_\omega^{2q-2}
    +\Lambda_{\rm V}V_\rho^2,\\
    \mathcal Z_\rho
    &=\frac{m_\rho^2}{g_\rho^2}
    +\Lambda_{\rm V}V_\omega^2,~~
    \mathcal Z_{\omega\rho}
    =2\Lambda_{\rm V}V_\omega V_\rho.
\end{align}

Introducing the determinants
\begin{equation}
       \Delta_{\rm S}
    =\mathcal Z_\sigma\mathcal Z_\delta
    -\mathcal Z_{\sigma\delta}^2,~~
       \Delta_{\rm V}
    =\mathcal Z_\omega\mathcal Z_\rho
    -\mathcal Z_{\omega\rho}^2,
\end{equation}
the solutions of Eqs.\,\eqref{eq:app_scalar_matrix} and
\eqref{eq:app_vector_matrix}, for non-singular response matrices, can
be written as
\begin{align}
    S_\sigma'&=S_{\sigma,0}+S_{\sigma,1}\zeta,~~
    S_\delta'=S_{\delta,0}+S_{\delta,1}\zeta,\label{eq:app_field_response-1}\\
    V_\omega'&=V_{\omega,0}+V_{\omega,1}\zeta,~~
    V_\rho'=V_{\rho,0}+V_{\rho,1}\zeta,\label{eq:app_field_response-2}
\end{align}
where
\begin{align}
    S_{\sigma,0}
    &=\frac{
    u_{\rm n}(\mathcal Z_\delta+\mathcal Z_{\sigma\delta})
    }{    \Delta_{\rm S}},\\
    S_{\delta,0}
    &=-\frac{
    u_{\rm n}(\mathcal Z_\sigma+\mathcal Z_{\sigma\delta})
    }{    \Delta_{\rm S}},\\
    S_{\sigma,1}
    &=\frac{
    \mathcal Z_\delta(u_{\rm p}-u_{\rm n})
    -\mathcal Z_{\sigma\delta}(u_{\rm p}+u_{\rm n})
    }{    \Delta_{\rm S}},\\
    S_{\delta,1}
    &=\frac{
    \mathcal Z_\sigma(u_{\rm p}+u_{\rm n})
    -\mathcal Z_{\sigma\delta}(u_{\rm p}-u_{\rm n})
    }{    \Delta_{\rm S}},\\
    V_{\omega,0}
    &=\frac{\mathcal Z_\rho+\mathcal Z_{\omega\rho}}
    {\Delta_{\rm V}},
    ~~
    V_{\omega,1}
    =-\frac{2\mathcal Z_{\omega\rho}}{\Delta_{\rm V}},\\
    V_{\rho,0}
    &=-\frac{\mathcal Z_\omega+\mathcal Z_{\omega\rho}}
    {\Delta_{\rm V}},
   ~~
    V_{\rho,1}
    =\frac{2\mathcal Z_\omega}{\Delta_{\rm V}}.
\end{align}
The numerical subscripts $0$ and $1$ label the parts independent of
and proportional to $\zeta$, respectively. Thus, once $\zeta$ is
known, all four mean field responses are determined.

It remains to determine the composition response $\zeta$. We do so by
differentiating the charge neutrality and $\beta$-equilibrium
conditions. For ideal leptons at zero temperature,
\begin{equation}
    \mu_{\rm e}'=\ell\zeta,
    ~~
    \ell=
    \left[
    \frac{\d(\rho_{\rm e}+\rho_\mu)}{\d\mu_{\rm e}}
    \right]^{-1}
    =\frac{\pi^2}
    {\mu_{\rm e}(k_{\rm F}^{\rm e}+k_{\rm F}^{\mu})},
    \label{eq:app_lepton_response}
\end{equation}
where
\begin{equation}
    k_{\rm F}^{\rm e}=\sqrt{\mu_{\rm e}^2-m_{\rm e}^2},
    ~~
    k_{\rm F}^{\mu}=
    \begin{cases}
        0, & \mu_{\rm e}\leq m_\mu,\\
        \sqrt{\mu_{\rm e}^2-m_\mu^2}, & \mu_{\rm e}>m_\mu.
    \end{cases}
\end{equation}
Using Eqs.\,\eqref{eq:app_mun_prime} and \eqref{eq:app_mup_prime},
together with $\mu_{\rm n}'-\mu_{\rm p}'=\mu_{\rm e}'$, gives
\begin{equation}
    0=F_{\rm n}
    -(F_{\rm n}+F_{\rm p}+\ell)\zeta
    +(u_{\rm p}-u_{\rm n})S_\sigma'
    +(u_{\rm p}+u_{\rm n})S_\delta'
    -2V_\rho'.
\end{equation}
Substituting the field responses in
Eqs.\,\eqref{eq:app_field_response-1} and (\ref{eq:app_field_response-2}), the composition response is
therefore
\begin{equation}
    \zeta=
    \frac{
        F_{\rm n}
        +(u_{\rm p}-u_{\rm n})S_{\sigma,0}
         +(u_{\rm p}+u_{\rm n})S_{\delta,0}
        -2V_{\rho,0}
    }{
        F_{\rm n}+F_{\rm p}+\ell
        -(u_{\rm p}-u_{\rm n})S_{\sigma,1}
         -(u_{\rm p}+u_{\rm n})S_{\delta,1}
        +2V_{\rho,1}
    }.
    \label{eq:app_composition_response}
\end{equation}

We now have all the ingredients required to evaluate
Eq.\,\eqref{eq:app_s2_thermo}. Combining
Eqs.\,\eqref{eq:app_s2_thermo}, \eqref{eq:app_mun_prime}, and
\eqref{eq:app_composition_response}, we finally obtain
\begin{align}
    s^2=\frac{\rho}{\mu_{\rm n}}
    \bigg[
    &F_{\rm n}
    -u_{\rm n}S_{\sigma,0}
    +u_{\rm n}S_{\delta,0}
    +V_{\omega,0}-V_{\rho,0}\notag\\
    &+\left(
    -F_{\rm n}
    -u_{\rm n}S_{\sigma,1}
    +u_{\rm n}S_{\delta,1}
    \right.
    \left.
    +V_{\omega,1}-V_{\rho,1}
    \right)\zeta
    \bigg].\label{eq:app_s2_closed}
\end{align}
All quantities entering Eq.\,\eqref{eq:app_s2_closed} are evaluated at
the same equilibrium state. For fixed model parameters, the baryon
density $\rho$ is the only independent input, while the mean fields,
particle fractions, composition response, and their density
dependences are determined self consistently by the field equations,
charge neutrality, and $\beta$-equilibrium. Eq.\,\eqref{eq:app_s2_closed} therefore provides a closed analytical
expression for the SSS along the cold $\beta$-equilibrium EOS used
throughout this work.

\section{Asymptotic Behavior of the Vector Mean Fields for $c_\omega=0$}
\label{app:vector_fields}

For the case of $c_\omega=0$, the vector field equations
(\ref{eq:omega_high}) and (\ref{eq:rho_high}) reduce to
\begin{align}
\rho
&=
\left(
\frac{m_\omega^2}{g_\omega^2}
+\Lambda_{\rm V}V_\rho^2
\right)V_\omega ,
\label{eq:app_omega}
\\
\rho\delta
&=
-\left(
\frac{m_\rho^2}{g_\rho^2}
+\Lambda_{\rm V}V_\omega^2
\right)V_\rho .
\label{eq:app_rho}
\end{align}
We first consider asymptotic solutions with
$V_\omega\sim\rho^a$ and $a>0$, for which
$\delta\sim\rho^0$. Writing further $V_\rho\sim\rho^b$,
a dominant balance analysis of Eqs.\,(\ref{eq:app_omega}) and
(\ref{eq:app_rho}) gives two candidate branches,
\begin{equation}
(a,b)=(1,-1),~~ (1/3,1/3).
\end{equation}
As shown below, the second branch requires values of
$\Lambda_{\rm V}$ far outside the range relevant to the present ensembles.
We then consider separately the alternative possibility in which
$V_\omega$ decreases with density, for which $\delta$ no longer
approaches a constant asymptotically.

For the branch $(a,b)=(1/3,1/3)$, we write
$V_\omega=A\rho^{1/3}$ and
$V_\rho=-B\rho^{1/3}$ with $A,B>0$.
The leading terms in Eqs.\,(\ref{eq:app_omega}) and
(\ref{eq:app_rho}) yield
$1=\Lambda_{\rm V}AB^2$ and
$\delta=\Lambda_{\rm V}A^2B$, giving
$A=\delta B$ and
$B=(\Lambda_{\rm V}\delta)^{-1/3}$.
In $\beta$-equilibrium, the neutron-proton chemical potential
difference satisfies
\begin{equation}
\mu_{\rm e}
=
\mu_{\rm n}-\mu_{\rm p}
=
\varepsilon_{\rm F}^{*,{\rm n}}
-\varepsilon_{\rm F}^{*,{\rm p}}
-2V_\rho .
\end{equation}
For neutron-rich matter, $\delta>0$ and hence
$k_{\rm F}^{\rm n}>k_{\rm F}^{\rm p}$, so that
$\varepsilon_{\rm F}^{*,{\rm n}}
-\varepsilon_{\rm F}^{*,{\rm p}}\geq0$.
Since $V_\rho<0$, it follows that
\begin{equation}
\mu_{\rm e}\geq-2V_\rho
=
2\left(\Lambda_{\rm V}\delta\right)^{-1/3}\rho^{1/3}.
\label{eq:app_mue_lower}
\end{equation}
At asymptotically high densities, both electrons and muons become
ultra-relativistic. Using $\mu_{\rm e}=\mu_\mu$ together with charge
neutrality, one obtains
\begin{equation}
\rho_{\rm p}
=
\frac{1-\delta}{2}\rho
=
\rho_{\rm e}+\rho_\mu
=
\frac{2\mu_{\rm e}^3}{3\pi^2}
\leftrightarrow
\mu_{\rm e}^3
=
\frac{3\pi^2}{4}(1-\delta)\rho.
\label{eq:app_mue_CN}
\end{equation}
Combining Eqs.\,(\ref{eq:app_mue_lower}) and
(\ref{eq:app_mue_CN}) gives
\begin{equation}
\Lambda_{\rm V}
\geq
\frac{32}{3\pi^2\delta(1-\delta)}
\geq
\frac{128}{3\pi^2}
\approx4.32,
\label{eq:app_lambda_bound}
\end{equation}
where $\delta(1-\delta)\leq1/4$ has been used. Such a large
$\Lambda_{\rm V}$ is also incompatible with the empirical symmetry
energy around saturation. In SNM, $V_\rho=0$,
and for $c_\omega=0$ the $\omega$-field equation gives
$V_\omega=(g_\omega^2/m_\omega^2)\rho$. Since the $\rho$-meson
contribution to the symmetry energy is positive and its denominator
contains the additional positive term $m_\rho^2/g_\rho^2$, one obtains
the upper bound
\begin{equation}
E_{\rm{sym}}(\rho)
<
\frac{k_{\rm F}^2}{6E_{\rm F}^*}
+
\frac{1}
{2\Lambda_{\rm V}
(g_\omega^2/m_\omega^2)^2\rho}.
\label{eq:app_Esym_bound}
\end{equation}
For $\Lambda_{\rm V}\gtrsim4.32$, the potential contribution is
therefore strongly suppressed. Taking typical RMF values
$V_{\omega,0}=(g_\omega^2/m_\omega^2)\rho_0\approx250$-$300$\,MeV,
its upper bound at saturation is only about $1.5$-$2.3$\,MeV.
Together with the kinetic contribution, this gives typically
$E_{\rm{sym}}(\rho_0)\lesssim18$-$21$\,MeV, well below the empirical
range $29$-$35$\,MeV adopted in the present sampling. Consistently,
the sampled values of $\Lambda_{\rm V}$ are all well below unity,
typically $\lesssim\mathcal{O}(0.1)$. Hence, the branch
$V_\omega\sim V_\rho\sim\rho^{1/3}$ is not realized in the present
ensembles.

We next consider the alternative possibility in which $V_\omega$
decreases asymptotically with density. To determine the corresponding
density dependence self-consistently, let
$V_\omega\sim\rho^{-a}$ with $a>0$. In this limit,
$\Lambda_{\rm V}V_\omega^2\to0$, and hence the $\rho$-meson mass term
dominates Eq.\,(\ref{eq:app_rho}). Correspondingly, the $\rho$-meson
contribution to the symmetry energy scales as
$E_{\rm{sym}}^{\rho}(\rho)\sim\rho$.
Since this contribution dominates the symmetry energy asymptotically,
the $\beta$-equilibrium condition,
$\mu_{\rm n}-\mu_{\rm p}
\approx4E_{\rm{sym}}(\rho)\delta
=\mu_{\rm e}$,
together with $\mu_{\rm e}\sim\rho^{1/3}$, gives
\begin{equation}
\delta\sim\rho^{-2/3}.
\end{equation}
Eq.\,(\ref{eq:app_rho}) then implies
$V_\rho\sim\rho\delta\sim\rho^{1/3}$.
Substituting this scaling back into
Eq.\,(\ref{eq:app_omega}) further gives
\begin{equation}
V_\omega
\sim
{\rho}/{V_\rho^2}
\sim\rho^{1/3},
\end{equation}
which contradicts the assumed decreasing behavior
$V_\omega\sim\rho^{-a}$. Therefore, no self-consistent asymptotic
solution with decreasing $V_\omega$ exists.

Consequently, the physically relevant asymptotic branch is
$(a,b)=(1,-1)$, for which $\delta\sim\rho^0$.
Indeed, Eq.\,(\ref{eq:app_omega}) gives
\begin{equation}
\rho\approx
{m_\omega^2V_\omega}/{g_\omega^2}
\to
V_\omega\sim\rho,
\end{equation}
while $\Lambda_{\rm V}V_\omega^2\sim\rho^2$ dominates
Eq.\,(\ref{eq:app_rho}), yielding
\begin{equation}
V_\rho\approx
-\frac{\rho\delta}{\Lambda_{\rm V}V_\omega^2}
\sim\rho^{-1}.
\end{equation}
Thus, for $c_\omega=0$ and the physically relevant parameter sets
considered here, we have
$
V_\omega\sim\rho$ together with $
V_\rho\sim\rho^{-1}$.

\end{document}